\documentclass[12pt]{article}
\pdfoutput=1
\usepackage{epsfig,amssymb,amsmath,psfrag,multirow,epstopdf,color}
\usepackage{breqn,placeins,hyperref}

\numberwithin{equation}{section}

\newcommand{\be}{\begin{equation}}
\newcommand{\ee}{\end{equation}}
\newcommand{\bea}{\begin{eqnarray}}
\newcommand{\eea}{\end{eqnarray}}
\newcommand{\eqn}[1]{eq.~\eqref{#1}}

\def\sect#1{section~{\ref{#1}}}
\def\eqn#1{eq.~(\ref{#1})}

\def\eqns#1#2{eqs.~(\ref{#1}) and~(\ref{#2})}

\def\eqn#1{eq.~(\ref{#1})}
\def\eqns#1#2{eqs.~(\ref{#1}) and~(\ref{#2})}

\def\Li{{\rm Li}}

\def\cO{{\mathcal O}}

\def\ws{{w^\ast}}
\def\Su{{\mathcal{S}_u}}

\def\to{\rightarrow}
\def\lr{\leftrightarrow}

\def\e{\epsilon}
\newcommand{\cP}{{\cal P}}

\def\PhiTilde{{\tilde{\Phi}_6}}
\def\Omegauvw{{\Omega^{(2)}(u,v,w)}}
\def\Omegavwu{{\Omega^{(2)}(v,w,u)}}
\def\Omegawuv{{\Omega^{(2)}(w,u,v)}}

\def\Fuvw{{F_1(u,v,w)}}
\def\Fwuv{{F_1(w,u,v)}}
\def\Qep{Q_{\rm ep}}

\def\beq{\begin{equation}}
\def\eeq{\end{equation}}

\def\bsp#1\esp{\begin{split}#1\end{split}}

\newcommand{\dE}[1]{{D_\nu^#1 E_{\nu,n}}}

\newcommand{\hf}{{\tfrac{1}{2}}}
\newcommand{\thf}{{\tfrac{3}{2}}}
\newcommand{\psip}{\psi^{\prime}}
\newcommand{\psipp}{\psi^{\prime\prime}}

\newcommand{\psipppp}{\psi^{\prime\prime\prime\prime}}

\newcommand{\dnu}[0]{{D_{\nu}}}
\newcommand{\NeqFour}{{\cal N}=4}
\newcommand{\Ord}{{\cal O}}

\newfont{\scyr}{wncyr10 scaled 550}

\def\beq{\begin{equation}}
\def\eeq{\end{equation}}
 
\def\bsp#1\esp{\begin{split}#1\end{split}}

\begin{document}

\catcode`\@=11
\font\manfnt=manfnt
\def\Watchout{\@ifnextchar [{\W@tchout}{\W@tchout[1]}}
\def\W@tchout[#1]{{\manfnt\@tempcnta#1\relax%
  \@whilenum\@tempcnta>\z@\do{%
    \char"7F\hskip 0.3em\advance\@tempcnta\m@ne}}}
\let\foo\W@tchout
\def\dubious{\@ifnextchar[{\@dubious}{\@dubious[1]}}
\let\enddubious\endlist
\def\@dubious[#1]{%
  \setbox\@tempboxa\hbox{\@W@tchout#1}
  \@tempdima\wd\@tempboxa
  \list{}{\leftmargin\@tempdima}\item[\hbox to 0pt{\hss\@W@tchout#1}]}
\def\@W@tchout#1{\W@tchout[#1]}
\catcode`\@=12


\thispagestyle{empty}

\null\vskip-10pt \hfill
\begin{minipage}[t]{42mm}
SLAC--PUB--15970\\
\end{minipage}
\vspace{5mm}

\begingroup\centering
{\Large\bfseries\mathversion{bold}
Bootstrapping an NMHV amplitude through three loops\par}%
\vspace{7mm}

\begingroup\scshape\large
Lance~J.~Dixon$^{(1)}$ and Matt von Hippel$^{(2)}$\\
\endgroup
\vspace{5mm}
\begingroup\small
$^{(1)}$\emph{SLAC National Accelerator Laboratory,
Stanford University, Stanford, CA 94309, USA} \\
$^{(2)}$\emph{Simons Center for Geometry and Physics,
Stony Brook University,
Stony Brook NY 11794 }
\endgroup

\vspace{0.4cm}
\begingroup\small
E-mails:\\
{\tt lance@slac.stanford.edu},\ \ \ {\tt matthew.vonhippel@stonybrook.edu}
\endgroup
\vspace{0.7cm}

\textbf{Abstract}\vspace{5mm}\par
\begin{minipage}{14.7cm}
We extend the hexagon function bootstrap to the
next-to-maximally-helicity-violating (NMHV) configuration
for six-point scattering in planar $\NeqFour$ super-Yang-Mills theory
at three loops.  Constraints from the $\bar{Q}$ differential equation,
from the operator product expansion (OPE)
for Wilson loops with operator insertions, and from multi-Regge factorization,
lead to a unique answer for the three-loop ratio function.
The three-loop result also predicts additional terms in the OPE
expansion, as well as the behavior of NMHV amplitudes in
the multi-Regge limit at one higher logarithmic accuracy (NNLL) than was
used as input.  Both predictions are in agreement with recent results
from the flux-tube approach.  We also study the multi-particle
factorization of multi-loop amplitudes for the first time.  We find
that the function controlling this factorization is purely logarithmic
through three loops.  We show that a function $U$, which is closely
related to the parity-even part of the ratio function $V$, is
remarkably simple; only five of the nine possible final entries in its
symbol are non-vanishing.  We study the analytic and numerical behavior of
both the parity-even and parity-odd parts of the ratio function
on simple lines traversing the space of cross ratios $(u,v,w)$,
as well as on a few two-dimensional planes.  Finally, we present
an empirical formula for $V$ in terms of elements of the coproduct
of the six-gluon MHV remainder function $R_6$ at one higher loop,
which works through three loops for $V$ (four loops for $R_6$).
\end{minipage}\par
\endgroup

\newpage

\tableofcontents

\newpage

\section{Introduction}

The maximally supersymmetric gauge theory in four dimensions,
$\mathcal{N}=4$ super Yang-Mills theory (SYM),
has been a valuable proving ground
for scattering amplitudes research, especially in the planar limit
of a large number of colors. Over the past two decades, calculations
in planar $\mathcal{N}=4$ SYM have pushed further,
in terms of loops, legs, and general understanding, than they have in
other gauge theories~\cite{Unitarity,BRY,BDS,Goncharov2010jf,%
ArkaniHamed2010kv,ArkaniHamed2013jha}.
In doing so, they have also offered insight into efficient methods
for handling other gauge theories, as well as into the general properties
of scattering amplitudes.  In addition, empirical results have led
to the discovery of many hidden properties of planar $\mathcal{N}=4$
SYM, such as dual (super)conformal
invariance~\cite{DualConformal,FourLoopNeq4,FiveLoopNeq4,%
AMStrong,Drummond2008vq},
and the amplitude-Wilson-loop duality~\cite{AMStrong,WilsonLoopWeak}.

Many of the more powerful approaches to $\mathcal{N}=4$ supersymmetric
scattering amplitudes compute the loop integrand of the 
theory~\cite{Unitarity,BRY,BDS,FourLoopNeq4,FiveLoopNeq4,%
Alday2008yw,ArkaniHamed2010kv,Carrasco2011hw,Adamo2011pv,ArkaniHamed2012nw}.
These approaches can produce the integrand at very high loop
order~\cite{Bourjaily2011hi,Eden2012tu,Bern2012di}, but the evaluation
of the loop integrals can be quite challenging, in part due to severe
infrared divergences.  Some of the methods for producing the integrands
are only valid exactly in four dimensions in the massless theory,
where the integrals are infinite.  Even when the integrands can be computed
with a regulator in place, it is difficult to isolate the infrared
divergences of high-loop order integrals directly at the integrand level.
Although there are exceptions, such as the energy-energy
correlation~\cite{EECNeq4},
finite observables typically require the explicit cancellation of infrared
divergences across different loop orders.

In this paper we will follow an alternative approach, the {\it hexagon
function bootstrap}~\cite{Dixon2011pw,Dixon2011nj,Dixon2013eka,%
Dixon2014voa,Dixon2014xca}.
The philosophy of this program is to bypass integrands altogether
and focus on infrared-finite quantities from the very beginning.
One such finite quantity is the remainder function~\cite{Bern2008ap},\
$R_n$, defined by dividing the maximally-helicity-violating (MHV) scattering
amplitude for $n$ gluons by the BDS ansatz~\cite{BDS}.
Another useful observable, starting with the next-to-MHV (NMHV)
helicity configuration, is the ratio function
${\cal P}$~\cite{Drummond2008vq},
in which super-amplitudes for other helicity configurations
are divided by the MHV super-amplitude.  An on-shell
superspace~\cite{Nair1988bq,Georgiou2004by,Drummond2008vq,ArkaniHamed2008gz}
is used to organize the external states into ${\cal N}=4$ supermultiplets,
and the amplitudes into super-amplitudes.

Such finite observables can be constrained directly
from their analytic properties, particularly their behavior in
kinematical limits where amplitudes factorize and can be computed by
other methods.  In the case of planar ${\cal N}=4$ super-Yang-Mills theory,
we are fortunate to have a rich abundance of such boundary data.
Perhaps the most powerful information comes in the near-collinear limit
where two of the external states are almost parallel.
Thanks to the equivalence between amplitudes and polygonal Wilson loops,
this limit corresponds to an operator product expansion
(OPE)~\cite{Alday2010ku,Gaiotto2010fk,Gaiotto2011dt,Sever2011da}.
The relevant operators, whose anomalous dimensions are known
exactly~\cite{Basso2010in}, generate excitations of a one-dimensional
flux tube.  These states have integrable $1+1$ dimensional scattering
matrices.  In the past year or so, Basso, Sever and Vieira (BSV) have shown
that the OPE is governed by ``pentagon transitions'', which they argue
can be expressed in terms of the integrable $S$
matrices~\cite{Basso2013vsa} to all orders in the 't Hooft coupling.
BSV have worked out the consequences of this picture in increasingly great
detail~\cite{Basso2013aha,Basso2014koa,BSVIV}.  The perturbative
expansions of their results provides valuable boundary data for
the hexagon function bootstrap.  Recently, aspects of the flux-tube
approach have been reformulated in terms of Baxter
equations~\cite{Belitsky}.

Another important limit is the multi-Regge limit, when the outgoing
gluons are well separated in rapidity.  In this limit,
Lipatov and collaborators have described the factorization
of the ${\cal N}=4$ amplitudes in a Fourier-Mellin transformed
space~\cite{Bartels2008ce,Bartels2008sc,Lipatov2010qg,Lipatov2010ad,%
Bartels2010tx,BLPCollRegge,Fadin2011we,Lipatov2012gk}.
Further perspectives on multi-Regge factorization have been
provided by Caron-Huot~\cite{CaronHuot2013fea}.
The factorization limit has a logarithmic ordering, which allows for
the efficient recycling of lower-loop information to higher
loops~\cite{Dixon2012yy,Pennington2012zj,Dixon2013eka,Dixon2014voa}.
The recycling is aided by the recognition~\cite{Dixon2012yy}
that in the six-point case the functions relevant for the multi-Regge
limit are single-valued harmonic polylogarithms
(SVHPLs)~\cite{BrownSVHPLs}.

Very recently, a proposal for the multi-Regge limit has been
made~\cite{BCHS} that predicts {\it all} subleading
logarithmic orders.  This proposal is based on an analytic
continuation from the near-collinear limit, which is similar in spirit
to earlier work~\cite{BLPCollRegge,Hatsuda2014oza}, but now
provides much more detailed information.

The near-collinear, multi-Regge, and other physical constraints are
most effective in determining an amplitude when they are combined with
a suitable ansatz for the space of functions in which the solution lies.
For the case of six-point amplitudes, dual conformal invariance implies
that the amplitudes depend essentially on only three variables, the dual
conformal cross ratios $(u,v,w)$. The analytic solution for the two-loop
remainder function $R_6^{(2)}(u,v,w)$~\cite{DelDuca2009au}, 
after it was simplified dramatically using the
symbol~\cite{Goncharov2010jf}, provided the inspiration for
an ansatz for the symbol of the remainder function
at higher loops~\cite{Dixon2011pw}.
The same ansatz could also be applied to the symbols of
a pair of functions $V(u,v,w)$ and $\tilde{V}(u,v,w)$
entering the NMHV ratio function~\cite{Dixon2011nj}.
Those symbols define a class of functions of three variables,
iterated integrals called hexagon functions~\cite{Dixon2013eka}.
The number of iterated integrations defines the {\it weight} of
the hexagon function, which should be $2L$ for the $L$-loop
contributions to $R_6$, $V$ and $\tilde{V}$.
Given the hexagon-function ansatz, the near-collinear limit,
multi-Regge behavior, and a few other physical constraints uniquely
determine the full six-point remainder function at both
three~\cite{Dixon2013eka} and four loops~\cite{Dixon2014voa}.
The uniquenes of the solution, despite the existence of around 6000
unknown parameters in the inital four-loop ansatz, is a testament to
the power of the boundary data.

The aim of this paper is to apply the hexagon function bootstrap
to the six-gluon NMHV amplitude.  In particular, we will
compute $V$ and $\tilde{V}$ through three loops, entirely
from physical constraints.  A similar exercise was performed
previously at two loops~\cite{Dixon2011nj}.  However, at that time fewer
constraints were available, and so an explicit evaluation
of two-loop integrals for special kinematics had to be performed as well,
in order to fix all the unknown parameters.  Now the bootstrap works
unassisted at both two and three loops.  The increasing amount
of powerful, higher-twist OPE data~\cite{Basso2014koa,BSVIV}
suggests that it can be carried out to much higher loop order,
with the main limiting factor likely to be computing power.

At three loops, the weight (number of iterated integrations)
of $V$ and $\tilde{V}$ is six.
We characterize the functions in terms of their weight-five
$\{5,1\}$ coproduct components~\cite{Duhr2011zq,Duhr2012fh},
which are essentially their first derivatives.
This characterization makes use of a previous
classification of hexagon functions through weight five~\cite{Dixon2013eka}.
There are several hundred free parameters (unknown rational numbers)
in our initial ansatz.  We then apply a series of constraints to reduce the
number of parameters.  These constraints include fairly simple and
obvious ones, such as symmetry, spurious pole cancellations
and vanishing collinear limits. Other constraints incorporate more
sophisticated information, such as:
\begin{itemize}
\item a final-entry condition (a characterization of the first derivative)
which comes from~\cite{CaronHuot2011kk} the $\bar{Q}$ differential equation
in the super-Wilson loop approach~\cite{BullimoreSkinner,CaronHuot2011kk};
\item the near-collinear limits, which are required to match the 
OPE results of 
refs.~\cite{Sever2011da,Basso2013vsa,Basso2013aha,Basso2014koa,BSVIV}
in particular;
\item the multi-Regge limits, where we match to a formula that is a
natural generalization of one proposed for the MHV
amplitude~\cite{Fadin2011we,CaronHuot2013fea}, and for the leading-logarithmic
terms in the NMHV amplitude~\cite{Lipatov2012gk}.
\end{itemize}
Together, these constraints are more than enough to fully determine
$V$ and $\tilde{V}$.   Indeed, we have powerful cross checks
of the consistency of our assumptions, as well as those made by other groups
providing these constraints.  With the parity-even and parity-odd functions
fully determined, we discuss some of their limiting behaviors,
plot them, and showcase their interesting features.
 
This article is organized as follows.  In \sect{nmhvfirstconstr} we explain
our setup further, and then apply the first constraints:
(anti)symmetry in $u\lr w$; vanishing of $\tilde{V}$ under
cyclic permutations of $u,v,w$; the final-entry condition;
and the vanishing of spurious poles. These constraints reduce the
number of parameters in the ansatz down to 142.
In \sect{nmhvcoll} we apply constraints in the collinear limit, at leading
order and at the first near-collinear order, which together determine all but
two parameters. In \sect{MRKsection} we inspect the multi-Regge limits, which
fix the remaining two parameters in $V(u,v,w)$ and $\tilde{V}(u,v,w)$.
The next term in the near-collinear limit is then determined uniquely and
agrees precisely with the OPE predictions of ref.~\cite{Basso2014koa}.
We also extract the NMHV impact factor for the multi-Regge limit
through next-to-next-to-leading-logarithm (NNLL), and compare
it to the recent predictions of ref.~\cite{BCHS}.
In \sect{multiparticle}, we inspect the multi-particle
factorization limit of the NMHV amplitude.  We introduce a function $U$,
closely related to $V$, that plays an important role in this limit.
We show that $U$ collapses to a simple polynomial in $\ln(uw/v)$
in the factorization limit.  In \sect{coprels}, we find that $U$ has
additional simplicity across the entire space of cross ratios: it has
a restricted set of only five final symbol entries, which leads to
a simple form for one of its three derivatives.
In \sect{plotssect} we derive formulae for
$U$ and $\tilde{V}$ on various lines through the space of cross ratios
where they simplify.  We also investigate the numerical behavior
of $V$ and $\tilde{V}$ on these lines and on some two-dimensional planes.
In \sect{curiousrelsect}, we explore an intriguing empirical
relation between $V$ and coproduct components of the remainder
function $R_6$ at one higher loop order.
In \sect{conclusions} we discuss our
conclusions and directions for future work.  
In appendix~\ref{Ucoprod}, we give the $\{2L-1,1\}$
coproduct elements that characterize the weight $2L$ functions $U$
(from which $V$ can be derived) and $\tilde{V}$ through three loops.

We also provide ancillary files containing machine-readable expressions
for the near-collinear and multi-Regge limits of the ratio function.
Additional machine-readable results are available
elsewhere~\cite{V3webpage}.


\section{Setup and first constraints}
\label{nmhvfirstconstr}

As in ref.~\cite{Dixon2011nj}, we introduce an on-shell superspace
(see {\it e.g.} refs.~\cite{%
Nair1988bq,Georgiou2004by,Drummond2008vq,ArkaniHamed2008gz}).
We arrange the different on-shell states of the theory into an
on-shell superfield $\Phi$ which depends on Grassmann variables $\eta^A$
transforming in the fundamental representation of $su(4)$,
\be
\Phi\ =\ G^+ + \eta^A \Gamma_A + \tfrac{1}{2!} \eta^A \eta^B S_{AB}
+ \tfrac{1}{3!} \eta^A \eta^B \eta^C \epsilon_{ABCD} \overline{\Gamma}^D
+ \tfrac{1}{4!} \eta^A \eta^B \eta^C \eta^D \epsilon_{ABCD} G^-.
\label{onshellmultiplet}
\ee
Here $G^+$, $\Gamma_A$,
$S_{AB}=\tfrac{1}{2}\epsilon_{ABCD}\overline{S}^{CD}$,
$\overline{\Gamma}^A$, and $G^-$ are the positive-helicity gluon,
gluino, scalar, anti-gluino, and negative-helicity gluon states,
respectively.

We then consider superamplitudes, $\mathcal{A}(\Phi_1,\Phi_2,\ldots,\Phi_n)$,
which are functions of the superfields $\Phi_i$.  The ratio function is the
ratio of the full superamplitude to the MHV superamplitude, defined as
follows~\cite{Drummond2008vq},
\be
\mathcal{A}\ =\ \mathcal{A}_{\rm MHV} \, \times \, \cP\,.
\label{ratiodef}
\ee
By expanding in the Grassmann degree, {\it i.e.}~powers of $\eta$,
we can select out different values of $k$ in the N$^k$MHV expansion:
\be
\cP\ =\ 1 + \cP_{\rm NMHV}
+ \cP_{\rm N^2MHV} + \ldots
+ \cP_{\overline{\rm MHV}}\,,
\ee
where successive terms in the expansion carry four more powers of $\eta$.
For the six-point superamplitude, the only nontrivial term in this expansion
is the NMHV one, because N$^2$MHV is ${\rm \overline{MHV}}$, which is
related to MHV by parity (reversal of all helicities).

At tree level, the six-point NMHV ratio function is best described in
terms of $R$-invariants, which in turn are defined in terms of dual
coordinates $(x_i, \theta_i)$: 
\be
p_i^{\alpha \dot\alpha} = 
\lambda_i^\alpha \tilde{\lambda}_i^{\dot\alpha}
= x_i^{\alpha \dot\alpha} - x_{i+1}^{\alpha \dot\alpha}, \qquad
q_i^{\alpha A} = \lambda_i^{\alpha} \eta_i^A
= \theta_i^{\alpha A} - \theta_{i+1}^{\alpha A}\,.
\label{xthetadef}
\ee
The usual dual conformal cross ratios are denoted by
\beq\label{uvw_def}
u = u_1 = \frac{x_{13}^2\,x_{46}^2}{x_{14}^2\,x_{36}^2}\,, 
\qquad v = u_2 = \frac{x_{24}^2\,x_{51}^2}{x_{25}^2\,x_{41}^2}\,, \qquad
w = u_3 = \frac{x_{35}^2\,x_{62}^2}{x_{36}^2\,x_{52}^2}\,,
\eeq
where $x_{ij}^2 \equiv (x_i^\mu - x_j^\mu)^2$.

Using the coordinates $(x_i, \theta_i)$ we may define momentum
(super)twistors~\cite{Hodges2009hk,Mason2009qx}
\be
\mathcal{Z}_i = (Z_i \, | \, \chi_i), \qquad 
Z_i^{R=\alpha,\dot\alpha} = 
(\lambda_i^\alpha , x_i^{\beta \dot\alpha}\lambda_{i\beta}),
\qquad
\chi_i^A= \theta_i^{\alpha A}\lambda_{i \alpha} \,.
\ee
The momentum (super)twistors $\mathcal{Z}_i$ transform linearly under
dual (super)conformal symmetry, so that
$\langle abcd\rangle = \epsilon_{RSTU} Z_a^R Z_b^S Z_c^T Z_d^U$
is a dual conformal invariant. If we label our six external lines as
${a,b,c,d,e,f}$, then the $R$-invariants can be written as
\be
(f)\ \equiv\ [abcde]\ =\ 
\frac{\delta^4\bigl(\chi_a \langle bcde\rangle + {\rm cyclic}\bigr)}
{\langle abcd\rangle\langle bcde\rangle
\langle cdea\rangle\langle deab\rangle\langle eabc\rangle}\,.
\label{five_bracket_def}
\ee

In general, $R$-invariants obey many identities; see for example
refs.~\cite{Drummond2008vq,Drummond2008bq}. At six points, the only
identity we need is~\cite{Drummond2008vq}
\be
(1)-(2)+(3)-(4)+(5)-(6)\ =\ 0.
\label{5bracketidentity}
\ee
Using this identity, the NMHV tree amplitude may be written as
\be
\cP^{(0)}_{\rm NMHV}\ =\ [12345] + [12356] + [13456]
\ =\ (6) + (4) + (2)\ =\ (1) + (3) + (5).
\label{NewPtree}
\ee
Beyond tree level, the $R$-invariants will be dressed with
transcendental functions of the dual conformal cross ratios $(u,v,w)$,
which we will assume are hexagon functions.

Hexagon functions are a particular class of iterated
integrals~\cite{Chen} or multiple polylogarithms~\cite{FBThesis,Gonch},
which we will also refer to as pure (transcendental) functions.
When a weight-$n$ pure function $f$ is differentiated, the result
can be written as
\be
df = \sum_{s_k\in {\cal S}} f^{s_k} d \ln s_k \,,
\label{df}
\ee
where ${\cal S}$ is a finite set of rational expressions, called
the letters of the symbol of $f$, and $f^{s_k}$ are weight-$(n-1)$ 
pure functions.  The functions $f^{s_k}$ describe the $\{n-1,1\}$ component
of a coproduct $\Delta$ associated with a Hopf algebra for
iterated integrals~\cite{Gonch3,Gonch2,Brown2011ik}.
Similarly, each $f^{s_k}$ can be differentiated,
\be
df^{s_k} = \sum_{s_j\in {\cal S}} f^{s_j \, s_k} d \ln s_j \,,
\label{dfk}
\ee
thereby defining the weight-$(n-2)$ functions $f^{s_j \, s_k}$, which describe
the $\{n-2,1,1\}$ components of $\Delta$.  The maximal iteration
of this procedure defines the symbol of $f$, an $n$-fold tensor product
of elements of ${\cal S}$ (each standing for a $d\ln$).

Hexagon functions are functions whose symbols have letters drawn from
a particular nine-letter set:
\be
{\cal S} = \{u,v,w,1-u,1-v,1-w,y_u,y_v,y_w\}\,,
\label{nineletters}
\ee
where
\be
y_u = \frac{u-z_+}{u-z_-}\,, \qquad y_v = \frac{v-z_+}{v-z_-}\,, 
\qquad y_w = \frac{w - z_+}{w - z_-}\,,
\label{yfromu}
\ee
and
\be
z_\pm = \frac{1}{2}\Bigl[-1+u+v+w \pm \sqrt{\Delta}\Bigr]\,, 
\qquad \Delta = (1-u-v-w)^2 - 4 uvw\,.
\label{z_definition}
\ee
These nine letters are related to the 15 projectively invariant
ratios of momentum-twistor four-brackets $\langle abcd\rangle$,
which can be factored into nine independent combinations.

Hexagon functions are defined by one additional property:
their branch cuts should only be at 0 or $\infty$ in the variables
$u,v,w$, which means that the first entry of their symbol
is restricted to just these three letters~\cite{Gaiotto2011dt}.

We note that a cyclic permutation of the six external legs sends 
$u \to v \to w \to u$, while the $y_i$ variables transform as
$y_u \to 1/y_v \to y_w \to 1/y_u$. 
A three-fold cyclic rotation amounts to a space-time parity 
transformation, under which the cross ratios are invariant
while the $y_i$ variables invert.  It is useful to classify hexagon
functions by their transformation properties under parity.
Many additional properties of hexagon functions, and methods
for constructing them, are detailed in refs.~\cite{Dixon2013eka,Dixon2014xca}.

The six-point NMHV ratio function can be written in terms of two functions,
a parity-even function $V(u,v,w)$ and a parity-odd function
$\tilde{V}(y_u,y_v,y_w)$ as follows~\cite{Drummond2008vq,Dixon2011nj}:
\be
\bsp
&\cP_{\rm NMHV}\ =\ \frac{1}{2}\Bigl[
 [(1) + (4)] V(u,v,w) + [(2) + (5)] V(v,w,u) + [(3) + (6)] V(w,u,v)  \\
&\hskip1.8cm 
+ [(1) - (4)] \tilde{V}(y_u,y_v,y_w) - [(2)-(5)] \tilde{V}(y_v,y_w,y_u)
  + [(3) - (6)] \tilde{V}(y_w,y_u,y_v) \Bigr] \,.
\label{PVform}
\esp
\ee
It is better to think of the parity-odd function $\tilde{V}$
as a function of the $y_i$ variables, because its properties under
cyclic permutations are then captured correctly.  
The loop expansions of $V$ and $\tilde{V}$ are given by
\bea
V &=& 1 + \sum_{L=1}^\infty a^L V^{(L)} \,, \qquad \label{Vexpand}\\
\tilde{V} &=& \sum_{L=1}^\infty a^L \tilde{V}^{(L)} \,, 
\label{Vtexpand}
\eea
where $a=g_{\textrm{YM}}^2N_c/(8\pi^2)$ is our loop expansion parameter,
in terms of the Yang-Mills coupling constant $g_{\textrm{YM}}$ and
the number of colors $N_c$.  We remark that the expansion parameter 
conventionally used for the Wilson loop, $g^2$, is related to our
parameter by $g^2=a/2$.

The fundamental assumption in this paper, which was also used at two
loops~\cite{Dixon2011nj}, is that $V^{(L)}$ and $\tilde{V}^{(L)}$ are
weight $2L$ hexagon functions, with even and odd parity respectively.
The same basic assumption for the (parity-even) remainder function
$R_6^{(L)}$~\cite{Dixon2011pw,Dixon2013eka} results in a consistent solution
through four loops~\cite{Dixon2013eka,Dixon2014voa}.  

In this paper, we will work directly with hexagon functions, rather
than their symbols.  Through three loops, we only need hexagon functions
through weight six.  According to \eqn{df}, the $\{5,1\}$ coproduct
elements of a weight-six function $f$ completely specify the function
in terms of the weight-five functions $f^{s_k}$ up to a single constant of
integration, which we can take to be the value of $f$ at the point
$(u,v,w)=(1,1,1)$.  In ref.~\cite{Dixon2013eka}, all the hexagon functions
were classified through weight five.  We use this information to
construct the space of weight-six hexagon functions, by writing
the most general $\{5,1\}$ coproduct elements leading to consistent
mixed partial derivatives, {\it i.e.}~$d^2f=0$.  Including lower-weight
functions multiplied by Riemann $\zeta$ values, there are a total
of 639 parity-even weight-six hexagon functions, and 122 parity-odd ones.
Our initial ansatz for $V^{(3)}$ is the most general linear combination
of the parity-even functions with 639 unknown rational-number coefficients.
Similarly, the ansatz for $\tilde{V}^{(3)}$ is constructed from the
122 parity-odd functions.   We then impose constraints on $V^{(3)}$
and $\tilde{V}^{(3)}$, as described in the remainder of this section
and in the following two sections, until all 761 parameters are fixed.

Before carrying out this procedure at three loops, we recall what is
known about the functions $V^{(L)}$ and $\tilde{V}^{(L)}$ at lower loop orders.
At one loop, the parity-odd function vanishes, while the parity-even one
is nontrivial~\cite{Drummond2008vq}:
\bea
V^{(1)}(u,v,w) &=& \frac{1}{2} \Bigl[ H_2^u + H_2^v + H_2^w 
 + (\ln u + \ln w) \ln v - \ln u \ln w  - 2 \zeta_{2} \Bigr] \,,~~~~
\label{Voneloop}\\
\tilde{V}^{(1)}(u,v,w) &=& 0.
\label{Vtoneloop}
\eea
The vanishing of the weight-two parity-odd function $\tilde{V}^{(1)}$
can be understood simply from the fact that there are no such hexagon
functions.  The first parity-odd hexagon function, $\tilde\Phi_6$, is
related to the one-loop massless hexagon integral in six
dimensions~\cite{Dixon2011ng,DelDuca2011ne}, and it has weight three.

In ref.~\cite{Dixon2011nj}, the two-loop ratio function was determined up to
ten symbol-level parameters and one beyond-the-symbol parameter, using
general constraints, including the leading-discontinuity part of
the NMHV OPE~\cite{Sever2011da}.  These eleven parameters were then
fixed via an explicit evaluation of the relevant loop integrals on
the line in which all three cross ratios are equal, $(u,u,u)$.  This
procedure led to the following expressions for $V^{(2)}(u,v,w)$ and
$\tilde{V}^{(2)}(u,v,w)$:
\bea
V^{(2)} &=& -\frac{1}{4} \Bigl\{
 \Omegauvw + \Omegavwu + 2 \, \Omegawuv
 + 5 \, ( H_4^u + H_4^w ) + H_{3,1}^u + H_{3,1}^w
\nonumber\\ &&\null\hskip0.55cm
 - 3 \, ( H_{2,1,1}^u + H_{2,1,1}^w )
 - 2 \, \Bigl[ (H_2^u)^2 + (H_2^w)^2 \Bigr]
 - 4 \, ( \ln u \, H_3^u + \ln w \, H_3^w )
\nonumber\\ &&\null\hskip0.55cm
 + \frac{1}{2} \, ( \ln^2\!u \, H_2^u + \ln^2\!w \, H_2^w )
 + 4 H_4^v - 2 H_{3,1}^v - \frac{3}{2} (H_2^v)^2
 - 2 \, \ln v \, ( 2 H_3^v - H_{2,1}^v )
 + \ln^2v \, H_2^v
\nonumber\\ &&\null\hskip0.55cm
 - 2 \, \Bigl[ ( H_2^u + H_2^w ) \, H_2^v + H_2^u \, H_2^w \Bigr]
 + \ln(u/v) \, ( H_3^w + H_{2,1}^w )
 + \ln(w/v) \, ( H_3^u + H_{2,1}^u )
\nonumber\\ &&\null\hskip0.55cm
 - \Bigl[ \ln u \, \ln(v/w) + 2 \, \ln v \, \ln w \Bigr] \, H_2^u
 - \Bigl[ \ln w \, \ln(v/u) + 2 \, \ln v \, \ln u \Bigr] \, H_2^w
\nonumber\\ &&\null\hskip0.55cm
 - \biggl[ \frac{1}{2} \, \ln^2(u/w) + \ln(uw) \, \ln v \biggr] \, H_2^v
 - \frac{1}{2} \, \ln(uw) \, \ln v 
   \, \Bigl[ \ln(uw) \, \ln v - \ln u \, \ln w \Bigr]
\nonumber\\ &&\null\hskip0.55cm
 - \frac{1}{4} \, \ln^2u \, \ln^2w
 + \zeta_2 \, \Bigl[ 4 \, ( H_2^u + H_2^w ) + 2 \, H_2^v
             - \ln^2u - \ln^2w - 2 \, \ln^2v
\nonumber\\ &&\null\hskip4.2cm
             + 6 \, ( \ln(uw) \, \ln v - \ln u \, \ln w ) \Bigr]
 - 12 \, \zeta_4 \Bigr\} \,,
\label{V2}\\
\tilde{V}^{(2)} &=& \frac{1}{8} \Bigl[ - \Fuvw + \Fwuv
  + \ln(u/w) \PhiTilde(u,v,w) \Bigr] \,.
\label{Vt2}
\eea
Here we have rewritten the results in terms of 
harmonic polylogarithms (HPLs)~\cite{Remiddi1999ew}, as well as the
other functions constituting the basis of hexagon functions through
weight four, namely $\Omega^{(2)}$, $\PhiTilde$ and $F_1$~\cite{Dixon2013eka}.

The HPLs we need have weight vectors containing only 0 and 1.
They can be defined recursively by
\be
H_{0,\vec{w}}(u) = \int_0^u \frac{dt}{t} H_{\vec{w}}(t), \quad
H_{1,\vec{w}}(u) = \int_0^u \frac{dt}{1-t} H_{\vec{w}}(t), 
\label{Hdef}
\ee
except for $H_{0_n}(u)$ which is defined by
$H_{0_n}(u) = \tfrac{1}{n!} \log^n u$.
We choose a basis for the HPLs in which the point $u=1$ is regular, by
letting the argument be $1-u$, and restricting to weight vectors whose last
entry is 1.  We also use a compressed notation where $(k-1)$ 0's followed
by a 1 is replaced by $k$ in the weight vector, and the argument $(1-u)$ 
is replaced by the superscript $u$~\cite{Dixon2013eka}.
So, for example, $H_{3,1}^u = H_{0,0,1,1}^u = H_{0,0,1,1}(1-u)$,
and similarly for when the argument is $v$ or $w$.

In ref.~\cite{Dixon2011nj}, only the leading-discontinuity terms in the OPE
were available~\cite{Sever2011da}.  Now, thanks to the work of
BSV~\cite{Basso2013vsa,Basso2013aha,Basso2014koa,BSVIV},
who have used integrability to determine the OPE expansion exactly
in the coupling, we have access to enough data to fix not only the two-loop,
but also the three-loop six-point NMHV ratio function, without resorting to
evaluating any loop integrals.  The starting ansatz at two loops
involves 50 parity-even weight-four hexagon functions for $V^{(2)}$,
and 2 parity-odd ones for $\tilde{V}^{(2)}$.  (At one loop, there are
7 parity-even weight-two hexagon functions for $V^{(1)}$,
and no parity-odd ones for $\tilde{V}^{(1)}$.)

We now begin to determine the various unknown rational numbers
by applying many of the same constraints as in ref.~\cite{Dixon2011nj}.
Specifically, the constraints we inherit from that paper are as follows:
\begin{itemize}
\item {\bf Symmetry:}
Under the exchange of $u$ and $w$, the function $V$ is symmetric
while $\tilde{V}$ is antisymmetric:
\be
V(w,v,u)\ =\ V(u,v,w),\qquad
\tilde{V}(y_w,y_v,y_u)\ =\ -\tilde{V}(y_u,y_v,y_w).
\label{VVtsym}
\ee
At three loops, this constraint reduces the $639 + 122 = 761$ parameters
to $363 + 49 = 412$.
\item {\bf Spurious Pole Constraints:} Scattering amplitudes have
  poles corresponding to sums of color-adjacent momenta, of the form
  $(p_i+p_{i+1}+\ldots+p_{j-1})^2\equiv x_{ij}^2$. These are produced
  by four-brackets of the form $\langle i-1,i,j-1,j\rangle$. Poles in
  other four-brackets do not correspond to sums of color-adjacent
  momenta, and should not be present in the full amplitude. While such
  poles never appear in hexagon functions, they are present in the
  $R$-invariants. In order for
  such spurious poles to vanish in the full function, the coefficients
  of the $R$-invariants must be such that these poles cancel.
  The $R$-invariants $(1)$ and $(3)$ contain poles as
  $\langle2456\rangle\rightarrow 0$, with equal and opposite residues.
  In order for them to cancel, we see from~\eqn{PVform} that
\be
[ V(u,v,w) - V(w,u,v) + \tilde{V}(y_u,y_v,y_w) - \tilde{V}(y_w,y_u,y_v)
]_{\langle 2456\rangle \rightarrow 0} \ =\ 0.
\label{spurious}
\ee
The $\langle 2456\rangle\rightarrow 0$ limit can be implemented by taking
\be
w\rightarrow 1 \,,
\quad y_u\rightarrow(1-w)\frac{u(1-v)}{(u-v)^2} \,,
\quad y_v\rightarrow\frac{1}{(1-w)}\frac{(u-v)^2}{v(1-u)} \,,
\quad y_w\rightarrow\frac{1-u}{1-v} \,.
\ee
\item {\bf Collinear Limit:}
As two external particles become collinear, the six-point NMHV amplitude
should reduce to either the five-point MHV or
$\overline{\textrm{MHV}}$ amplitude times a splitting function. The
five-point ratio function is equal to its tree-level value due to parity
(NMHV/MHV is $\overline{\textrm{MHV}}$/MHV at the five-point level).
Therefore, at any nonzero loop order the collinear limit of the six-point
ratio function must vanish. In particular, taking $w\rightarrow 0$ and
$v\rightarrow 1-u$ gives a collinear limit in which all $R$-invariants
vanish except for $(6)$ and $(1)$, which become equal.
Inserting this condition into~\eqn{PVform}, we find the collinear constraint,
\be
[ V(u,v,w) + V(w,u,v) + \tilde{V}(y_u,y_v,y_w)
- \tilde{V}(y_w,y_u,y_v) ]_{w\rightarrow 0,\ v\rightarrow 1-u}\ =\ 0.
\label{collvanish}
\ee
Parity-odd functions always vanish in the collinear limit~\cite{Dixon2011nj},
so the constraint is really just that $V(u,v,w) + V(w,u,v)$ vanishes
in the limit.
\end{itemize}

In addition to these constraints, we impose several new constraints,
here in rough order of simplicity:
\begin{itemize}
\item {\bf Cyclic Vanishing:}
It turns out that not all of the apparent freedom in $\tilde{V}$ is
physically meaningful. It is possible to add a cyclicly symmetric
function to $\tilde{V}$ that is consistent with its other symmetries,
but such a contribution $\tilde{f}(u,v,w)$ vanishes in the full ratio
function~(\ref{PVform}):
\be
\bsp
&\hskip0.7cm \frac{1}{2}\left[ [(1)-(4)] \tilde{f}(u,v,w)
  -[(2)-(5)] \tilde{f}(u,v,w) + [(3)-(6)] \tilde{f}(u,v,w) \right]\\
&=\ \frac{1}{2}\Bigl[ [(1)+(3)+(5)]-[(2)+(4)+(6)] \Bigr]\ \tilde{f}(u,v,w)\\
&=\ 0,
\label{cyclicvanish}
\esp
\ee
using \eqn{NewPtree}.
A function $\tilde{f}$ of this sort cannot contribute to the ratio function,
and so it will never be constrained by any physical limits.  Therefore,
we might as well set any such contribution to zero.  This constraint did not
appear at two loops, because there are no cyclicly invariant parity-odd
hexagon functions at weight four.  However, at weight six there are
10 such functions.  We remove them using this constraint, right after
imposing the $u \lr w$ symmetry constraints.
\item {\bf Final-Entry Condition:}
Caron-Huot and He have observed that supersymmetry constrains the
possible final entries of the symbols of finite quantities in planar 
$\NeqFour$ SYM~\cite{CaronHuot2011kk,SimonSongPrivate}.
Specifically, they express the
action of certain dual superconformal generators on the N${}^k$MHV
amplitude in terms of lower-loop N${}^{k+1}$MHV quantities. For the
MHV remainder function, these constraints imply a set of six possible
final entries. For the NMHV ratio function, expressed in our
variables, the constraints are that $V(u,v,w)$ and $\tilde{V}(u,v,w)$,
the functions multiplying the $R$-invariant $(1)$, can only
have final entries from the following seven-element set:
\be
\left\{ \frac{u}{1-u}, \frac{v}{1-v}, \frac{w}{1-w},
   y_u, y_v, y_w, \frac{uw}{v} \right\} \,.
\label{finalentries}
\ee
The other $R$-invariants multiply functions with final entries
from sets related by the appropriate cyclic permutations of the variables. 
(Technically, these constraints apply to the NMHV amplitude from which 
infrared divergences have been subtracted using the BDS ansatz, rather
than to the ratio function itself.  However, these quantities differ by the
MHV remainder function, which has final entries in a subset of
the NMHV set~(\ref{finalentries}) ($uw/v$ is not present).  Therefore, this
final-entry condition can be applied to the ratio function without
modification.)  We impose this constraint right after the cyclic-vanishing
constraint.  It reduces the $363+39$ free parameters down to $166 + 16 = 182$.
Then we impose the vanishing of the spurious poles, which fixes
another 40 parameters, and mixes the parity-even and parity-odd sectors
so that we can no longer count their parameters separately.
\item {\bf Near-Collinear Limits:}
BSV use integrability to evaluate the OPE for
Wilson loops nonperturbatively in the coupling.  They proceed order by order
in the number of flux-tube excitations, which corresponds to powers of
an expansion parameter $T$.  This parameter is proportional to the
square root of a vanishing cross ratio (see section~\ref{nmhvcoll}).
By inserting states on the boundaries of the Wilson loop they are able
to replicate particular components of the NMHV amplitude. Constraining our
results to agree with their expansions at first order in
$T$~\cite{Basso2013aha} constrains many parameters.
Two parameters that remain can be constrained using BSV's more recent
results at order $T^2$~\cite{Basso2014koa,BSVPrivate}.
\item {\bf Multi-Regge Kinematics:}
The multi-Regge limit is a generalization of the Regge limit in which
the outgoing particles of a $2\rightarrow n$ scattering process are
strongly ordered in rapidity. Lipatov, Prygarin, and
Schnitzer~\cite{Lipatov2012gk} have investigated the multi-Regge limit of NMHV
amplitudes in $\NeqFour$ SYM, creating an ansatz for their
behavior at leading-logarithmic order that mirrors previous results for the MHV
amplitude. In this paper we generalize their results beyond
leading-log order, along the lines of
refs.~\cite{Fadin2011we,CaronHuot2013fea}.
These generalizations are fully consistent with the near-collinear boundary
conditions, and thereby serve as an independent check of them. 
Also, we can derive the NMHV impact factor in the factorization we propose,
through NNLL.  The NMHV and MHV impact factors are strikingly similar.
Our results are completely consistent with the recent all-orders
multi-Regge proposal~\cite{BCHS}.
\end{itemize}

In practice it can be useful to constrain the symbol of the ratio
function first, and then constrain the full function, making use of
the coproduct to characterize the beyond-the-symbol terms.
Indeed, this was our first approach to obtaining $V^{(3)}$ and
$\tilde{V}^{(3)}$.  However, as mentioned earlier in this section,
it is straightforward to dispense with the symbol altogether, and begin
with a function-level ansatz characterized by various coproduct components.
We then apply all constraints directly at function level, using the
coproduct information to compute the necessary limiting behavior.
Because such an approach may well scale better computationally
to higher loops than
a symbol-level approach, we describe the results of using that
approach here.  After applying each set of constraints the number of
parameters in the ansatz is reduced, as
shown in table \ref{tab:NMHV_constr}.
This table also includes the corresponding numbers for lower loop
orders, so that one can appreciate the growth in the number
of parameters with loop order.

\renewcommand{\arraystretch}{1.25}
\begin{table}[!t]
\centering
\begin{tabular}[t]{|l|c|c|c|}
\hline\hline
Constraint & $L=1$ & $L=2$ & $L=3$\\\hline
1. (Anti)symmetry in $u$ and $w$ & 7 & 52 & 412\\\hline
2. Cyclic vanishing of $\tilde{V}$ & 7 & 52 & 402\\\hline
3. Final-entry condition & 4 & 25 & 182\\\hline
4. Spurious-pole vanishing & 3 & 15 & 142\\\hline
5. Collinear vanishing & 1 & 8 & 92\\\hline
6. $\Ord(T^1)$ OPE & 0 & 0 & 2\\\hline
7. $\Ord(T^2)$ OPE {\it or} multi-Regge kinematics & 0 & 0 & 0\\
\hline\hline
\end{tabular}
\caption{Remaining parameters in the function-level
ans\"{a}tze for $V^{(L)}$ and $\tilde{V}^{(L)}$ after each constraint is applied,
at each loop order.}
\label{tab:NMHV_constr}
\end{table}

As shown, after applying the constraints of $u\leftrightarrow w$
(anti)symmetry, cyclic vanishing of $\tilde{V}$, the final entry
condition, and the vanishing of spurious poles, we have 142 parameters
remaining in our ansatz. In the following sections, we use the
collinear constraints, OPE, and multi-Regge
limits to fix these final parameters.


\section{Collinear and near-collinear limits}
\label{nmhvcoll}

In this section, we consider the $w\rightarrow 0$ collinear limit.
In general, this limit may be expressed via a permutation of a map
between the cross ratios $(u,v,w)$ and the variables
$(F,S,T) \equiv (e^{i\phi},e^\sigma,e^{-\tau})$
defined in ref.~\cite{Basso2013vsa}:
\be
\bsp
u &= \frac{F}{F+FS^2+ST+F^2ST+FT^2} \,,\\
v &= \frac{FS^2}{(1+T^2)(F+FS^2+ST+F^2ST+FT^2)} \,,\\
w &= \frac{T^2}{1+T^2} \,,\\
y_u &= \frac{F+ST+FT^2}{F(1+FST+T^2)} \,,\\
y_v &= \frac{FS+T}{F(S+FT)} \,,\\
y_w &= \frac{(S+FT)(1+FST+T^2)}{(FS+T)(F+ST+FT^2)} \,.
\esp
\label{NMHVBSVparam}
\ee

As mentioned in section~\ref{nmhvfirstconstr}, the combination
$V(u,v,w)+V(w,u,v)$ should vanish in this limit. This is a fairly powerful
constraint, fixing 50 of the remaining 142 parameters, and leaving 92.
To determine the remaining parameters we will match to the OPE results
of Basso, Sever and Vieira.

Many features of BSV's approach to the OPE of
polygonal Wilson loops carry over to the NMHV helicity configuration
with only minor modifications~\cite{Basso2013aha}. In general, NMHV
scattering amplitudes are dual to Wilson loops dressed with insertions
of states that depend on the particular NMHV component being
investigated~\cite{Mason2010yk,CaronHuot2010ek}.
Two cases are explored by BSV, that of two scalar insertions, one on the
bottom cusp and one on the top, and that of a gluonic insertion on the
bottom cusp. We will consider each in turn.

BSV found that by inserting a scalar on the top and bottom cusps of
the Wilson loop they were able to probe the $\eta_6\eta_1\eta_3\eta_4$
(or ``6134'') component of the NMHV amplitude.  In this configuration,
the leading excitations are scalar ones.  Inspecting \eqn{five_bracket_def},
we see that all the $R$-invariants vanish for the $\eta_6\eta_1\eta_3\eta_4$
component except for $(2)$ and $(5)$.   Furthermore, the
identity~(\ref{5bracketidentity}) collapses for this component to
\be
(2)\ =\ (5)\ =\ \frac{1}{\langle 6 1 3 4\rangle}
\ =\ \frac{e^{-\tau}}{2\cosh\sigma} \,,
\label{DCI6134}
\ee
so that only the term multiplying $V(v,w,u)$ survives. Thus this component
of ${\cal P}$ has a particularly simple representation in terms of a single
pure function. Additionally, up to the first order in $T$ the Wilson loop
ratio investigated by BSV is equal to the ratio function. As such, we may
simply write
\be
\bsp
\mathcal{W}^{(6134)}&= \frac{e^{-\tau}}{2\cosh\sigma}
\sum_{L=0}^\infty \left(\frac{a}{2}\right)^{L} \sum_{n=0}^{L}
\tau^n F^{(L)}_n(\sigma) \,+\,\cO(e^{-2\tau})\\
&= \frac{T}{2\cosh\sigma}
\times V(v,w,u)|_{\cO(T^0)}\ +\ \cO(T^2) \,,
\esp
\label{OPE6134}
\ee
where the $F^{(L)}_n$ are given explicitly in appendix F of
ref.~\cite{Basso2013aha}.
Note that we only need the $T^0$ term in $V(v,w,u)$ as $w\to0$, because
the dual superconformal invariant prefactor carries a power of $T$
in this limit.
Applying the constraint~(\ref{OPE6134}) at three loops,
to the 92-parameter ansatz with vanishing collinear limits,
leaves 14 parameters unfixed.  In an ancillary file, we give the
near-collinear limit of ${\cal P}^{(6134)}$ through one higher order, $T^2$
(after all free parameters have been fixed).

Alternatively, one may insert a gluonic excitation at the bottom cusp
of the Wilson loop, probing the $(\eta_1)^4$ (or ``1111'') component.
Up to first order in $T$, the $R$-invariants in this component become
\be
\bsp
&(1)\rightarrow0,\qquad 
(2)\rightarrow \frac{F T}{S(1+S^2)}+\cO\left(T^2\right),\qquad 
(3)\rightarrow 1 - F S T +\cO\left(T^2\right),\\
&(4)\rightarrow 1 - \frac{F T}{S} + \cO\left(T^2\right), \qquad
(5)\rightarrow \frac{F S^3 T}{1+S^2}+\cO\left(T^2\right),\qquad
(6)\rightarrow 0 + \cO(T^4) \,.
\esp
\ee
The odd function $\tilde{V}$ vanishes in the collinear limit; it
is $\cO(T^1)$ for any permutation.  Also, we can use \eqn{collvanish}
to eliminate $V(w,u,v)$ in favor of $-V(u,v,w)$, up to terms suppressed
by a power of $T$.  Using such relations, we find that the $(\eta_1)^4$
component of the ratio function becomes,
\be
\bsp
\cP^{(1111)}\ =&\ \frac{1}{2} \biggl\{
V(u,v,w) + V(w,u,v) - \tilde{V}(u,v,w) + \tilde{V}(w,u,v) \\
&\hskip0.5cm
 + F T \biggl[ - \frac{1-S^2}{S} V(u,v,w) + \frac{1+S^4}{S(1+S^2)} V(v,w,u)
        \biggr] \biggr\}\ +\ \cO(T^2)\,.
\esp
\label{OPE1111}
\ee
We note that the terms without an explicit $T$ are also $\cO(T)$
due to the collinear-vanishing relations, except for the tree-level term,
which is $1+\cO(T)$.

We match the near-collinear limit of \eqn{OPE1111} to BSV's
computation~\cite{Basso2013aha} of the OPE, in terms of a single
gluonic excitation propagating across the Wilson loop.
The result is given as an integral over the excitation's rapidity $u$,
involving its anomalous dimension (or energy) $\gamma(u)$, its momentum $p(u)$,
a measure factor $\mu(u)$, and the NMHV dressing functions $h$ and $\bar{h}$.
The expansions of these quantities through $\cO(a^3)$ are given by,
\be
\bsp
\gamma(u) &= a \Bigl[ \psi(\hf-iu) + \psi(\hf+iu) - 2\psi(1) \Bigr] \\
&\hskip0.4cm\null
 - \frac{a^2}{4} \biggl[ 
    \psipp(\thf-iu) + \psipp(\thf+iu)
   + 4 \zeta_2 \Bigl[ \psi(\hf-iu) + \psi(\hf+iu) - 2\psi(1) \Bigr]
   + 12 \zeta_3 \biggr]\\
&\hskip0.4cm\null
 + \frac{a^3}{8} \biggl[ 
  \frac{1}{6} \Bigl[ \psipppp(\thf-iu) + \psipppp(\thf+iu) \Bigr]
     + 2 \zeta_2 \Bigl[ \psipp(\thf-iu) + \psipp(\thf+iu) \Bigr]\\
&\hskip1.4cm\null
     + 44 \zeta_4 \Bigl[ \psi(\hf-iu) + \psi(\hf+iu) - 2\psi(1) \Bigr]
             - 24 \zeta_2 \zeta_3 \tanh^2 \pi u 
             + 40 ( 2 \zeta_5 + \zeta_2 \zeta_3 ) \biggr]\\
&\hskip0.4cm\null
 + \cO(a^4)\,,
\esp
\label{energy}
\ee
\be
\bsp
p(u) &= 2 u - a \pi \tanh\pi u
 + \frac{a^2}{4} \pi^3 \biggl[ \frac{8}{3} \tanh\pi u - 2 \tanh^3\pi u \biggr]\\
&\hskip0.4cm\null
 + \frac{a^3}{8} \biggl[ \pi^5 \biggl( - \frac{172}{45} \tanh\pi u 
            + \frac{22}{3} \tanh^3\pi u - 4 \tanh^5\pi u \biggr)
            + 4 i \zeta_3 \Bigl[ \psip(\thf-iu) - \psip(\thf+iu) \Bigr]
                 \biggr] \\
&\hskip0.4cm\null
 + \cO(a^4)\,,
\esp
\label{momentum}
\ee
and
\be
h(u) = \frac{2 x^+(u) x^-(u)}{a} \,, \qquad
\bar{h}(u) = \frac{1}{h(u)} \,,
\label{hhbar}
\ee
where
\be
x^\pm(u) = x(u\pm\tfrac{i}{2})
\label{xpmu}
\ee
is given in terms of the Zhukovsky variable
\be
x(u) = \frac{1}{2} \Bigl[ u + \sqrt{ u^2 - 2 a } \Bigr] \,.
\label{Zhukovsky}
\ee
The perturbative expansion of the measure $\mu(u)$ can be found in
ref.~\cite{Basso2013vsa}.
It is a bit more complicated, but is still expressible in terms
of the function $\psi(x) = d\ln\Gamma(x)/dx$ and its derivatives,
as well as $\tanh\pi u$.  The rapidity $u$ should not be confused
with the cross ratio $u$.

In terms of these functions, the formula for the gluonic
flux-excitation contribution to the OPE is,
\be
\bsp
\cP^{(1111)} = 1 &
+ T F \int_{-\infty}^\infty 
\frac{du}{2\pi}\mu(u)(h(u)-1)e^{ip(u)\sigma-\gamma(u)\tau} \\
&+ \frac{T}{F} 
\int_{-\infty}^\infty 
\frac{du}{2\pi}\mu(u)(\bar{h}(u)-1)e^{ip(u)\sigma-\gamma(u)\tau} 
+ \cO(T^2) \,.
\esp
\label{OPE1111BSV}
\ee
We can carry out the integrals over $u$ by deforming the integral
into the lower half-plane, which converts it into a sum over residues
at $u=-im/2$ for positive integers $m$.  There are methods for performing
such sums exactly, see for example ref.~\cite{Papathanasiou2013uoa}.
We take a more mundane approach:  We truncate the series in $m$ at a
suitably large finite value (of order 100).  The truncation yields a
high-order Taylor expansion in $S$.  Then we write an ansatz for the 
exact result in terms of HPLs depending on $S^2$, and match the Taylor
expansion of the ansatz against the actual Taylor expansion, in order
to determine all of the rational-number coefficients in the ansatz.

After we have expressed the order $T$ term in \eqn{OPE1111BSV}
in terms of HPLs, in order to match it against our ansatz
we have to expand \eqn{OPE1111}, with the ansatz for
$V$ and $\tilde{V}$ inserted into it.  The ansatz has either 14 or 92
parameters in it (depending on whether or not we have already imposed the
order $T$ constraint on ${\cal P}^{(6134)}$).  We use the differential
equations method described in section 5 of ref.~\cite{Dixon2013eka}
to expand all the hexagon functions in this ansatz.
The resulting expressions for the expansion of \eqn{OPE1111BSV}
are too lengthy to display here, but we
provide them in a computer-readable ancillary file attached to this
article.  The file also includes the next order in the near-collinear
expansion of ${\cal P}^{(1111)}$, namely order $T^2$, after all free parameters
have been fixed.

After applying the constraints from the $T^1$ term in the OPE for the
1111 component, \eqn{OPE1111}, just two undetermined parameters remain.
These parameters multiply the functions $[\tilde\Phi_6]^2$ and
$V^{(1)} \, R_6^{(2)}$, where $\tilde\Phi_6$ is the pure function
associated with the $D=6$ one-loop hexagon
integral~\cite{Dixon2011ng,DelDuca2011ne},
$V^{(1)}$ is the one-loop ratio function given in \eqn{Voneloop},
and $R_6^{(2)}$ is the two-loop remainder function.
It is easy to see that the two parameters cannot be fixed by any
OPE information at $\cO(T^1)$:  Because $\tilde\Phi_6$ is parity odd,
it vanishes proportional to $T$, so its square vanishes like $T^2$.
Similarly, $V^{(1)}$ obeys the collinear vanishing
condition~(\ref{collvanish}), giving one power of $T$; and $R_6^{(2)}$
is totally symmetric and its vanishing provides an additional power of
$T$ in all channels.

Sever, Vieira and Wang~\cite{Sever2011da} have described the
leading-discontinuity OPE behavior of the ratio function.  This
behavior captures the leading $\ln^L T$ behavior at $L$ loops,
irrespective of the number of powers of $T$ multiplying it as $T\to0$.
Hence the leading-discontinuity OPE might contain complementary information
to the full $T^1$ OPE.  However, in the present case the leading-discontinuity
information cannot be used to fix the coefficients of either
$[\tilde\Phi_6]^2$ or $V^{(1)} \, R_6^{(2)}$.  That is because the
functions $\tilde\Phi_6$, $V^{(1)}$ and $R_6^{(2)}$ each have only a single
discontinuity, so the two weight-6 functions in question have only
double discontinuities, not the triple discontinuity which is the
leading one at three loops.

We also remark that at $\cO(T^1)$, the 1111 component of the OPE is more
powerful than the 6134 component:  We imposed the 1111 constraint directly
on the 92-parameter ansatz with vanishing collinear limits, and found
that it still fixed all but two of the parameters, even without any
assistance from the 6134 component. Recall that the 6134 component imposed
on the same 92-parameter ansatz still left 14 parameters unfixed.

Basso, Sever and Vieira have evaluated the two flux-excitation
contributions to the OPE for the ratio function~\cite{Basso2014koa}
and they have provided us with the small $S$ expansion of the resulting
$\cO(T^2)$ terms in the OPE~\cite{BSVPrivate}.  We can use these terms
to fix the two remaining parameters in our ansatz.  Alternatively, we can
use factorization in the multi-Regge limit, as described in the next section.
Either approach leads to the same values for the two parameters, providing
a very nice consistency check.


\section{Multi-Regge limits}
\label{MRKsection}

In this section we propose a factorization of the NMHV amplitude in the
limit of multi-Regge kinematics (MRK), which is a natural extension of
previous work by Fadin and Lipatov~\cite{Fadin2011we} in the MHV case,
and by Lipatov, Prygarin and Schnitzer~\cite{Lipatov2012gk} for
the leading-logarithmic behavior of the NMHV amplitude.
We use this factorization as one method for fixing the remaining
two parameters in our ansatz.  We are then able to extract from the
fully-fixed ansatz the NMHV impact factor, which we compare to
the previously-known MHV impact factor, through
next-to-next-to-leading-logarithmic accuracy.

We remind the reader that the multi-Regge limit of a $2\to (n-2)$ process
is the limit in which the $(n-2)$ outgoing particles are strongly ordered
in rapidity. For $2\to 4$ gluon scattering, this means that two of the
gluons are emitted at high energy almost parallel to the incoming gluons,
while the other two, while still emitted at small angles to the path of
the incoming gluons, have smaller energy.
Due to helicity conservation on the highest energy lines,
the MHV 6-gluon amplitude in the MRK limit can be viewed as having
two positive incoming helicities scattering into four positive outgoing
ones.  The appropriate color-ordering for the $2\to4$ process is to take
two diagonally opposite legs to be the incoming legs.
So we may consider the MHV helicity configuration to be
\be
3^+ 6^+ \ \to\ 2^+ 4^+ 5^+ 1^+ \,,
\label{MHVconfig}
\ee
where legs 1 and 2 are the highest-energy outgoing gluons.  
For an NMHV amplitude, one of the two lower-energy outgoing gluons
has its helicity reversed, say 
\be
3^+ 6^+ \, \to \, 2^+ 4^- 5^+ 1^+ \,.
\label{NMHVconfig}
\ee
In \eqns{MHVconfig}{NMHVconfig} we are {\it not} using the all-outgoing
helicity convention, in order to emphasize helicity conservation
on the high-energy lines.

In this MRK limit, the cross
ratios~$u_1$, $u_2$ and $u_3$ approach the values
\beq
u_1 \to 1\,,\qquad u_2,u_3 \to 0\,,
\eeq
with the ratios
\beq
\frac{u_2}{1-u_1}\equiv \frac{1}{(1+w)\,(1+\ws)} {\rm~~~~and~~~~} 
\frac{u_3}{1-u_1}\equiv \frac{w\ws}{(1+w)\,(1+\ws)}
\label{wdef}
\eeq
held fixed.  In this section, we use $(u_1,u_2,u_3)$ to denote the
three cross ratios~(\ref{uvw_def}), instead of $(u,v,w)$,
in order to minimize confusion
between the cross-ratio $w$ and the variable $w$ used to parametrize
the multi-Regge kinematics.

Fadin and Lipatov~\cite{Fadin2011we} proposed a precise factorization
relation for the MRK limit of the six-point MHV remainder function,
through at least next-to-leading-logarithmic (NLL) accuracy.
Caron-Huot~\cite{CaronHuot2013fea} suggested that, subject to some
reasonable assumptions, the same formula should hold in the planar
limit to all subleading logarithms.   Some additional evidence
for factorization beyond NLL was provided in ref.~\cite{Dixon2014voa},
where the four-loop remainder function was computed and found to be
consistent with the proposed MRK limit through at least
next-to-next-to-leading-logarithmic (NNLL) accuracy.

The proposal of Fadin and Lipatov is that the remainder function
$R_6$ obeys~\cite{Fadin2011we}:
\be
\bsp
e^{R_6+i\pi\delta}|_{\textrm{MRK}}\ =\ \cos\pi\omega_{ab} 
+ i  \frac{a}{2} \sum_{n=-\infty}^\infty
(-1)^n\left(\frac{w}{\ws}\right)^{\frac{n}{2}}
&\int_{-\infty}^{+\infty}
\frac{d\nu}{\nu^2+\frac{n^2}{4}}|w|^{2i\nu}
\, \Phi^{\textrm{MHV}}_{\textrm{Reg}}(\nu,n)\\
&\hskip0.5cm \times\left(-\frac{1}{1-u_1}\frac{|1+w|^2}{|w|}\right)^{\omega(\nu,n)}
\,,
\esp
\label{MHV_MRK}
\ee
where
\be
\bsp
\omega_{ab} &= \frac{1}{8}\,\gamma_K(a)\,\log|w|^2\,,\\
\delta &= \frac{1}{8}\,\gamma_K(a)\,\log\frac{|w|^2}{|1+w|^4}\,,
\esp
\ee
and $\gamma_K(a)$ is the cusp anomalous dimension.

The BFKL eigenvalue $\omega(\nu,n)$ and the MHV impact factor
$\Phi^{\textrm{MHV}}_{\textrm{Reg}}(\nu,n) = \Phi_{\textrm{Reg}}(\nu,n)$
may both be expanded perturbatively in $a$:
\beq\bsp
\omega(\nu,n) &\,= 
- a \left(E_{\nu,n} + a\,E_{\nu,n}^{(1)}+ a^2\,E_{\nu,n}^{(2)}+\cO(a^3)\right)\,,\\
\Phi_{\textrm{Reg}}(\nu,n)&\, = 1 + a \, \Phi_{\textrm{Reg}}^{(1)}(\nu,n)
 + a^2 \, \Phi_{\textrm{Reg}}^{(2)}(\nu,n)
 + a^3 \, \Phi_{\textrm{Reg}}^{(3)}(\nu,n)+\cO(a^4)\,.
\label{expandomegaPhi}
\esp\eeq
Because $\omega(\nu,n)$ starts at order $a$, while the impact
factor $\Phi_{\textrm{Reg}}(\nu,n)$ is unity at leading order, the highest
power of $\ln(1-u_1)$ that appears at loop order $L$ is $\ln^{L-1}(1-u_1)$.
This property allows the MRK limit to be organized in successive orders
of $\ln(1-u_1)$, beginning with the leading-log approximation, or LLA. 
At this order, only the leading BFKL eigenvalue $E_{\nu,n}$
contributes nontrivially to the remainder function.
The next order in the logarithmic expansion, the term of
order $\ln^{L-2}(1-u_1)$,
is called the next-to-leading-log approximation, or NLLA.
It is determined by $E_{\nu,n}^{(1)}$ and $\Phi_{\textrm{Reg}}^{(1)}$,
which were computed in ref.~\cite{Fadin2011we}.
Computations of the remainder function at three and four loops
have provided the BFKL eigenvalue through NNLLA ($E_{\nu,n}^{(2)}$),
and the MHV impact factor through N$^3$LLA ($\Phi_{\textrm{Reg}}^{(2)}$ 
and $\Phi_{\textrm{Reg}}^{(3)}$)~\cite{Dixon2012yy,Dixon2013eka,Dixon2014voa}.

The BFKL eigenvalue $\omega(\nu,n)$ is a property of the Reggeized gluon
ladder being exchanged in the $t$-channel.  It does not depend on the
external states attached to the end of the ladder.  For the six-point
amplitude, no states should be emitted from the middle of the
ladder~\cite{Bartels2008ce}.  At seven and higher points, there can be
such emission vertices~\cite{Bartels2011ge}.

Using the independence of the BFKL eigenvalue from the external states,
Lipatov, Prygarin, and Schnitzer proposed modifying the LLA
version of \eqn{MHV_MRK} for the NMHV case~\cite{Lipatov2012gk},
obtaining:
\be\label{NMHV_MRK}
R_{\textrm{NMHV}}^{\textrm{LLA}}\ =\ -\frac{ia}{2}\,\sum_{n=-\infty}^\infty(-1)^n
\,\int_{-\infty}^{+\infty}
\frac{d\nu  \, w^{i\nu+n/2} \, \ws^{i\nu-n/2} }{ (i\nu+\frac{n}{2})^2}
\, \Bigl[ (1-u_1)^{a\,E_{\nu,n}} - 1 \Bigr] \,.
\ee
Here $R_{\textrm{NMHV}}$ is the NMHV remainder function, a quantity which is
particularly convenient to work with in the MRK limit. It can be defined
as the product of the NMHV ratio function and the
(exponentiated) MHV remainder function:
\be 
R_{\textrm{NMHV}}\ =\ \frac{A_{\textrm{NMHV}}}{A_{\textrm{BDS}}}
\ =\ \frac{A_{\textrm{NMHV}}}{A_{\textrm{MHV}}} 
\times \frac{A_{\textrm{MHV}}}{A_{\textrm{BDS}}}
\ =\ {\cal P}_{\textrm{NMHV}} \times \exp(R_6) \,.
\label{NMHVtoRatio}
\ee
Clearly, the LLA NMHV formula~(\ref{NMHV_MRK}) is the same as the LLA
version of \eqn{MHV_MRK} for the MHV case, but with the substitution,
\be
\frac{1}{-i\nu+\frac{n}{2}}\to-\frac{1}{ i\nu+\frac{n}{2}} \,.
\label{LLANMHVsubstitution}
\ee

We wish to extend this relation beyond the LLA.  The same BFKL eigenvalue
will enter the NMHV formula, but in general the NMHV impact factor will receive
different loop corrections than in the MHV case.  We therefore propose the
following ansatz:
\be
\bsp
\cP_{\textrm{NMHV}}\times e^{R_6+i\pi\delta}|_{\textrm{MRK}}
\ =\ \cos\pi\omega_{ab} - i  \frac{a}{2} \sum_{n=-\infty}^\infty
(-1)^n\left(\frac{w}{\ws}\right)^{\frac{n}{2}}
&\int_{-\infty}^{+\infty}
\frac{d\nu}{(i\nu+\frac{n}{2})^2}|w|^{2i\nu}
\, \Phi^{\textrm{NMHV}}_{\textrm{Reg}}(\nu,n)\\
&\hskip0.5cm\times\left(-\frac{1}{1-u_1}\frac{|1+w|^2}{|w|}\right)^{\omega(\nu,n)}
\,.
\esp
\label{extMRK}
\ee
To investigate the validity of this ansatz, we expand $\cP_{\textrm{NMHV}}$
perturbatively in $a$, and then decompose the $L$-loop coefficient
in successive orders of $\ln(1-u_1)$, starting with the leading (LLA)
behavior proportional to $\ln^{L-1}(1-u_1)$.

First we recall the analogous decomposition of the
MHV remainder function used in ref.~\cite{Dixon2012yy}:
\be
R_6^{(L)}(1-u_1,w,\ws) = 2\pi i \, \sum_{r=0}^{L-1}\ln^r(1-u_1)\!\!
\left[g_r^{(L)}(w,\ws)+2\pi i h_r^{(L)}(w,\ws)\right] + \cO(1-u_1)\,.
\label{ghfromR}
\ee
Here $g_r^{(L)}(w,\ws)$ corresponds to the
leading-log approximation (LLA) for $r=L-1$, next-to-LLA (NLLA) for
$r=L-2$, and so on.  Both $g_r^{(L)}$ and $h_r^{(L)}$ are pure functions,
with weight $2L-r-1$ and $2L-r-2$ respectively.  In fact, they 
are single-valued harmonic polylogarithms
(SVHPLs)~\cite{BrownSVHPLs,Dixon2012yy}, particular linear combinations
of harmonic polylogarithms~\cite{Remiddi1999ew} in $w$ and in $\ws$
that are single-valued, or real-analytic, in the $(w,\ws)$ plane.

We take the multi-Regge limit of the $(\eta_4)^4$ component of the ratio
function.  This corresponds to flipping the helicity of outgoing gluon 4
from plus to minus, as we go from MHV to NMHV in the processes
$3^+ 6^+ \, \to \, 2^+ 4^\pm 5^+ 1^+$ displayed in \eqns{MHVconfig}{NMHVconfig}.
In this limit, the $R$-invariants become rational functions of $\ws$.
In particular, we have
\be
(1)\rightarrow\frac{1}{1+\ws}, \qquad (5)\rightarrow \frac{\ws}{1+\ws},
\qquad (6)\rightarrow  1,
\label{RinvsMRK}
\ee
and all of the other $R$-invariants vanish. 

Due to parity symmetry, the ratio function in the MRK limit,
$\cP_{\textrm{MRK}}$, should be 
invariant under $(w,\ws)\rightarrow(1/w,1/\ws)$. This leads us to divide 
up $\cP_{\textrm{MRK}}$ as follows:
\bea
\cP_{\textrm{MRK}}^{(L)} &=& 2\pi i \, \sum_{r=0}^{L-1}\ln^r(1-u_1)
\biggl\{
\frac{1}{1+\ws} \Bigl[ p_r^{(L)}(w,\ws) + 2\pi i\, q_r^{(L)}(w,\ws) \Bigr]
\nonumber\\
&&\hskip3.2cm\null +\frac{\ws}{1+\ws}
\Bigl[ p_r^{(L)}(w,\ws) + 2\pi i\, q_r^{(L)}(w,\ws) \Bigr] 
\Big|_{(w,w^*)\rightarrow(1/w,1/w^*)} \biggr\}
\nonumber\\
&&\hskip0.1cm\null + \cO(1-u_1) \,.
\label{NMHVMRKgeneral}
\eea
The functions $p_r^{(L)}$ and $q_r^{(L)}$ turn out to be pure functions,
in fact they are SVHPLs, just like $g_r^{(L)}$ and $h_r^{(L)}$. 

In order to extract $p_r^{(L)}$ and $q_r^{(L)}$ from the 
ratio function~(\ref{PVform}),
we use \eqn{RinvsMRK} to take the MRK limit of the $R$-invariants,
and then we compare with \eqn{NMHVMRKgeneral}.  We find that,
\be
\bsp
&2\pi i \, \Bigl[ p_r^{(L)}(w,\ws) + 2\pi i\, q_r^{(L)}(w,\ws) \Bigr] \\
&= \frac{1}{2} \Bigl[ V^{(L)}(u_1,u_2,u_3) + V^{(L)}(u_3,u_1,u_2)
+ \tilde{V}^{(L)}(u_1,u_2,u_3) - \tilde{V}^{(L)}(u_3,u_1,u_2)
\Bigr]_{\textrm{MRK},\ \ln^r(1-u_1)\ \textrm{term}} \,.
\esp
\label{pqfromVVt}
\ee
These equations relate the pure functions $p_r^{(L)}$ and $q_r^{(L)}$ to the
MRK limits of $V^{(L)}$ and $\tilde{V}^{(L)}$.
We can take the MRK limits of these functions (or ans\"{a}tze for them)
using their $\{2L-1,1\}$ coproduct components as input to the differential
equation method established in ref.~\cite{Dixon2013eka}.

On the other hand, $p_r^{(L)}$ and $q_r^{(L)}$, together with the MHV
coefficients $g_r^{(L)}$ and $h_r^{(L)}$, can also be related to the BFKL
eigenvalue $\omega(\nu,n)$ and the NMHV impact factor
$\Phi^{\textrm{NMHV}}_{\textrm{Reg}}(\nu,n)$ through the NMHV
master formula~(\ref{extMRK}).
In general, to determine $p_r^{(L)}$ and $q_r^{(L)}$, we have to
evaluate the sum over $n$ and the integral over $\nu$
in \eqn{extMRK}, for a given loop order, a given power of $\ln(1-u_1)$,
and either the real or imaginary part.
We will not give the details of how we perform the sum and integral,
because the general method was described in ref.~\cite{Dixon2012yy}:
We deform the $\nu$ integral into a sum over an integer $m$, and truncate
the sum over $n$ and $m$ at some large value.  Then we match the truncated
sum against the truncated Taylor expansion for a generic linear combination
of SVHPLs with the correct transcendental weight 
for the relevant $p_r^{(L)}$ or $q_r^{(L)}$ coefficient in \eqn{NMHVMRKgeneral},
and the appropriate rational prefactors of $1/(1+\ws)$ and $\ws/(1+\ws)$.
The matching determines the rational number coefficients in the linear
combination.  Once these coefficients are all fixed, we can check them
using higher-order terms in the truncated sum and Taylor expansion.

At LLA, for which $\omega(\nu,n)= -a E_{\nu,n}$ and
$\Phi^{\textrm{NMHV}}_{\textrm{Reg}}(\nu,n)=1$,
the functions $p_{L-1}^{(L)}$ were predicted in ref.~\cite{Lipatov2012gk}
through three loops. We find complete agreement with those predictions.
In our notation, which follows that of ref.~\cite{Dixon2012yy},
we have for the LLA coefficients at one loop,
\be
\bsp
p_0^{(1)}\ =&\ \frac{1}{2} \left[ \frac{1}{2} \, L_0^- - L_1^+ \right] \,, \\
q_0^{(1)}\ =&\ 0 \,,
\esp
\label{pqLLAoneloop}
\ee
at two loops,
\be
\bsp
p_1^{(2)}\ =&\ \frac{1}{4} \left[ L_2^- + \frac{1}{2} \, L_0^- \, L_1^+ 
  - (L_1^+)^2 \right] \,, \\
q_1^{(2)}\ =&\ 0 \,,
\esp
\label{pqLLAtwoloop}
\ee
and at three loops,
\be
\bsp
p_2^{(3)}\ =&\ \frac{1}{4} \left[ L_3^+ + L_{2,1}^-
+ \frac{1}{2} \, L_1^+ \, L_2^- - \frac{1}{16} \, (L_0^-)^3 
- \frac{1}{8} \, (L_0^-)^2 \, L_1^+ - \frac{1}{3} \, (L_1^+)^3
+ \zeta_3 \right] \,,\\
q_2^{(3)}\ =&\ 0 \,.
\esp
\label{pqLLAthreeloop}
\ee

In principle, these LLA results for $p_2^{(3)}$ and $q_2^{(3)}$ could be
used to fix parameters in our three-loop ansatz.  However, once we have imposed
all the previously-mentioned constraints, through the $\cO(T^1)$
terms in the OPE, we find that the two remaining parameters cannot be
fixed by the LLA information.  To see this, let's consider the MRK behavior
of the two functions multiplying these parameters.
These functions were $[\tilde\Phi_6]^2$ and $V^{(1)} \, R_6^{(2)}$.
From ref.~\cite{Dixon2013eka}, we know that the function $\tilde\Phi_6$
is totally symmetric, vanishes in the MRK limit before the analytic
continuation, and has a single discontinuity, with no logarithmic
($\ln(1-u_1)$) enhancement~\cite{Dixon2013eka}:
$\Phi_6 |_{\rm MRK}\ =\ -4\pi i \, L_2^-$.
Hence the square of this function has a double discontinuity,
and no logarithmic enhancement:
\be
[\tilde\Phi_6]^2 |_{\rm MRK}\ =\ (2\pi i)^2 \times 4 \, (L_2^-)^2 \,.
\label{PhisquaredMRK}
\ee
Comparing to \eqn{pqfromVVt}, we see that this function
only contributes to $q_0^{(3)}$, that is, to the NNLLA real part.

The other function with an undetermined coefficient is $V^{(1)} \, R_6^{(2)}$.
Recalling that $\tilde{V}^{(1)}$ vanishes, and inspecting
\eqns{ghfromR}{pqfromVVt}, we see that its behavior in the MRK limit
is
\be
V^{(1)} \times R_6^{(2)} |_{\rm MRK}\ \propto\ 
2\pi i \, p_0^{(1)} \times 
2\pi i \, \Bigl[ \ln(1-u_1) \, g_1^{(2)} + g_0^{(2)} \Bigr] \,.
\label{V1R62MRK}
\ee
Hence this function contributes to both $q_1^{(3)}$ and $q_0^{(3)}$,
which means that we can fix its coefficient using the NLLA real part.

In general, if one knows the N$^{k-1}$LLA imaginary part, 
one can also predict the N$^k$LLA real part from the master
formula~(\ref{extMRK}).  That is because $i\pi$
and $\ln(1-u_1)$ enter the formula in the same way, through $-(1-u_1)$.
Thus the LLA information also predicts the NLLA real part, as pointed
out also in ref.~\cite{Lipatov2012gk}.  The NLLA real part vanishes at
one loop, but it is given at two loops by,
\be
q_0^{(2)}\ =\ \frac{1}{8} \left[ L_2^- - \frac{1}{2} \, L_0^- \, L_1^+
+ (L_1^+)^2 \right] \,,
\label{qNLLAtwoloop}
\ee
and at three loops by,
\be
q_1^{(3)}\ =\ \frac{1}{4} \biggl[ L_3^+ + L_{2,1}^-
- \frac{1}{4} \, (L_0^-)^2 \, L_1^+ - \frac{1}{2} \, L_0^- \, (L_1^+)^2
+ \frac{2}{3} \, (L_1^+)^3 + \zeta_3  \biggr] \,.
\label{qNLLAthreeloop}
\ee
Matching the MRK behavior of our ansatz to the NLLA real part $q_1^{(3)}$
fixes the coefficient of $V^{(1)} \, R_6^{(2)}$.  It only remains to
fix the coefficient of $[\tilde\Phi_6]^2$,using the NNLLA real part $q_0^{(3)}$.

In order to predict both the NLLA imaginary part and the NNLLA real part,
we first need to determine the NLL NMHV impact factor
$\Phi^{\textrm{NMHV}, (1)}_{\textrm{Reg}}(\nu,n)$ entering the master
formula~(\ref{extMRK}).  This impact factor first contributes to the MRK
behavior of the ratio function at two loops, where it determines
$p_0^{(2)}$.  We take the MRK limits of $V^{(2)}$ and $\tilde{V}^{(2)}$ from
ref.~\cite{Dixon2011nj}, and use \eqn{pqfromVVt} to find
\bea
p_0^{(2)} &=& \frac{1}{8} \biggl[ 11 \, L_3^+ - 2 \, L_{2,1}^-
 + \biggl( \frac{3}{2} \, L_0^- + L_1^+ \biggr) \, L_2^-
 - \frac{1}{12} \, (L_0^-)^3 - \frac{3}{2} \, (L_0^-)^2 \, L_1^+
 + L_0^- \, (L_1^+)^2 - \frac{4}{3} \, (L_1^+)^3
\nonumber\\ &&\hskip0.5cm\null
 - 2 \, \zeta_2 \, ( L_0^- - 2 \, L_1^+ ) - 2 \, \zeta_3 \biggr] \,.
\label{pNLLAtwoloop}
\eea
Then we ask what NLL NMHV impact factor
$\Phi^{\textrm{NMHV}, (1)}_{\textrm{Reg}}(\nu,n)$
generates this expression for $p_0^{(2)}$,
via the NMHV master formula~(\ref{extMRK}) evaluated at two loops.

The answer can be expressed quite simply in terms of the corresponding
MHV impact factor, plus a simple rational function\footnote{We thank
Benjamin Basso for suggesting that we try an ansatz of this form.} 
of $\nu$ and $n$:
\be
\Phi^{\textrm{NMHV}, (1)}_{\textrm{Reg}}(\nu,n)
\ =\ \Phi^{\textrm{MHV}, (1)}_{\textrm{Reg}}(\nu,n)
+ \frac{i n \nu }
 {2 \left(-\frac{n}{2}+i \nu \right)^2 \left(\frac{n}{2}+i \nu \right)^2} \,,
\label{PhiNMHVNLLA}
\ee
where $\Phi^{\textrm{MHV}, (1)}_{\textrm{Reg}}(\nu,n)$ is the MHV NLL impact factor,
and is equal to~\cite{Fadin2011we}
\be\label{eq:Phi_1}
\Phi_{\textrm{Reg}}^{\textrm{MHV}, (1)}(\nu,n) 
= -\frac{1}{2}E_{\nu,n}^2 - \frac{3}{8}\,\frac{n^2}{ (\nu^2+\frac{n^2}{4})^2}
 - \zeta_2\,,
\ee
where
\be
E_{\nu,n} = -{1\over2}\,{|n|\over \nu^2+{n^2\over 4}}
+\psi\left(1+i\nu+{|n|\over2}\right) +\psi\left(1-i\nu+{|n|\over2}\right) 
- 2\psi(1)
\label{E_0}
\ee
is the leading-order BFKL eigenvalue.

In order to work out the NLL approximation at three loops, we also
need the NLL BFKL eigenvalue~\cite{Fadin2011we},
\be
E_{\nu,n}^{(1)} = - {1\over4} \, \dE{2}
 + {1\over2} \, V \, \dnu E_{\nu,n} - \zeta_2 \, E_{\nu,n} 
- 3 \, \zeta_3 \,, \label{E_1}
\ee
where 
\be
V \equiv - \frac{1}{2}
  \left[ \frac{1}{i\nu+\frac{|n|}{2}} - \frac{1}{-i\nu+\frac{|n|}{2}} \right]
  = \frac{i \nu}{\nu^2+\frac{|n|^2}{4}} \,,
\label{V_def}
\ee
and $\dnu\equiv-i\partial /\partial \nu$.

Using these functions in the master formula~(\ref{extMRK})
at three loops, we obtain both $p_1^{(3)}$ and $q_0^{(3)}$:
\bea
p_1^{(3)} &=& \frac{1}{16} \biggl[
- 4 \, L_4^- + 2 \, L_{3,1}^+ - 12 \, L_{2,1,1}^-
+ \biggl( - \frac{1}{2} \, L_0^- + 15 \, L_1^+ \biggr) \, L_3^+
+ \bigl( L_0^- + 2 \, L_1^+ \bigr) \, L_{2,1}^-
\nonumber\\ &&\hskip0.5cm\null
+ \Bigl( (L_0^-)^2 + L_0^- \, L_1^+ + 4 \, (L_1^+)^2 \Bigr) \, L_2^-
- \frac{1}{24} \, (L_0^-)^4 - \frac{5}{24} \, (L_0^-)^3 \, L_1^+
- \frac{9}{4} \, (L_0^-)^2 \, (L_1^+)^2
\nonumber\\ &&\hskip0.5cm\null
+ \frac{1}{2} \, L_0^- \, (L_1^+)^3
- \frac{5}{3} \, (L_1^+)^4
- 8 \, \zeta_2 \, \biggl( L_2^- + \frac{1}{2} \, L_0^- \, L_1^+
                       - (L_1^+)^2 \biggr)
\nonumber\\ &&\hskip0.5cm\null
+ \zeta_3 \, \bigl( L_0^- + 2 \, L_1^+ \bigr) \biggr] \,,
\label{p3_1}
\eea
\bea
q_0^{(3)} &=& \frac{1}{16} \biggl[ - 2 \, L_4^- + L_{3,1}^+ - 6 \, L_{2,1,1}^-
+ \biggl( \frac{15}{4} \, L_0^- - \frac{23}{2} \, L_1^+ \biggr) \, L_3^+
+ \biggl( \frac{1}{2} \, L_0^- + 3 \, L_1^+ \biggr) \, L_{2,1}^-
\nonumber\\ &&\hskip0.5cm\null
+ \biggl( \frac{1}{2} \, (L_0^-)^2 - L_0^- \, L_1^+ + (L_1^+)^2 \biggr) \, L_2^-
- \frac{1}{48} \, (L_0^-)^4 - \frac{25}{48} \, (L_0^-)^3 \, L_1^+
+ \frac{11}{8} \, (L_0^-)^2 \, (L_1^+)^2
\nonumber\\ &&\hskip0.5cm\null
- \frac{17}{12} \, L_0^- \, (L_1^+)^3 + \frac{11}{6} \, (L_1^+)^4
- 4 \, \zeta_2 \, \biggl( L_2^- - \frac{1}{2} \, L_0^- \, L_1^+
                       + (L_1^+)^2 \biggr)
\nonumber\\ &&\hskip0.5cm\null
+ \zeta_3 \, \biggl( \frac{1}{2} \, L_0^- + 3 \, L_1^+ \biggr) \biggr] \,.
\label{q3_0}
\eea
The MRK limit of our ansatz for $V^{(3)}$ and $\tilde{V}^{(3)}$,
inserted into \eqn{pqfromVVt}, yields complete agreement with
these expressions.  The agreement with $q_0^{(3)}$ fixes the one remaining
parameter in the ansatz, namely the coefficient of $[\tilde\Phi_6]^2$.

Finally, having fixed the ansatz, we turn to NNLLA.   Three loops is the
first order in which a truly NNLLA quantity appears, namely $p_0^{(3)}$.
Thus $p_0^{(3)}$ cannot be predicted using lower-loop information.
Extracting it from our function gives novel data. We find that
\bea
p_0^{(3)} &=&  \frac{1}{16} \biggl[
- 87 \, L_5^+ + 4 \, L_{4,1}^- - 14 \, L_{3,1,1}^+ + 12 \, L_{2,1,1,1}^-
- \bigl( 7 \, L_0^- + 2 \, L_1^+ \bigr) \, L_4^-
+ \biggl( \frac{1}{2} \, L_0^- + L_1^+ \biggr) \, L_{3,1}^+
\nonumber\\ &&\hskip0.5cm\null
- 3 \, \bigl( L_0^- + 2 \, L_1^+ \bigr) \, L_{2,1,1}^-
+ \biggl( \frac{45}{4} \, (L_0^-)^2 - \frac{1}{2} \, L_0^- \, L_1^+
        + 11 \, (L_1^+)^2 \biggr) \, L_3^+
\nonumber\\ &&\hskip0.5cm\null
+ \Bigl( - (L_0^-)^2 + 4 \, L_0^- \, L_1^+ - 2 \, (L_1^+)^2 \Bigr) \, L_{2,1}^-
\nonumber\\ &&\hskip0.5cm\null
+ \biggl( \frac{17}{16} \, (L_0^-)^3 + \frac{3}{8} \, (L_0^-)^2 \, L_1^+ 
  + \frac{5}{4} \, L_0^- \, (L_1^+)^2 + \frac{3}{2} \, (L_1^+)^3 \biggr) \, L_2^-
+ \frac{3}{80} \, (L_0^-)^5 - \frac{5}{4} \, (L_0^-)^4 \, L_1^+
\nonumber\\ &&\hskip0.5cm\null
+ \frac{1}{24} \, (L_0^-)^3 \, (L_1^+)^2
- \frac{13}{6} \, (L_0^-)^2 \, (L_1^+)^3 + \frac{1}{2} \, L_0^- \, (L_1^+)^4
- \frac{8}{15} \, (L_1^+)^5
\nonumber\\ &&\hskip0.5cm\null
+ \zeta_2 \, \biggl( - 32 \, L_3^+ - 16 \, L_{2,1}^-
            - 4 \, ( L_0^- - 2 \, L_1^+ ) \, L_2^-
            + \frac{5}{6} \, (L_0^-)^3 + 5 \, (L_0^-)^2 \, L_1^+
            - 4 \, L_0^- \, (L_1^+)^2
\nonumber\\ &&\hskip0.5cm\null
            + \frac{40}{3} \, (L_1^+)^3 \biggr)
+ \zeta_3 \, \biggl( \frac{5}{2} \, (L_0^-)^2 + 3 \, L_0^- \, L_1^+
                   - 6 \, (L_1^+)^2 \biggr)
+ 22 \, \zeta_4 \, ( L_0^- - 2 \, L_1^+ )
\nonumber\\ &&\hskip0.5cm\null
+ 30 \, \zeta_5 - 16 \, \zeta_2 \, \zeta_3 \biggr] \,.
\label{p3_0}
\eea
In an ancillary file, we provide computer-readable expressions for
all the $p_r^{(L)}$ and $q_r^{(L)}$ functions for $L=1,2,3$.

Knowledge of $p^{(3)}_0$ allows us to fix the NNLL impact factor
$\Phi^{\textrm{NMHV}, (2)}_{\textrm{Reg}}(\nu,n)$, in the same way that
we used $p^{(2)}_0$ to determine the NMHV impact factor at NLL.
Again we find that the NMHV impact factor can be expressed simply
in terms of the MHV impact factor and rational functions of $\nu$ and $n$:
\be
\bsp
\Phi^{\textrm{NMHV}, (2)}_{\textrm{Reg}}(\nu,n)
\ =&\ \Phi^{\textrm{MHV}, (2)}_{\textrm{Reg}}(\nu,n)
\ +\ \left(\Phi^{\textrm{MHV}, (1)}_{\textrm{Reg}}(\nu,n)+\zeta_2\right)
 \frac{i n \nu }{2 \left(-\frac{n}{2}+i \nu \right)^2
   \left(\frac{n}{2}+i \nu \right)^2}\\
&\null -\frac{i n \nu \left(n^2 - i n \nu -2 \nu^2\right)}
       {8 \left(-\frac{n}{2}+i \nu \right)^4
   \left(\frac{n}{2}+i \nu \right)^4} \,.
\esp
\label{PhiNMHVNNLL}
\ee
This formula has recently been reproduced, and extended
to all orders, using a kind of analytic continuation from the
near-collinear limit~\cite{BCHS}.

The all-orders formula is expressed
in terms of the Zhukovsky variable $x(u)$ defined in
\eqn{Zhukovsky}.  It reads,\footnote{We thank Benjamin Basso for discussions
on these points prior to the appearance of ref.~\cite{BCHS}.}
in our definition of $\Phi$,
\be
\Phi^{\textrm{NMHV}}_{\textrm{Reg}}(\nu,n) = \Phi^{\textrm{MHV}}_{\textrm{Reg}}(\nu,n)
\times \frac{\nu-\frac{in}{2}}{\nu+\frac{in}{2}}
\, \frac{x(u+\frac{in}{2})}{x(u-\frac{in}{2})} \,.
\label{BCHSallorders}
\ee
The rapidity $u$ in this expression is related to the variable $\nu$
by an integral expression~\cite{BCHS}.  (Our $\nu$ is defined to
be precisely $1/2$ of the $\nu$ defined in ref.~\cite{BCHS}, while
our $n$ is just their $m$.)
The integrals can be performed in the weak coupling expansion, and
the equation for $\nu(u,n)$ can be inverted to solve for $u(\nu,n)$,
order-by-order in the coupling.
The first three orders are enough for us here,
\be 
u = \nu - \frac{i}{2} \, a \, V
+ \frac{i}{8} \, a^2 \, V \, ( N^2 + 4 \, \zeta_2 )
+ {\cal O}(a^3), 
\label{u_nu}
\ee
where $V=i\nu/(\nu^2+n^2/4)$ is defined in \eqn{V_def} and
$N=n/(\nu^2+n^2/4)$.  Through this order,
the relation between $u$ and $\nu$ only involves rational functions
of $\nu$ and $n$.  Inserting the expansion~(\ref{u_nu}) into
\eqn{BCHSallorders} yields both \eqn{PhiNMHVNLLA} for 
$\Phi^{\textrm{NMHV}, (1)}_{\textrm{Reg}}(\nu,n)$ and \eqn{PhiNMHVNNLL} for 
$\Phi^{\textrm{NMHV}, (2)}_{\textrm{Reg}}(\nu,n)$.  At the next loop order,
the relation~(\ref{u_nu}) begins to contain $\psi$ functions, which
should then enter the formula for $\Phi^{\textrm{NMHV}, (3)}_{\textrm{Reg}}(\nu,n)$
in terms of $\Phi^{\textrm{MHV}}_{\textrm{Reg}}(\nu,n)$.  It would be
interesting to check this statement once the four-loop ratio
function is determined.

The ratio in \eqn{BCHSallorders} might appear to be upside-down with
respect to ref.~\cite{BCHS}.  However, we defined $\Phi^{\textrm{NMHV}}$
for the $(\eta_4)^4$ Grassmann component of the NMHV super-amplitude,
while it was defined
for the $(\eta_1)^4$ component in ref.~\cite{BCHS}.  The two components
are related by the cyclic permutation that inverts all the $y_i$ variables,
which exchanges $w \lr w^*$ and therefore takes $n \lr -n$.

Using \eqn{PhiNMHVNNLL} and the known BFKL NNLL eigenvalue, we have the
information necessary to find the NNLLA imaginary part and N${}^3$LLA
real part to all loop orders. (Of course, the very recent
all-orders formulae~\cite{BCHS} could be used to go well beyond this.) 
While we do not pursue this exercise here, such fixed-order data in
the $(w,\ws)$ space will prove quite useful during the construction of the
ratio function at four loops and beyond.


\section{Multi-particle factorization}
\label{multiparticle}

\begin{figure}
\begin{center}
\includegraphics[width=6.5in]{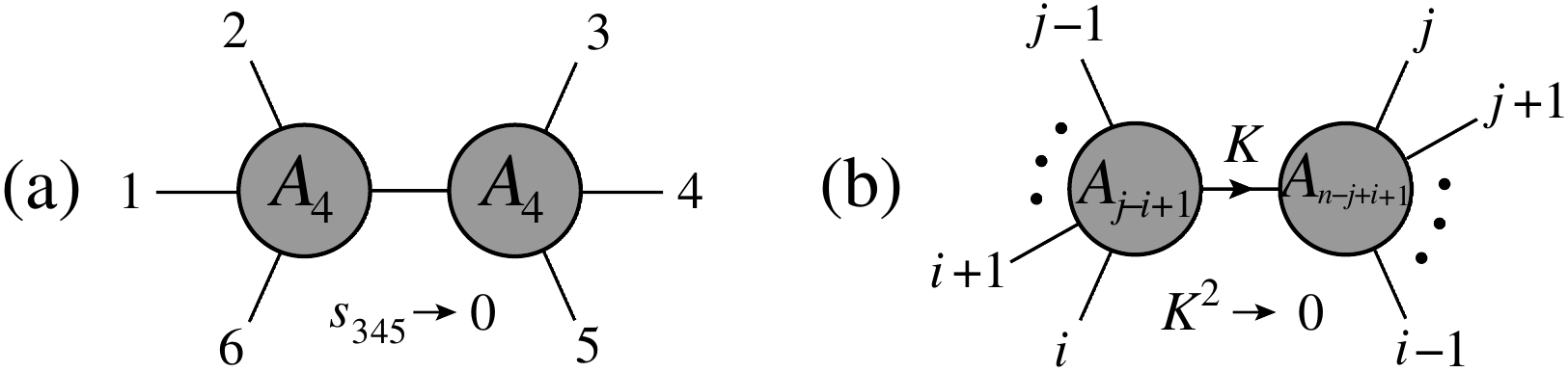}
\end{center}
\caption{(a) Multi-particle factorization of a six-point amplitude into
two four-point amplitudes, in the limit $s_{345}\to0$. (b) The most general
multi-particle factorization of an $n$-point amplitude into a $(j-i+1)$-point
amplitude and an $(n-j+i+1)$-point amplitude, in the limit that
$K^2 = s_{i,i+1,\ldots,j-1}\to0$.}
\label{fig:multi}
\end{figure}

A six-point amplitude can factorize onto a product of four-point
amplitudes in the limit that a three-particle momentum invariant
goes on shell, $s_{i,i+1,i+2} \equiv (k_i+k_{i+1}+k_{i+2})^2 \to 0$.
This limit is called a multi-particle factorization limit,
in order to distinguish it from the two-particle factorization limits,
or collinear limits.  The multi-particle factorization limit
of the six-gluon amplitude, in which the invariant $s_{345} \to 0$,
is shown in figure~\ref{fig:multi}(a).  We will discuss this limit
first, and later consider the most general multi-particle factorization
of an $n$-point amplitude, shown in figure~\ref{fig:multi}(b).

In supersymmetric theories, all multi-particle
poles of MHV amplitudes have zero residue, because of the same helicity
counting rules that apply at tree level~\cite{ParkeTaylor}:
Each four-point amplitude needs
to have two negative and two positive external helicities.  One negative
and one positive helicity must be assigned to the virtual gluon crossing the
pole, leaving three negative- and three positive-helicity external gluons,
{\it i.e.}~the NMHV helicity configuration.  Using the three-loop
ratio function, we can extract the multi-particle factorization
behavior of six-point amplitudes in planar $\NeqFour$ SYM through three loops.
We will find that it is remarkably simple, containing no function more
complicated than the logarithm.   The simplicity of the six-point
factorization leads to a natural conjecture for the $n$-point case.

First we review the general factorization behavior and what is known
at one loop from the work of Bern and Chalmers~\cite{BernChalmers}.
For definiteness, we will factorize
the amplitude in the limit that $s_{345} = K^2 \to 0$, where 
$K = k_3 + k_4 + k_5$.  The only two dual superconformal invariants
``$(i)$'' that contain a pole in  the $s_{345}$ channel are $(1)$ and $(4)$.
They become equal in this limit.
Furthermore the dual conformal cross ratios $u$ and $w$ contain
$s_{345}$ in the denominator, while $v$ does not contain it.
Therefore the factorization limit of the ratio function $\cP$
in the $s_{345}$ channel will be obtained by letting $u,w\to\infty$ in 
$V(u,v,w)$, with $u/w$ and $v$ held fixed.  The odd part $\tilde{V}(u,v,w)$
will not contribute at any loop order in this limit, because
it multiplies $(1)-(4)$, which is power-suppressed in the limit.

We also assume that the component of the NMHV amplitude has
been chosen so that there are two negative
helicities on one side of the pole, and one negative-helicity
on the other side, so that the multi-particle factorization is non-trivial
at tree-level. Then we can define an all loop order factorization function
$F_6$ by,
\be
A_6^{\rm NMHV}(k_i)\ \mathop{\longrightarrow}^{s_{345} \to 0}\
A_4(k_6,k_1,k_2,K) \, \frac{F_6(K^2,s_{i,i+1})}{K^2} \, A_4(-K,k_3,k_4,k_5) \,,
\label{facts345}
\ee
where $A_6$ and $A_4$ are all-orders amplitudes.
For the given choice of external helicities, there is only one
nontrivial assignment of the intermediate gluon helicity.

When we expand \eqn{facts345} out to one loop,
we obtain~\cite{BernChalmers}
\bea
A_6^{\rm NMHV\ (1)}(k_i)\ \mathop{\longrightarrow}^{s_{345} \to 0} &&
A_4^{(1)}(k_6,k_1,k_2,K) \frac{1}{K^2} A_4^{(0)}(-K,k_3,k_4,k_5)
\nonumber\\ &&\null\hskip-0.5cm
+ A_4^{(0)}(k_6,k_1,k_2,K) \frac{1}{K^2} A_4^{(1)}(-K,k_3,k_4,k_5)
\nonumber\\ &&\null\hskip-0.5cm
+ A_4^{(0)}(k_6,k_1,k_2,K) \frac{F_6^{(1)}}{K^2} A_4^{(0)}(-K,k_3,k_4,k_5)\,,
\label{facts345oneloop}
\eea
which defines the one-loop factorization function, $F_6^{(1)}$.
This function was computed in $\NeqFour$ SYM in ref.~\cite{BernChalmers}.
Setting $\mu=1$, multiplying by $(4\pi)^2/2$ to account for a difference 
in expansion parameters, and permuting the indices appropriately for the 
$s_{345}$ channel, the function is,
\bea
F_6^{(1)} &=&
- \frac{1}{\e^2}
   \Bigl[ (-s_{23})^{-\e} - (-s_{61})^{-\e} - (-s_{45})^{-\e} \Bigr]
 - \frac{1}{2\e^2}\frac{(-s_{61})^{-\e}(-s_{45})^{-\e}}{(-s_{23})^{-\e}}
\nonumber\\ &&\null\hskip0.0cm
 + \frac{1}{2} \ln^2\Bigl( \frac{-s_{23}}{-s_{345}} \Bigr)
 - \frac{1}{2} \ln^2\Bigl( \frac{-s_{45}}{-s_{345}} \Bigr) 
 - \frac{1}{2} \ln^2\Bigl( \frac{-s_{61}}{-s_{345}} \Bigr) 
 - 2 \zeta_2
\nonumber\\ &&\null\hskip0.0cm
+ \{ k_3 \lr k_6, \, k_4 \lr k_1, \, k_5 \lr k_2 \}
\label{BCfact1}\\
&=& \frac{1}{2} \biggl\{ 
 \frac{1}{\e^2} \biggl[
     \biggl(\frac{(-s_{12})(-s_{34})}{(-s_{56})}\biggr)^{-\e}
   + \biggl(\frac{(-s_{45})(-s_{61})}{(-s_{23})}\biggr)^{-\e} \biggr]
\nonumber\\ &&\null\hskip0.5cm
  - \frac{1}{2} \biggl[
      \ln\biggl(\frac{(-s_{12})(-s_{34})}{(-s_{56})}\biggr)
    - \ln\biggl(\frac{(-s_{45})(-s_{61})}{(-s_{23})}\biggr) \biggr]^2
  - \frac{1}{2} \ln^2(uw/v) - 8 \, \zeta_2 \biggr\} \,.
\label{BCfact2}
\eea
Using this function, together with the one-loop MHV four-point and six-point
amplitudes (which also enter the BDS ansatz), we will be able to predict
the behavior of the ratio function in the factorization limit
at one loop.  Later we will turn the argument around, and use
the factorization behavior of the two- and three-loop NMHV amplitudes
to determine the higher-loop factorization functions $F_6^{(L)}$.

First we record the required one-loop amplitudes, after dividing
by their respective tree amplitudes, which can be factored out of
\eqn{facts345oneloop}.  Ref.~\cite{Unitarity} gives
the sum of the two required four-point amplitudes, in
terms of functions called $V_4$ there,
\bea
\frac{1}{2} \Bigl[ V_4 + V_4^\prime \Bigr]
&=& \frac{1}{2} \biggl\{ -\frac{2}{\e^2} 
   \Bigl[ (-s_{34})^{-\e} + (-s_{45})^{-\e} + (-s_{61})^{-\e} + (-s_{12})^{-\e} \Bigr]
\nonumber\\ &&\null\hskip0.5cm
+ \ln^2\Bigl(\frac{-s_{34}}{-s_{45}}\Bigr)
+ \ln^2\Bigl(\frac{-s_{61}}{-s_{12}}\Bigr) + 12 \, \zeta_2 \biggr\} \,.
\label{V4sum}
\eea
The six-point amplitude is also given in terms of the function $V_6$,
\bea
\frac{1}{2} V_6 &=&
\frac{1}{2} \Biggl\{ \sum_{i=1}^6 \biggl[
  -\frac{1}{\e^2} (-s_{i,i+1})^{-\e}
 - \ln\Bigl(\frac{-s_{i,i+1}}{-s_{i,i+1,i+2}}\Bigr)
   \ln\Bigl(\frac{-s_{i+1,i+2}}{-s_{i,i+1,i+2}}\Bigr)
 + \frac{1}{4} \ln^2\Bigl(\frac{-s_{i,i+1,i+2}}{-s_{i+1,i+2,i+3}}\Bigr) \biggl]
\nonumber\\ &&\null\hskip0.5cm
 - \Li_2(1-u)  - \Li_2(1-v) - \Li_2(1-w) + 6 \, \zeta_2 \Biggr\}
\label{V6A}\\
&=& \frac{1}{2} \Biggl\{ \sum_{i=1}^6 \biggl[
  - \frac{1}{\e^2} \Bigl( 1 - \e \ln(-s_{i,i+1}) \Bigr)
  - \ln(-s_{i,i+1})\ln(-s_{i+1,i+2})
  + \frac{1}{2} \ln(-s_{i,i+1})\ln(-s_{i+3,i+4}) \biggr]
\nonumber\\ &&\null\hskip0.5cm
  - Y(u,v,w) + 6 \, \zeta_2 \Biggr\} \,,
\label{V6B}
\eea
where
\be
Y(u,v,w)\ \equiv\ H_2^u + H_2^v + H_2^w
         + \frac{1}{2} \Bigl( \ln^2u + \ln^2v + \ln^2w \Bigr) \,.
\label{Ydef}
\ee

We will be interested in the combination,
\bea
\frac{1}{2} \Bigl[ V_4 + V_4^\prime - V_6 \Bigr]
&=& \frac{1}{2} \biggl\{
 - \frac{1}{\e^2} \biggl[ \biggl(\frac{(-s_{12})(-s_{34})}{(-s_{56})}\biggr)^{-\e}
   + \biggl(\frac{(-s_{45})(-s_{61})}{(-s_{23})}\biggr)^{-\e} \biggr]
\nonumber\\ &&\null\hskip0.3cm
   + \frac{1}{2} \biggl[
      \ln\biggl(\frac{(-s_{12})(-s_{34})}{(-s_{56})}\biggr)
    - \ln\biggl(\frac{(-s_{45})(-s_{61})}{(-s_{23})}\biggr) \biggr]^2
\nonumber\\ &&\null\hskip0.3cm
   + Y(u,v,w) + 6 \, \zeta_2 \biggr\} \,.  \label{V46sum}
\eea
We note that the sum of this quantity with $F_6^{(1)}$ is
dual conformal invariant, even before we enter the factorization limit:
\be
F_6^{(1)} + \frac{1}{2} \Bigl[ V_4 + V_4^\prime - V_6 \Bigr]
= \frac{1}{2} \, Y(u,v,w) - \frac{1}{4} \ln^2(uw/v) - \zeta_2 \,.
\label{FV46sum}
\ee

At one loop, when we divide the left-hand side of \eqn{facts345} by
$A_6^{\rm MHV}$, and expand out to first order, the tree factors correspond
to the superconformal invariant $(1)$. The limiting behavior of $V^{(1)}(u,v,w)$
is then given by, using also \eqn{FV46sum},
\bea
V^{(1)}(u,v,w)\bigr|_{u,w\to\infty} &=& F_6^{(1)}
   + \frac{1}{2} \Bigl[ V_4 + V_4^\prime - V_6 \bigr|_{u,w\to\infty} \Bigr]
\nonumber\\
&=& - \frac{1}{4} \ln^2(uw/v) - 2 \, \zeta_2
 + \frac{1}{2} \Bigl( \Li_2(1-v) + \frac{1}{2} \ln^2 v \Bigr) \,.
\label{V1fromFact}
\eea
The result is manifestly finite and dual conformally invariant.
It also matches perfectly against the limit $u,w\to\infty$ of
the known one-loop expression, in the form~(\ref{Voneloop}) given in
ref.~\cite{Dixon2011nj}.
In fact, we note from \eqns{Voneloop}{FV46sum} that
\be
V^{(1)}(u,v,w)
= F_6^{(1)} + \frac{1}{2} \Bigl[ V_4 + V_4^\prime - V_6 \Bigr]
\label{V1equalsFV46sum}
\ee
even outside of the factorization limit.

Now we proceed to higher loops.  At this point we should be careful
to consider the actual NMHV amplitude, not the ratio function.
The ratio function does not have a simple factorization limit because
it treats the MHV amplitude on the same footing as the NMHV
amplitude.  However, there is no tree-level pole for the MHV amplitude,
so there is no reason for the transcendental function multiplying
the tree amplitude to have a simple form in the factorization limit.
In order to do this, and still deal with a finite, dual conformally invariant
quantity for a while longer, we multiply the ratio function by
the (exponentiated) remainder function.  It is also convenient to take
the logarithm.  Whereas the remainder function $R=R_6$ is defined by
\be 
\frac{A^{\rm MHV}}{A^{\rm BDS}} = \exp(R),
\label{Rdef}
\ee
here we define
\be 
\frac{A^{\rm NMHV}}{A^{\rm BDS}} 
= \frac{A^{\rm NMHV}}{A^{\rm MHV}} \times \frac{A^{\rm MHV}}{A^{\rm BDS}}
= \cP \times \exp(R) 
 \equiv (1) \times \exp(\hat{U}).
\label{Uhatdef}
\ee
We call the function $\hat{U}$ because it will be useful to
adjust it slightly later.
In the factorization limit, the tree-(super)amplitude prefactor
in $\cP$ collapses to $(1)$ and we can identify $Ve^R = e^{\hat{U}}$,
or
\be
\hat{U}(u,v,w) = \ln V(u,v,w) + R_6(u,v,w),
\label{UhatdefALT}
\ee
so that the perturbative expansion of $\hat{U}$ is,
\bea
&&\hat{U}^{(1)} = V^{(1)} \,, \label{Uh1}\\
&&\hat{U}^{(2)} = V^{(2)} - \frac{1}{2} [V^{(1)}]^2 + R_6^{(2)} \,, \label{Uh2}\\
&&\hat{U}^{(3)} = V^{(3)} + \frac{1}{3} [V^{(1)}]^3 - V^{(1)} V^{(2)} + R_6^{(3)} \,,
\label{Uh3}
\eea
where we used the fact that the remainder function only becomes
nonvanishing starting at two loops.

We also need to evaluate the BDS ansatz~\cite{BDS},
\be
\ln A_n^{\rm BDS}\ =\ 
\sum_{L=1}^\infty a^L \Bigl( f^{(L)}(\e) \frac{1}{2} V_n(L\e) + C^{(L)} \Bigr) \,,
\label{BDSAnsatz}
\ee
where
\be
f^{(L)}(\e) \equiv f_0^{(L)} + \e \, f_1^{(L)} + \e^2 \, f_2^{(L)} \,.
\label{fepsdef}
\ee
Two of the constants,
\be
f_0^{(L)}\ =\ \frac{1}{4} \, \gamma_K^{(L)} \,, \qquad
f_1^{(L)}\ =\ \frac{L}{2} \, {\cal G}_0^{(L)} \,,
\label{cuspG}
\ee
are given in terms of the planar cusp anomalous dimension $\gamma_K$
(see \eqn{cuspdef} below)
and the ``collinear'' anomalous dimension ${\cal G}_0$, 
while $f_2^{(L)}$ and $C^{(L)}$ are other (zeta-valued) constants.
The $L$-loop coefficient of the combination we need,
$\ln(A_4^{\rm BDS}\times A_4^{\rm BDS'}/A_6^{\rm BDS})$,
where $A_4^{\rm BDS^{(\prime)}}$ are the ans\"{a}tze for the two four-point
subprocesses, is closely related to \eqn{V46sum}:
\bea
\ln\biggl(\frac{A_4^{\rm BDS} \, A_4^{\rm BDS'}}{A_6^{\rm BDS}}\biggr)^{(L)}
&=& - \frac{\gamma_K^{(L)}}{8\e^2L^2}
 \biggl( 1 + 2 \, \e \, L \, \frac{{\cal G}_0^{(L)}}{\gamma_K^{(L)}} \biggr)
  \biggl[ \biggl(\frac{(-s_{12})(-s_{34})}{(-s_{56})}\biggr)^{-L\e}
   + \biggl(\frac{(-s_{45})(-s_{61})}{(-s_{23})}\biggr)^{-L\e} \biggr]
\nonumber\\ &&\null\hskip0.0cm
  + \frac{\gamma_K^{(L)}}{8}\biggl[
     \frac{1}{2} \ln^2 \biggl(\frac{(-s_{12})(-s_{34})}{(-s_{56})}
        \bigg/ \frac{(-s_{45})(-s_{61})}{(-s_{23})}\biggr)
   + Y(u,v,w) + 6 \, \zeta_2 \biggr]
\nonumber\\ &&\null\hskip0.0cm
   - \frac{f_2^{(L)}}{L^2} - C^{(L)} \,.
\label{BDScombination}
\eea

Because of the appearance of the function $Y(u,v,w)$ in $A_6^{\rm BDS}$
and in \eqn{BDScombination},
it is useful to define a function $U(u,v,w)$ that absorbs this
function:
\be
U(u,v,w) = \hat{U}(u,v,w) - \frac{\gamma_K}{8} Y(u,v,w) \,.
\label{Udef}
\ee
We will see that $U$ has simpler analytic properties than $\hat{U}$,
even outside of the factorization limit.  At one loop, we have
\be
U^{(1)}(u,v,w) = - \frac{1}{4} \ln^2(uw/v) - \zeta_2 \,,
\label{U1}
\ee
so the polylogarithms have cancelled from $U^{(1)}$.

In ref.~\cite{Dixon2011nj},
$V^{(2)}(u,v,w)$ was given in terms of one-dimensional HPLs,
plus the three independent permutations of the function $\Omega^{(2)}(u,v,w)$.
The two-loop remainder function $R_6^{(2)}$ was also given, in a similar
form.  In the sum of $V^{(2)}$ and $R_6^{(2)}$ entering $U^{(2)}$,
two of the permutations cancel, and only the permutation
$\Omega^{(2)}(w,u,v)$ survives.
In total, before taking the factorization limit, $U^{(2)}$ as
defined by \eqns{Uh2}{Udef} is given by,
\bea
U^{(2)}(u,v,w) &=& \frac{1}{4} \biggl\{ - \Omega^{(2)}(w,u,v)
 - H_{4}^u - H_{4}^w
 - 3 \, \Bigl( H_{3,1}^u + H_{3,1}^w -  H_{2,1,1}^u - H_{2,1,1}^w \Bigr)
\nonumber\\ &&\null\hskip0.5cm
 + \frac{1}{2} \, \Bigl[ (H_{2}^u)^2 + (H_{2}^w)^2 \Bigr]
 + 2 \, \Bigl( \ln u \, H_{2,1}^u + \ln w \, H_{2,1}^w \Bigr)
 - \ln(w/v) \, ( H_{3}^u + H_{2,1}^u )
\nonumber\\ &&\null\hskip0.5cm
 - \ln(u/v) \, ( H_{3}^w + H_{2,1}^w )
 + \frac{1}{2} \, \ln(uw/v)
       \, \Bigl( \ln(uv/w) \, H_{2}^u + \ln(wv/u) \, H_{2}^w \Bigr)
\nonumber\\ &&\null\hskip0.5cm
 - \frac{1}{2} \, \Bigl( \ln u \, \ln w  - 8 \, \zeta_2 \Bigr)
       \, \Bigl( \ln u \, \ln w - \ln v \, \ln(uw) \Bigr)
 - \frac{1}{4} \, \Bigl( \ln^2 u + \ln^2 w \Bigr) \, \ln^2 v
\nonumber\\ &&\null\hskip0.5cm
 - \zeta_2 \, \Bigl[ 2 \, ( H_{2}^u + H_{2}^w )
           - \ln^2 u - \ln^2 w - 2\, \ln^2 v \Bigr]
 + 15 \, \zeta_4 \biggr\} \,.
\label{U2}
\eea
Note that the dependence on $v$ is particularly simple; 
aside from $\Omega^{(2)}(w,u,v)$, the only
function of $v$ that appears is $\ln v$.

We wish to use the coproduct formalism to extract the
behavior of $\Omega^{(2)}(w,u,v)$ in the factorization limit.
This exercise will be a useful warmup for obtaining the limit
of the NMHV amplitude at three loops.  First we recall~\cite{Dixon2013eka}
the formula for the $u$-derivative of a generic hexagon function $F$,
holding $v$ and $w$ fixed:
\be
\label{eq:diff_basis}
\frac{\partial F}{\partial u}\bigg|_{v,w} 
\ =\ \frac{F^u}{u} - \frac{F^{1-u}}{1-u} 
+ \frac{1-u-v-w}{u\sqrt{\Delta}} F^{y_u}
+ \frac{1-u-v+w}{(1-u)\sqrt{\Delta}}F^{y_v}
+ \frac{1-u+v-w}{(1-u)\sqrt{\Delta}} F^{y_w}\,.
\ee
We can permute this relation cyclicly in order to obtain the $v$-
and $w$-derivatives.
Now we take the limit of \eqn{eq:diff_basis} as $u,w\to\infty$,
finding
\bea
\partial_u F &=& \frac{1}{u} \biggl[ F^u + F^{1-u}
- \frac{1}{r} ( F^{y_u} - F^{y_w} ) + \frac{u-w}{(u+w)r} F^{y_v} \biggr] \,,
\label{duFfact}\\
\partial_v F &=& \frac{1}{v} \biggl[ F^v - \frac{1}{r} F^{y_v} \biggr]
 - \frac{1}{1-v} \biggl[ F^{1-v} + \frac{u-w}{(u+w)r} ( F^{y_u} - F^{y_w} ) \biggr]
\,,
\label{dvFfact}
\eea
where
\be
r = \sqrt{ 1 - \frac{4uvw}{(u+w)^2} } \,.
\label{rdef}
\ee
The $w$-derivative is obtained from the $u$-derivative simply by exchanging
$u$ and $w$ labels everywhere.

We see from \eqns{duFfact}{dvFfact} that the factorization limit of
a hexagon function is likely to be simple, with all the occurrences of $r$
dropping out, if two conditions on the $\{n-1,1\}$ coproduct
elements are met:
\be
F^{y_u} = F^{y_w}\quad\hbox{and}\quad F^{y_v} = 0.
\label{simplefact}
\ee
We will see that this condition is satisfied by the specific combinations
of nontrivial hexagon functions that we need for taking the limits
of $U$, through three loops.

First consider the function $\Omega^{(2)}(w,u,v)$.
By permuting eqs.~(B.10) and (B.11) of ref.~\cite{Dixon2013eka},
we see that
\be
[\Omega^{(2)}(w,u,v)]^{y_u} = [\Omega^{(2)}(w,u,v)]^{y_w}
\quad\hbox{and}\quad [\Omega^{(2)}(w,u,v)]^{y_v} = 0,
\label{simplefactOm2}
\ee
so $\Omega^{(2)}(w,u,v)$ should have a simple limit.
However, we see further that 
\bea
&&[\Omega^{(2)}(w,u,v)]^{v} = [\Omega^{(2)}(w,u,v)]^{1-v} = 0, 
\label{zerofactOm2A} \\
&&[\Omega^{(2)}(w,u,v)]^{u} + [\Omega^{(2)}(w,u,v)]^{1-u}
= [\Omega^{(2)}(w,u,v)]^{w} + [\Omega^{(2)}(w,u,v)]^{1-w} = 0,
\label{zerofactOm2B}
\eea
which means that all the derivatives of $\Omega^{(2)}(w,u,v)$
vanish in the factorization limit.
Therefore $\Omega^{(2)}(w,u,v)$ can at most be a constant in this
limit.

To fix the constant, we consider the line $(u,1,u)$ as
$u\to\infty$.  On this line, all hexagon functions collapse
to one-dimensional HPLs, so it is easy to take the large $u$ limit.
Here we need,
\bea
\Omega^{(2)}(u,u,1) &=& - 2 \, H_4^u - 2 \, H_{3,1}^u + 6 \, H_{2,1,1}^u
  + 2 \, ( H_2^u )^2
  + 2 \, \ln u \, ( H_3^u + H_{2,1}^u )
\nonumber\\ &&\null
  + \ln^2 u \, H_2^u  + \frac{1}{4} \, \ln^4 u - 6 \, \zeta_4 \,.
\label{Om2_uu1}
\eea
Using standard identities for inverting the arguments of the HPLs,
we find that this function vanishes as $u\to\infty$.
Therefore
\be
\Omega^{(2)}(w,u,v)\bigr|_{u,w\to\infty} = 0.
\label{Om2factvanish}
\ee

Aside from $\Omega^{(2)}(w,u,v)$, the remaining terms in \eqn{U2}
for $U^{(2)}$ are one-dimensional HPLs with arguments $u$, $v$ and $w$.
The same HPL argument-inversion identities allow us to extract
the limiting behavior of the HPLs in $u$ and $w$ terms.
The final result has the simple form,

\bea
U^{(2)}(u,v,w)\bigr|_{u,w\to\infty} &=&
\frac{3}{4} \, \zeta_2 \, \ln^2(uw/v)
- \frac{1}{2} \, \zeta_3 \, \ln(uw/v)
+ \frac{71}{8} \, \zeta_4 \,.
\label{U2_fact}
\eea
Remarkably, the limit of $U^{(2)}$ is simply a polynomial in $\ln(uw/v)$
with zeta-valued coefficients.

Turning now to three loops, we find that the $\{5,1\}$ coproducts
of $U^{(3)}$ obey the relations~(\ref{simplefact})
required for a simple factorization limit.
For example,
\bea
[U^{(3)}]^{y_u} &=&
\frac{1}{32} \Bigl[ 3 \, H_1(u,v,w) + H_1(v,w,u) + H_1(w,u,v) \Bigr]
\nonumber\\&&\null
- \frac{1}{128} \Bigl[ 11 \, J_1(u,v,w) + J_1(v,w,u) + J_1(w,u,v) \Bigr]
\nonumber\\&&\null
+ \frac{1}{32} \tilde\Phi_6(u,v,w) \Bigl[ \ln^2 u + \ln^2 w + \ln^2 v
   + 2 \, \Bigl( \ln u \, \ln w - \ln(uw) \, \ln v \Bigr)
   - 22 \, \zeta_2 \Bigr]\,.
\nonumber\\&~&
\label{U3yu}
\eea
Because the functions $H_1$ and $J_1$ are symmetric under exchange
of their first and third argument~\cite{Dixon2013eka},
and $\tilde\Phi_6$ is totally symmetric, we see that \eqn{U3yu}
is symmetric under $u\lr w$.  But $U^{(3)}(u,v,w)$ is symmetric
under the exchange of $u\lr w$.  Together, these two properties
imply that
\be
[U^{(3)}]^{y_u} = [U^{(3)}]^{y_w} \,,
\label{U3yueqU3yw}
\ee
as desired by
\eqn{simplefact}.  Note that this ``bonus'' relation holds even outside
of the factorization limit, a property to which we will return in
the following section.

We also find that the $y_v$ coproduct element of $U^{(3)}$ is proportional
to the weight-5 parity odd function $H_1(u,v,w)$:
\be
[U^{(3)}]^{y_v} = \frac{1}{8} H_1(u,v,w) \,.
\label{U3yv}
\ee
One can check that $H_1$ obeys all the same coproduct relations that
$\Omega^{(2)}(w,u,v)$ does~\cite{Dixon2013eka}, 
so that all of its derivatives vanish in the factorization limit.  
(Its $y_v$ coproduct element is in fact proportional to $\Omega^{(2)}(w,u,v)$.)
In the case of $H_1$, the vanishing of the constant of integration, 
{\it i.e.} the fact that
\be
H_1(u,v,w)\bigr|_{u,w\to\infty} = 0,
\label{H1factvanish}
\ee
follows simply because all parity-odd functions vanish on the surface
$\Delta(u,v,w)=0$, which contains the line $(u,1,u)$.

The $F^u+F^{1-u}$, $F^v$ and $F^{1-v}$ coproduct elements contributing
to \eqns{duFfact}{dvFfact} also
simplify dramatically for $F=U^{(3)}$ as $u,w\to\infty$;
the only functions they contain are logarithms of $u$, $v$ and $w$.
We find that in the factorization limit,
\bea
&&[U^{(3)}]^u + [U^{(3)}]^{1-u} = \zeta_3 \, \ln^2(uw/v)
  - \frac{75}{4} \, \zeta_4 \, \ln(uw/v)
  + 7 \, \zeta_5 + 8 \, \zeta_2 \, \zeta_3
\, =\, - [U^{(3)}]^{v}  \,,~~
\label{U3uomusumU3vfact}\\
&&[U^{(3)}]^{1-v} = 0 \,.
\label{U3omv}
\eea
It is quite fascinating that \eqn{U3omv} actually holds for any $(u,v,w)$.
We will explore the consequences of this second bonus relation in the
next section.
 
Unlike \eqn{U3omv}, the first relation does not hold for arbitrary
$(u,v,w)$, but it does hold in the factorization limit.
Using eqs.~(\ref{U3yueqU3yw}), (\ref{U3uomusumU3vfact}), and (\ref{U3omv}),
as well as the vanishing of $[U^{(3)}]^{y_v}$ in the factorization limit,
it is trivial to solve the differential
equations~(\ref{duFfact}) and (\ref{dvFfact}) for $F=U^{(3)}$ in this limit.
The result is,
\bea
U^{(3)}(u,v,w)\bigr|_{u,w\to\infty} &=& 
\frac{1}{3} \, \zeta_3 \, \ln^3(uw/v)
- \frac{75}{8} \, \zeta_4 \, \ln^2(uw/v)
+ ( 7 \, \zeta_5  + 8 \, \zeta_2 \, \zeta_3 ) \ln(uw/v)
\nonumber\\ &&\null
- \frac{161}{2} \, \zeta_6 - 3 \, (\zeta_3)^2 \,.
\label{U3_fact}
\eea
Again we fixed the constant of integration using the limiting behavior
on the line $(u,1,u)$ as $u\to\infty$.
Remarkably, at all three loop orders studied so far, the quantity $U$
approaches a simple polynomial in $\ln(uw/v)$ in the factorization limit.

Now we go back to construct the factorization function $F_6$ for
the NMHV six-point amplitude in terms of $U$.
To do this, we observe from \eqn{facts345}
that (apart from the trivial tree-level term), the log of
the factorization function is, using \eqn{Uhatdef},
\be
\ln F_6\ =\ 
\ln\biggl(\frac{A_6^{\rm NMHV}}{A_4^{\rm BDS} \, A_4^{\rm BDS'}}\biggr)
\ =\ \ln\biggl(\frac{A_6^{\rm NMHV}}{A_6^{\rm BDS}}
          \frac{A_6^{\rm BDS}}{A_4^{\rm BDS} \, A_4^{\rm BDS'}}\biggr)
\ =\ \hat{U}(u,v,w)
- \ln \biggl(\frac{A_4^{\rm BDS} \, A_4^{\rm BDS'}}{A_6^{\rm BDS}}\biggr) \,,
\label{F6Ldef}
\ee
or, using \eqns{BDScombination}{Udef},
\bea
[\ln F_6]^{(L)} &=& \frac{\gamma_K^{(L)}}{8\e^2L^2}
 \biggl( 1 + 2 \, \e \, L \, \frac{{\cal G}_0^{(L)}}{\gamma_K^{(L)}} \biggr)
  \biggl[ \biggl(\frac{(-s_{12})(-s_{34})}{(-s_{56})}\biggr)^{-L\e}
   + \biggl(\frac{(-s_{45})(-s_{61})}{(-s_{23})}\biggr)^{-L\e} \biggr]
\nonumber\\ &&\null\hskip0.0cm
  - \frac{\gamma_K^{(L)}}{8}\biggl[
     \frac{1}{2} \ln^2 \biggl(\frac{(-s_{12})(-s_{34})}{(-s_{56})}
        \bigg/ \frac{(-s_{45})(-s_{61})}{(-s_{23})}\biggr)
     + 6 \, \zeta_2 \biggr]
\nonumber\\ &&\null\hskip0.0cm
   + U^{(L)}(u,v,w)\bigr|_{u,w\to\infty}
   + \frac{f_2^{(L)}}{L^2} + C^{(L)} \,.
\label{lnF6L}
\eea

From the explicit formulae for $U^{(1)}$ (\eqn{U1}),
$U^{(2)}$ (\eqn{U2_fact}) and $U^{(3)}$ (\eqn{U3_fact}) in
the factorization limit, we see that the dependence of the factorization
function $F_6$ on the vanishing three-particle invariant $s_{345}$ only
appears through the logarithm of the ratio,
\be
\frac{uw}{v} = \frac{s_{12} s_{34}}{s_{56}} \cdot \frac{s_{45} s_{61}}{s_{23}}
               \cdot \frac{1}{s_{345}^2} \,.
\label{uw_v}
\ee
We note that the same ratios that we have assembled into the divergent
factors in \eqn{lnF6L} also appear in \eqn{uw_v}.

Consider now the more general multi-particle factorization of an
$n$-point amplitude in planar ${\cal N}=4$ SYM,
in which $s_{i,i+1,\ldots,j-1} \to 0$ as shown in figure~\ref{fig:multi}(b).
The corresponding factorization formula is,
\be
A_n^{\rm NMHV}(k_i) \longrightarrow
A_{j-i+1}(k_i,k_{i+1},\ldots,k_{j-1},K) \, \frac{F_n(K^2,s_{l,l+1})}{K^2} 
\, A_{n-(j-i)+1}(-K,k_{j},k_{j+1},\ldots,k_{i-1}) \,,
\label{factgen}
\ee
where $K = k_{j}+k_{j+1}+\cdots +k_{i-1}$ and all indices are mod $n$.
We conjecture that $\ln F_n$ can be extracted from formula~(\ref{lnF6L})
for $\ln F_6$ by the simple replacements,
\bea
\frac{s_{12} s_{34}}{s_{56}} &\to& \frac{s_{i,K} \, s_{i-1,-K}}{s_{i-1,i}} \,, \\
\frac{s_{45} s_{61}}{s_{23}} &\to& \frac{s_{j,-K} \, s_{j-1,K}}{s_{j-1,j}} \,, \\
\frac{uw}{v} &\to& \frac{s_{i,K} \, s_{i-1,-K} \cdot s_{j,-K} \, s_{j-1,K}}
                        {s_{i-1,i} \cdot s_{j-1,j} \cdot (K^2)^2}
\ =\ \frac{ x_{i+1,j}^2 \, x_{j,i-1}^2 \, x_{j+1,i}^2 \, x_{i,j-1}^2 }
          { x_{i-1,i+1}^2 \, x_{j-1,j+1}^2 \, (x_{i,j}^2)^2} \,.
\eea
Note that $uw/v$ gets replaced by a dual conformal cross ratio
for the $n$-point amplitude.  The other two ratios involve
the invariants near the factorization channel, and their
appearance in the singular terms in $\e$ is dictated by
the general structure of the infrared divergences.


\section{Coproduct relations for $U$ and $\tilde{V}$}
\label{coprels}

In the course of inspecting the coproducts of $U^{(3)}$, even {\it before}
taking the factorization limit, we found the following three relations,
\bea
U^u + U^{1-u} &=& U^w + U^{1-w}\ =\ - ( U^{v} + U^{1-v} ),
\label{Uuomuvomv_exact}\\
U^{1-v} &=& 0 \,,
\label{Uomv_exact}\\
U^{y_u} &=& U^{y_w} \,,
\label{Uyuyw_exact}
\eea
which hold for {\it any} $(u,v,w)$, at least through three loops.

The first relation is not unexpected.  It also holds for the 
parity-even part of the ratio function $V$, and it corresponds to the
existence of the seventh final entry $uw/v$ in \eqn{finalentries}.
Suppose such an entry were not present.  Then the final entry $u/(1-u)$ 
would correspond to the coproduct condition $V^u + V^{1-u} = 0$,
and similarly $V^v+V^{1-v}=0$ and $V^w+V^{1-w}=0$.  With the seventh final
entry, these conditions are violated, but they have to be violated
in the particular form shown in \eqn{Uuomuvomv_exact},
namely
\be
V^u + V^{1-u}\ =\ V^w + V^{1-w}\ =\ - ( V^{v} + V^{1-v} ).
\label{Vuomuvomv_exact}
\ee
Taking the logarithm of $V$ does not spoil \eqn{Vuomuvomv_exact},
and neither does adding $R_6$, since it obeys $R_6^{u_i} + R_6^{1-u_i} = 0$
for all three $u_i$.  Finally, the function $Y$ also obeys
$Y^{u_i} + Y^{1-u_i} = 0$.   Because \eqn{Vuomuvomv_exact}
follows from the analysis of Caron-Huot and He~\cite{CaronHuot2011kk},
so does \eqn{Uuomuvomv_exact}.

However, the other two relations, (\ref{Uomv_exact}) and 
(\ref{Uyuyw_exact}), are rather unexpected.
One virtue of these relations is that they
simplify the derivative of $U$ with respect to $y_u/y_w$.

Recall~\cite{Dixon2013eka} the formula for this derivative,
\bea
\sqrt{\Delta}\frac{\partial U}{\partial \ln(y_u/y_w)} \!\! &=& \!\!
(1-u)(1-v) U^u - (u-w)(1-v) U^v - (1-v)(1-w)U^w - u(1-v)U^{1-u} 
\nonumber\\
&&\quad\null + (u-w)v\,U^{1-v} + w(1-v)\,U^{1-w}
+ \sqrt{\Delta}\, U^{y_u}-\sqrt{\Delta}\, U^{y_w} \,.
\label{yuyw_diffeq}
\eea
Using eqs.~(\ref{Uyuyw_exact}), (\ref{Uuomuvomv_exact}) and
(\ref{Uomv_exact}), the differential equation~(\ref{yuyw_diffeq})
can be simplified dramatically to,
\be
\sqrt{\Delta}\frac{\partial U}{\partial \ln(y_u/y_w)}
\ =\ (1-v) ( U^u - U^w ) \,.
\label{U_yuyw_diffeq}
\ee
Only a single nontrivial coproduct combination, $U^u - U^w$, 
enters this equation, at any loop order!

We can find these coproduct combinations using the results
in appendix~\ref{Ucoprod}.  The combination $U^u - U^w$
is generally simpler than either $U^u$ or its $u\lr w$ image $U^w$.
At one loop, we have trivially,
\be
[U^{(1)}]^u - [U^{(1)}]^w = 0.
\label{U1_u_w}
\ee
At two loops, the combination is,
\bea
[U^{(2)}]^u - [U^{(2)}]^w &=&
  H_{2,1}^u + \frac{1}{2} \, \ln u \, H_{2}^u
- H_{2,1}^w - \frac{1}{2} \, \ln w \, H_{2}^w
- \frac{1}{4} \, \Bigl( \ln(v/w) H_{2}^u - \ln(v/u) H_{2}^w \Bigr)
\nonumber\\ &&\null\hskip0.0cm
+ \frac{1}{8} \, \ln(u/w) \, \Bigl( 2 \, H_{2}^v - \ln(uvw) \, \ln v
                     + 3 \, \ln u \, \ln w  - 8 \, \zeta_2 \Bigl) \,.
\label{U2_u_w}
\eea
At three loops, it is,
\be
[U^{(3)}]^u - [U^{(3)}]^w = A^{u-w}(u,v,w) - A^{u-w}(w,v,u),
\label{U3_u_w}
\ee
where
\bea
&&A^{u-w}(u,v,w) = \frac{1}{16} \, \biggl\{
- M_1(w,u,v) + \frac{128}{3} \, \Qep(v,u,w)
\nonumber\\&&\null
 - \frac{1}{2} \, \ln(u/w) \, \Bigl( 2 \, \Omegauvw - \Omegawuv \Bigr)
 + 12 \, H_{4,1}^u + 10 \, H_{3,2}^u - 72 \, H_{3,1,1}^u - 26 \, H_{2,2,1}^u
\nonumber\\&&\null
 - 40 \, H_{2,1,1,1}^u - 2 \, H_2^u \, ( 3 \, H_3^u - 7 \, H_{2,1}^u )
 - 2 \, H_2^u \, ( 2 \, H_{2,1}^v + \ln v \, H_2^v )
 - 2 \, ( \ln u + 2 \, \ln v - 3 \, \ln w ) \, H_{3,1}^u
\nonumber\\&&\null
 - \frac{1}{2} \, \ln u \, \Bigl( 4 \, H_4^v + 40 \, H_{3,1}^v 
   + 4 \, H_{2,1,1}^v - 11 \, (H_2^v)^2
   - 4 \, \ln v \, ( H_3^v - H_{2,1}^v ) \Bigr)
 - 12 \, \ln(uw/v) \, H_4^u
\nonumber\\&&\null
 - 2 \, ( 13 \, \ln u - 6 \, \ln(v/w) ) \, H_{2,1,1}^u
 + \Bigl( 8 \, \ln u \, \ln(uw/v) 
   - 2 \, \ln^2 v + 4 \, \ln v \, \ln w \Bigr) \, H_3^u
\nonumber\\&&\null
 + \frac{1}{2} \, ( 11 \, \ln u - \ln w ) \, (H_2^u)^2 
 - \Bigl( 8 \, ( \ln^2 u + \ln^2 v ) - 2 \, \ln u \, (\ln v-\ln w)
   - 14 \, \ln v \, \ln w \Bigr) \, H_{2,1}^u
\nonumber\\&&\null
 - \frac{11}{3} \, \ln^2 w \, ( H_3^u + H_{2,1}^u )
 - \frac{2}{3} \, H_2^w
         \, \Bigl( 5 \, ( H_3^u - \ln u \, H_2^u ) - 7 \, H_{2,1}^u \Bigr)
\nonumber\\&&\null
 - \frac{1}{6} \, \Bigl( 8 \, \ln^3 u + 4 \, \ln^3 v - 3 \, \ln^3 w 
  - 7 \, \ln u \, \ln^2 w
  + 3 \, \ln(u/w) \, \ln v \, ( 3 \, \ln v - 4 \, \ln w ) \Bigr) \, H_2^u
\nonumber\\&&\null
 - \frac{1}{2} \, \ln(u/w)
     \, \Bigl( \ln^2 u + 4 \, \ln u \, \ln v
 - 5 \, \ln w \, \ln u \Bigr) \, H_2^v
\nonumber\\&&\null
 + \frac{1}{12} \, \Bigl( 
   \ln^3 u \, ( 11 \, \ln^2 v - 28 \, \ln v \, \ln w + 35 \, \ln^2 w )
          + \ln^2 u \, \ln^2 v \, ( 8 \, \ln v - 27 \, \ln w ) \Bigr)
\nonumber\\&&\null
 - \zeta_2 \, \Bigl[ H_3^u + 36 \, H_{2,1}^u + 13 \, \ln u \, H_2^u 
        - \frac{5}{3} \, \ln^3 u
             + 2 \, \ln u \, ( 9 \, H_2^v - 4 \, \ln^2 v )
             + 2 \, \ln w \, ( H_2^u + 12 \, \ln^2 u )
\nonumber\\&&\hskip0.8cm\null
             - 4 \, \ln v \, ( H_2^u + 2 \, \ln^2 u ) \Bigr]
 - 2 \, \zeta_3 \, ( 3 \, \ln^2 u + 4 \, H_2^u )
 + 122 \, \zeta_4 \, \ln u \biggr\} \,.
\label{Auw}
\eea

The $y_u/y_w$ differential equation~(\ref{U_yuyw_diffeq}) is
relatively simple analytically.  In ref.~\cite{Dixon2013eka}
it was discussed how this differential equation has natural
boundary conditions at $(u,v,w)=(1,0,0)$ and $(0,0,1)$.  They
are natural from the point of view that they correspond to surfaces
in the coordinates $(y_u,y_v,y_w)$; therefore only a boundary condition
at one point is needed to integrate up to any $(u,v,w)$.
However, it was also mentioned in ref.~\cite{Dixon2013eka} that
there could be issues of regularization for an even function like $U$
near the endpoints.  Indeed, for $U^{(2)}$ or $U^{(3)}$ there are such
issues, which would have to be cured by subtracting a suitable
function in order for the endpoints $(1,0,0)$ and $(0,0,1)$ to
be usable.

Another strategy is to use the $y_u/y_w$ differential equation
to integrate not off a single point, but off a surface.
For example, since it is odd in $u\lr w$, one could use
this differential equation to move off the surface $u=w$,
once one has determined the function on the surface using a different
strategy.

Independently of the best numerical approach, the coproduct
relations for $U$ indicate a simplified analytic structure
for this function.  In terms of a final-entry condition,
the coproduct relations $U^{1-v}=0$ and $U^{y_u}=U^{y_w}$
reduce the seven member set~(\ref{finalentries})
to only five entries:
\be
\left\{ \frac{u}{1-u}, \frac{w}{1-w}, y_uy_w, y_v, \frac{uw}{v} \right\} \,.
\label{fivefinalentries}
\ee
It would be very interesting to try to find an explanation
for this property, which at the moment has only been observed
empirically through three loops.  In the next section we will see
that the function $U$ is in many ways simpler than $V$,
and even simpler than the remainder function $R_6$.

Before doing that, we close this section by remarking that
the potential seventh final entry $uw/v$, which we also
allowed for the parity-odd function $\tilde{V}(u,v,w)$,
does not actually appear.  In other words, at least through
three loops, $\tilde{V}$ obeys the coproduct relations,
\be
\tilde{V}^u + \tilde{V}^{1-u} = \tilde{V}^v + \tilde{V}^{1-v}
= \tilde{V}^w + \tilde{V}^{1-w} = 0 \,.
\label{Vt_coprod_rels}
\ee
The corresponding set of final entries for $\tilde{V}$ is
\be
\left\{ \frac{u}{1-u}, \frac{v}{1-v}, \frac{w}{1-w}, y_u, y_v, y_w \right\} \,,
\label{sixfinalentries}
\ee
which is the same set as for the remainder function.  It would be
interesting to understand this property better as well.


\section{Quantitative behavior}
\label{plotssect}

In this section we examine the analytical behavior of the components
of the ratio function on special lines through the three-dimensional
space of cross ratios.  On some of these lines,
$V$ and $\tilde{V}$ collapse to simpler functions, such as HPLs
of a single argument.  Some of the analytic formulas, for the function $U$
in particular, exhibit intriguing simplicity.
We also plot $V$ and $\tilde{V}$, or various
ratios, on these special lines and on some two-dimensional surfaces,
such as the plane $u+v+w=1$, and as a function of $u$ and $w$ for
particular values of $v$.

After the imposition of the MRK constraints, the coproducts of $V^{(3)}$ and
$\tilde{V}^{(3)}$ are fully fixed. Of course, given
the remainder function $R_6$ and the function $Y$ defined in \eqn{Ydef},
we can go back and forth between $V$ and $U$, using the relations
\bea
U(u,v,w) &=& \ln V(u,v,w) + R_6(u,v,w) - \frac{\gamma_K}{8} Y(u,v,w) \,,
\label{UfromV}\\
V(u,v,w) &=& \exp\biggl[ U(u,v,w) - R_6(u,v,w)
                       + \frac{\gamma_K}{8} Y(u,v,w) \biggr] \,.
\label{VfromU}
\eea
The $\{n-1,1\}$ coproduct elements for $U$ through three loops
are given in appendix~\ref{Ucoprodsubappendix}.  This information
completely specifies the first derivatives of $U^{(L)}$ and $\tilde{V}^{(L)}$.

We should also fix the functions by giving their values at one point,
say $(u,v,w)=(1,1,1)$.   This point is on the surface $\Delta=0$,
on which all parity-odd hexagon functions vanish.
Hence
\be
\tilde{V}^{(L)}(1,1,1) = 0 \qquad\hbox{for all $L$.}
\label{Vt_111}
\ee
On the other hand, parity-even functions such as $V$ have nontrivial
values at this point.  The constant term in $V^{(3)}(1,1,1)$ is
fixed when the collinear vanishing constraints are applied.
It is actually fixed in the vicinity of the collinear limit lines,
such as $v=0$, $u+w=1$.  We can use the simple analytic behavior
of hexagon functions on the lines $(u,u,1)$ and $(u,1,u)$ (see
the next subsection) to carry the information about the constant
out to the point $(1,1,1)$.  We find that
\be
V^{(3)}(1,1,1) = - \frac{243}{4} \, \zeta_6 \,.
\label{V3_111}
\ee
This value can be compared to the corresponding values for the one-
and two-loop ratio functions,
\be
\bsp
V^{(1)}(1,1,1) & = - \zeta_2 \,, \\
V^{(2)}(1,1,1) & = 9 \, \zeta_4 \,.
\esp
\label{V1V2_111}
\ee
We also quote the values of $U$ at this point:
\be
\bsp
U^{(1)}(1,1,1) & = - \zeta_2 \,, \\
U^{(2)}(1,1,1) & = \frac{21}{4} \, \zeta_4 \,, \\
U^{(3)}(1,1,1) & = - \frac{117}{4} \, \zeta_6 + (\zeta_3)^2 \,,
\esp
\label{U1U2U3_111}
\ee
where the $(\zeta_3)^2$ term in $U^{(3)}(1,1,1)$ comes entirely
from $R_6^{(3)}(1,1,1)$~\cite{Dixon2013eka}.

With $V$ and $\tilde{V}$ now completely fixed at three loops,
we can investigate their analytic and numerical behavior.  In the
remainder of this section, we describe lines on which the analytic
behavior simplifies, and then we plot the functions $V$ and $\tilde{V}$
on these lines and on various planes in the space of cross ratios
$(u,v,w)$.

\subsection{The lines $(u,u,1)$ and $(u,1,u)$}

When one of the cross ratios is equal to unity and the other two are equal
to each other, the hexagon functions collapse to pure HPLs.
Because $\Delta(u,u,1)=0$, all parity-odd functions vanish on this line,
including $\tilde{V}$.  On the other hand, $V$ is nontrivial but
relatively simple.  The simplest way to present $V$ is to give $U$,
and then $V$ can be obtained using \eqn{VfromU}.
We also use the ``linearized'' representation for the HPLs discussed in
ref.~\cite{Dixon2014voa}, in which we expand all products of HPLs in
terms of a linear combination of HPLs of maximum weight using the shuffle
algebra.  In that reference a compressed notation for the HPLs was
also introduced.  Here we will not need that notation, because
the formulas are not too lengthy through three loops, and because
it obscures some of the patterns in which HPL weight vectors occur.

In the linearized representation, we have,
\bea
U^{(1)}(u,u,1) &=& - \zeta_2 \,, \label{U1_uu1}\\
U^{(2)}(u,u,1) &=& \frac{1}{2} \, ( H_{0,1,0,1}^u + H_{1,1,0,1}^u )
- \frac{3}{2} \, ( H_{0,1,1,1}^u + H_{1,1,1,1}^u )
- \zeta_2 \, ( H_{0,1}^u + H_{1,1}^u ) + \frac{21}{4} \, \zeta_4 
\,, \label{U2_uu1}\\
U^{(3)}(u,u,1) &=& H_{0,1,0,1,0,1}^u + H_{1,1,0,1,0,1}^u
- 4 \, ( H_{0,1,0,1,1,1}^u + H_{1,1,0,1,1,1}^u )
-  5 \, ( H_{0,1,1,0,1,1}^u + H_{1,1,1,0,1,1}^u )
\nonumber\\&&\null
-  4 \, ( H_{0,1,1,1,0,1}^u + H_{1,1,1,1,0,1}^u )
+ 10 \, ( H_{0,1,1,1,1,1}^u + H_{1,1,1,1,1,1}^u )
\nonumber\\&&\null
- 2 \, \zeta_2  \, \Bigl[ H_{0,1,0,1}^u + H_{1,1,0,1}^u
           - 4 \, ( H_{0,1,1,1}^u + H_{1,1,1,1}^u ) \Bigr]
+ 8 \, \zeta_4 \, ( H_{0,1}^u + H_{1,1}^u )
\nonumber\\&&\null
- \frac{117}{4} \, \zeta_6 + (\zeta_3)^2  \,. \label{U3_uu1}
\eea
For reference, we also give
\be
Y(u,u,1) = 2 \, ( H_{0,1}^u + H_{1,1}^u ) \,, \label{Y_uu1}
\ee
and the remainder function in the same notation is,
\bea
R_6^{(2)}(u,u,1) &=& H_{0,0,0,1}^u + H_{1,0,0,1}^u
                - H_{0,0,1,1}^u - H_{1,0,1,1}^u
- \frac{5}{2} \, \zeta_4 \,, \label{R62_uu1}\\
R_6^{(3)}(u,u,1) &=& 
- \frac{1}{2} \, \Bigl[
  H_{0,0,1,0,0,1}^u + H_{1,0,1,0,0,1}^u
+ H_{0,1,0,1,0,1}^u + H_{1,1,0,1,0,1}^u
+ 2 \, ( H_{0,0,1,1,0,1}^u + H_{1,0,1,1,0,1}^u )
\nonumber\\&&\null
+ 3 \, ( H_{0,0,1,0,1,1}^u + H_{1,0,1,0,1,1}^u
      + H_{0,1,0,0,0,1}^u + H_{1,1,0,0,0,1}^u
      + H_{0,1,0,1,1,1}^u + H_{1,1,0,1,1,1}^u
\nonumber\\&&\null
      - H_{0,0,0,1,0,1}^u - H_{1,0,0,1,0,1}^u
      - H_{0,1,0,0,1,1}^u - H_{1,1,0,0,1,1}^u )
+ 6 \, ( H_{0,0,0,0,0,1}^u + H_{1,0,0,0,0,1}^u ) 
\nonumber\\&&\null
+ 9 \, ( H_{0,0,0,1,1,1}^u + H_{1,0,0,1,1,1}^u )
- 10 \, ( H_{0,0,0,0,1,1}^u + H_{1,0,0,0,1,1}^u ) \Bigr]
\nonumber\\&&\null
- \zeta_2 \, \Bigl[ H_{0,0,0,1}^u + H_{1,0,0,1}^u
        - 3 \, ( H_{0,0,1,1}^u + H_{1,0,1,1}^u )
        - 2 \, ( H_{0,1,0,1}^u + H_{1,1,0,1}^u ) \Bigr]
\nonumber\\&&\null
- 2 \, \zeta_4 \, ( H_{0,1}^u + H_{1,1}^u )
+ \frac{413}{24} \, \zeta_6 + (\zeta_3)^2 \,. \label{R63_uu1}
\eea
Although $U^{(2)}$ is slightly lengthier than $R_6^{(2)}$ in this
representation, $U^{(3)}$ is considerably shorter than $R_6^{(3)}$. 

Note that in the formulas for both $U$ and $R_6$,
the HPL weight vectors always end in 1.  This restriction simply
guarantees that there are no branch cuts developing at $u=1$.
Also, for both $U$ and $R_6$,
there is a pairing of terms of the form $H_{0,\vec{m}} + H_{1,\vec{m}}$, where
$\vec{m}$ is a sequence of 0's and 1's.  This pairing is a consequence
of the final-entry condition, as discussed in ref.~\cite{Dixon2014voa}
for $R_6$.   It also holds for $U$ on the line $(u,u,1)$, basically
because the extra entry $uw/v$ reduces to 1 on this line.

On the other hand, the function $U$ exhibits two patterns not found in $R_6$.
First of all, the second weight-vector entry $m_2$ in $H_{m_1 m_2\ldots m_n}$
is always 1 for $U$. Second of all, $U$ has a symmetry on reversing the
order of $m_3\ldots m_{n-1}$; the coefficients of the two HPLs with the
weight-vectors swapped in this way are always equal.
It will be interesting to see if these patterns are accidents
of the first three orders, or hold up in further orders in perturbation
theory; and if the latter, what they signify. 

Next we turn to the line $(u,1,u)$.   On this line, $R_6$,
which is cyclically symmetric, is still given by \eqns{R62_uu1}{R63_uu1},
but the formulas for $U$ are different:
\bea
U^{(1)}(u,1,u) &=& - 2 \, H_{1,1}^u - \zeta_2 \,, \label{U1_u1u}\\
U^{(2)}(u,1,u) &=& \frac{1}{2} \, H_{0,1,0,1}^u - \frac{3}{2} \, H_{0,1,1,1}^u
     +  H_{1,0,0,1}^u -       H_{1,0,1,1}^u
- \frac{1}{2} \, H_{1,1,0,1}^u - \frac{9}{2} \, H_{1,1,1,1}^u
\nonumber\\&&\null
- \zeta_2 \, ( H_{0,1}^u - 3 \, H_{1,1}^u )
+ \frac{21}{4} \, \zeta_4 \,, \label{U2_u1u}\\
U^{(3)}(u,1,u) &=& 
  H_{0,1,0,1,0,1}^u - H_{0,1,0,1,1,1}^u
- H_{0,1,1,1,0,1}^u - H_{0,1,1,0,1,1}^u
- 20 \, H_{0,1,1,1,1,1}^u
- 2 \, ( H_{1,0,0,0,1,1}^u
\nonumber\\&&\null
      + H_{1,0,0,1,0,1}^u
      + H_{1,0,1,0,0,1}^u + H_{1,0,1,0,1,1}^u
      + H_{1,1,0,0,0,1}^u + H_{1,1,0,1,0,1}^u
      + H_{1,1,1,1,0,1}^u )
\nonumber\\&&\null
- 3 \, ( H_{1,0,0,0,0,1}^u - H_{1,0,0,1,1,1}^u
      - H_{1,1,1,0,0,1}^u )
- 4 \, ( H_{1,0,1,1,0,1}^u + H_{1,1,1,0,1,1}^u )
\nonumber\\&&\null
+ 5 \, H_{1,1,0,0,1,1}^u - 9 \, H_{1,1,0,1,1,1}^u
- 16 \, H_{1,0,1,1,1,1}^u - 55 \, H_{1,1,1,1,1,1}^u
\nonumber\\&&\null
- 2 \, \zeta_2 \, ( H_{0,1,0,1}^u - H_{0,1,1,1}^u
   - 3 \, H_{1,0,1,1}^u - 3 \, H_{1,1,0,1}^u
   - 8 \, H_{1,1,1,1}^u )
\nonumber\\&&\null
+ \zeta_4 \, ( 8 \, H_{0,1}^u - 21 \, H_{1,1}^u )
- \frac{117}{4} \, \zeta_6 + (\zeta_3)^2 \,. \label{U3_u1u}
\eea
On this line, $U$ is not as simple as it is on $(u,u,1)$.

\begin{figure}
\begin{center}
\includegraphics[width=4.5in]{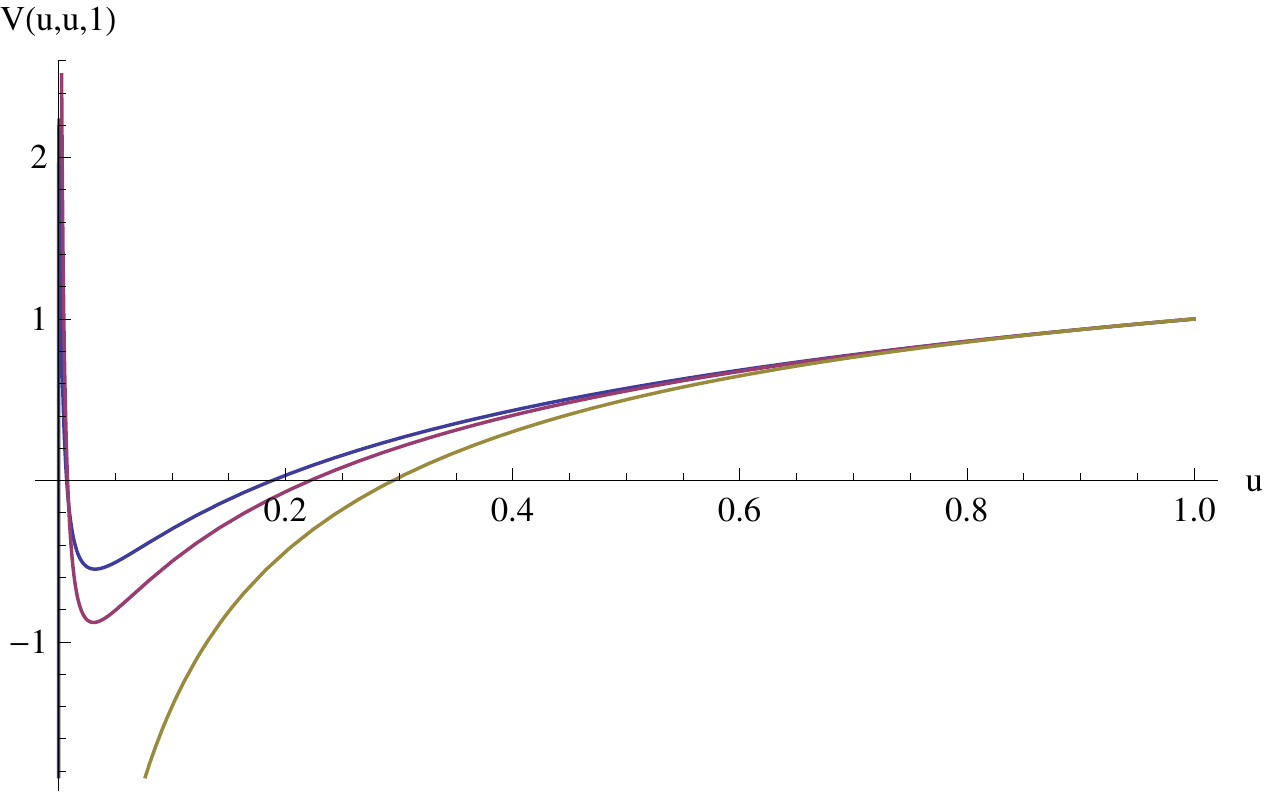}
\end{center}
\caption{$V^{(1)}(u,u,1)$, $V^{(2)}(u,u,1)$ and $V^{(3)}(u,u,1)$, normalized 
to unity at $(1,1,1)$. One loop is in green, two loops is in purple,
and three loops is in blue.}
\label{fig:vuu1}
\end{figure}

\begin{figure}
\begin{center}
\includegraphics[width=4.5in]{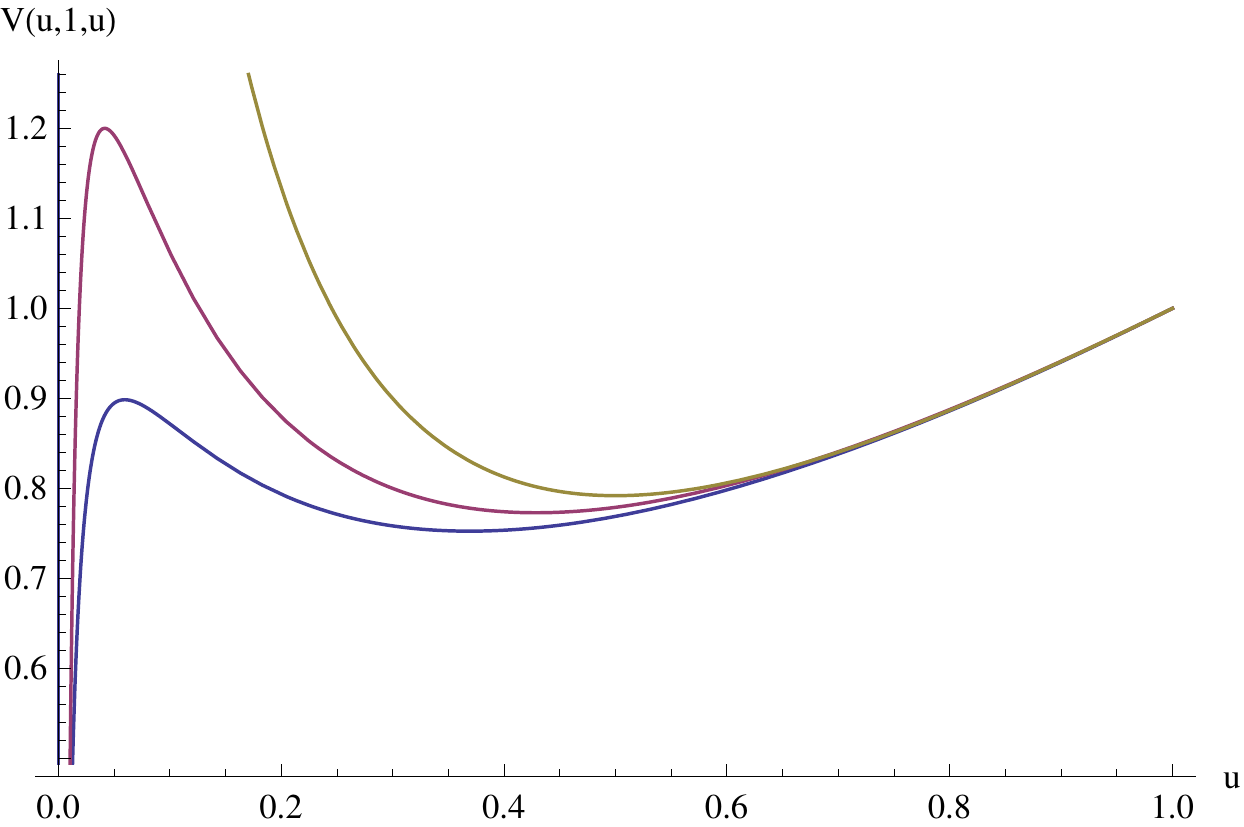}
\end{center}
\caption{$V^{(1)}(u,1,u)$, $V^{(2)}(u,1,u)$ and $V^{(3)}(u,1,u)$, normalized
to unity at $(1,1,1)$. One loop is in green, two loops is in purple, and
three loops is in blue.}
\label{fig:vu1u}
\end{figure}

In figure~\ref{fig:vuu1} and figure~\ref{fig:vu1u} we plot $V^{(1)}$,
$V^{(2)}$ and $V^{(3)}$ on the lines $(u,u,1)$ and $(u,1,u)$, respectively.
We normalize the functions by dividing by their values at $(1,1,1)$.
Note that for small $u$, the functions' values on $(u,1,u)$ are approximately
the negative of those on $(u,u,1)$.  This approximate relation
becomes exact if we drop power-suppressed terms as $u\to0$.
Then we find,
\bea
V^{(1)}(u,u,1) &\sim& \frac{1}{2} \, \ln^2 u \,, \label{V1_uu1_u0} \\
V^{(2)}(u,u,1) &\sim& \frac{1}{16} \, \ln^4 u
- \frac{3}{4} \, \zeta_2 \, \ln^2 u + \zeta_3 \, \ln u
+ \frac{5}{8} \, \zeta_4 \,, \label{V2_uu1_u0} \\
V^{(3)}(u,u,1) &\sim& \frac{1}{288} \, \ln^6 u
- \frac{5}{24} \, \zeta_2 \, \ln^4 u
+ \frac{71}{16} \, \zeta_4 \, \ln^2 u
- 2 \, ( 2 \, \zeta_5 + \zeta_2 \, \zeta_3 ) \, \ln u
\nonumber\\&&\null
- \frac{77}{16} \, \zeta_6 + \frac{1}{2} \, (\zeta_3)^2 \,, \label{V3_uu1_u0}\\
V^{(1)}(u,1,u) &\sim& - V^{(1)}(u,u,1) \,, \label{V1_u1u_u0} \\
V^{(2)}(u,1,u) &\sim& - V^{(2)}(u,u,1) \,, \label{V2_u1u_u0} \\
V^{(3)}(u,1,u) &\sim& - V^{(3)}(u,u,1) \,. \label{V3_u1u_u0}
\eea
The small value of the coefficient of $\ln^6 u$ in $V^{(3)}(u,u,1)$,
relative to that of $\ln^4 u$, causes the blue curve in figure~\ref{fig:vuu1}
to oscillate as $u\to0$:  it reaches a maximum around $u=0.0005$ and then
goes negative for even smaller $u$.

\subsection{The lines $(u,1,1)$ and $(1,v,1)$}

With two of the cross ratios equal to unity, the hexagon functions also
collapse to pure HPLs. On these lines, $\Delta$ is nonzero,
and so the function $\tilde{V}(u,1,1)$ is nonvanishing.
The function $\tilde{V}(1,v,1)$ vanishes, though, due to
$u\leftrightarrow w$ antisymmetry.  Again we give $U$ rather
than $V$, and use the linearized HPL representation.

On $(u,1,1)$, the function $\tilde{V}$ becomes,
\bea
\tilde{V}^{(2)}(u,1,1) &=& 
\frac{1}{4} \, ( H_{1,0,0,1}^u + H_{1,0,1,1}^u + H_{1,1,0,1}^u )
- \frac{1}{2} \, \zeta_2 \, H_{1,1}^u \,, \label{Vt2_u11}\\
\tilde{V}^{(3)}(u,1,1) &=& \frac{1}{8} \, \Bigl[
 H_{0,1,0,1,0,1}^u - H_{1,0,1,1,0,1}^u - H_{1,1,0,1,0,1}^u
+ 2 \, ( H_{0,1,0,0,1,1}^u + H_{0,1,1,0,0,1}^u
      + H_{1,0,1,0,1,1}^u
\nonumber\\&&\hskip0.4cm\null
      + H_{1,1,0,0,1,1}^u
      - H_{1,0,0,1,0,1}^u )
+ 3 \, ( H_{0,1,0,1,1,1}^u + H_{0,1,1,1,0,1}^u
      + H_{1,0,0,1,1,1}^u + H_{1,1,0,1,1,1}^u
\nonumber\\&&\hskip0.4cm\null
      + H_{1,1,1,0,1,1}^u - H_{1,1,1,0,0,1}^u )
+ 4 \, ( H_{0,1,1,0,1,1}^u - H_{1,0,1,0,0,1}^u )
\nonumber\\&&\hskip0.4cm\null
- 6 \, ( H_{1,0,0,0,0,1}^u + H_{1,1,0,0,0,1}^u ) \Bigr]
\nonumber\\&&\null
- \frac{1}{4} \, \zeta_2 \, \Bigl[ 3 \, ( H_{0,1,1,1}^u + H_{1,0,1,1}^u )
                  + H_{1,0,0,1}^u + 2 \, H_{1,1,0,1}^u \Bigr]
+ \frac{21}{4} \, \zeta_4 \, H_{1,1}^u \,. \label{Vt3_u11}
\eea
The function $U$ becomes,
\bea
U^{(1)}(u,1,1) &=& - \frac{1}{2} \, H_{1,1}^u - \zeta_2 \,, \label{U1_u11}\\
U^{(2)}(u,1,1) &=&
\frac{1}{4} \, ( H_{0,1,0,1}^u + H_{1,0,0,1}^u 
                       + H_{1,0,1,1}^u + H_{1,1,0,1}^u )
- \frac{1}{2} \, \zeta_2 \, ( H_{0,1}^u - H_{1,1}^u ) + \frac{21}{4} \, \zeta_4
\,,~~~ \label{U2_u11}\\
U^{(3)}(u,1,1) &=& - \frac{1}{8} \, \Bigl[ H_{1,0,1,0,1,1}^u
- 2 \, ( H_{0,1,0,1,1,1}^u + H_{0,1,1,1,0,1}^u
      - H_{1,0,1,1,0,1}^u - H_{1,1,0,0,1,1}^u )
\nonumber\\&&\hskip0.6cm\null
+ 3 \, ( H_{1,0,0,1,1,1}^u + H_{1,1,0,1,0,1}^u
      + H_{1,1,1,0,0,1}^u
      - H_{1,1,0,1,1,1}^u - H_{1,1,1,0,1,1}^u )
\nonumber\\&&\hskip0.6cm\null
- 4 \, ( H_{0,1,0,1,0,1}^u + H_{0,1,1,0,1,1}^u ) 
+ 5 \, ( H_{1,0,0,1,0,1}^u + H_{1,0,1,0,0,1}^u )
\nonumber\\&&\hskip0.6cm\null
+ 6 \, ( H_{1,0,0,0,0,1}^u + H_{1,0,0,0,1,1}^u + H_{1,1,0,0,0,1}^u ) \Bigr]
\nonumber\\&&\null
- \frac{1}{2} \, \zeta_2 \, ( 2 \, H_{0,1,0,1}^u
        + H_{0,1,1,1}^u + H_{1,0,1,1}^u )
+ 4 \, \zeta_4 \, ( H_{0,1}^u - H_{1,1}^u )
\nonumber\\&&\null
- \frac{117}{4} \, \zeta_6 + (\zeta_3)^2 \,, \label{U3_u11}
\eea
and
\bea
U^{(1)}(1,u,1) &=& - \frac{1}{2} \, H_{1,1}^u - \zeta_2 \,, \label{U1_1u1}\\
U^{(2)}(1,u,1) &=& \frac{1}{4} \, H_{0,1,0,1}^u
- \frac{1}{2} \, \zeta_2 \, ( H_{0,1}^u - 2 \, H_{1,1}^u )
 + \frac{21}{4} \, \zeta_4 \,, \label{U2_1u1}\\
U^{(3)}(1,u,1) &=& \frac{1}{2} \, ( H_{0,1,0,1,0,1}^u + H_{0,1,1,0,1,1}^u )
+ \frac{1}{4} \, ( H_{0,1,0,1,1,1}^u + H_{0,1,1,1,0,1}^u
        + H_{1,0,1,0,1,1}^u + H_{1,0,1,1,0,1}^u )
\nonumber\\&&\null
- \frac{1}{2}\, \zeta_2 \, ( 2 \, H_{0,1,0,1}^u + H_{0,1,1,1}^u + H_{1,0,1,1}^u )
+ 4 \, \zeta_4 \, ( H_{0,1}^u - 2 \, H_{1,1}^u )
\nonumber\\&&\null
- \frac{117}{4} \, \zeta_6 + (\zeta_3)^2 \,. \label{U3_1u1}
\eea
We see that $U$ is simpler on the line $(1,u,1)$ than on 
the line $(u,1,1)$.

A combination that seems exceptionally simple, at least through
three loops, is the difference between $U$ on the line $(u,1,1)$
and on the line $(1,u,1)$.  Defining
\be
\Delta U\ \equiv\ U(u,1,1) - U(1,u,1)\,,
\label{DeltaUdef}
\ee
we find
\bea
\Delta U^{(1)} &=& 0\,, \label{DeltaU1}\\
\Delta U^{(2)} &=& \frac{1}{4} \, ( H_{1,0,0,1}^u +  H_{1,0,1,1}^u +  H_{1,1,0,1}^u )
- \frac{1}{2} \, \zeta_2 \, H_{1,1}^u \,, \label{DeltaU2}\\
\Delta U^{(3)} &=& - \frac{1}{8} \, \Bigl[ 2 \, H_{1,1,0,0,1,1}^u
+ 3 \, ( H_{1,0,0,1,1,1}^u + H_{1,1,1,0,0,1}^u 
      + H_{1,0,1,0,1,1}^u + H_{1,1,0,1,0,1}^u
\nonumber\\&&\hskip0.6cm\null
      - H_{1,1,0,1,1,1}^u - H_{1,1,1,0,1,1}^u )
+ 4 \, H_{1,0,1,1,0,1}^u
+ 5 \, ( H_{1,0,0,1,0,1}^u + H_{1,0,1,0,0,1}^u )
\nonumber\\&&\hskip0.6cm\null
+ 6 \, ( H_{1,0,0,0,0,1}^u
      + H_{1,0,0,0,1,1}^u + H_{1,1,0,0,0,1}^u ) \Bigr]
+ 4 \, \zeta_4 \, H_{1,1}^u \,. \label{DeltaU3}
\eea
We observe a similar pattern to that for $U(u,u,1)$, with the role
of the second weight vector entry in $U(u,u,1)$ played by the
first weight vector entry in $\Delta U$.  In other words, the first
entry as well as the last entry in $\Delta U$ is always 1.  Also,
$\Delta U$ is a {\it palindrome}:  reversing the ordering of the letters
(weight vector entries) leaves it invariant.

\begin{figure}
\begin{center}
\includegraphics[width=4.5in]{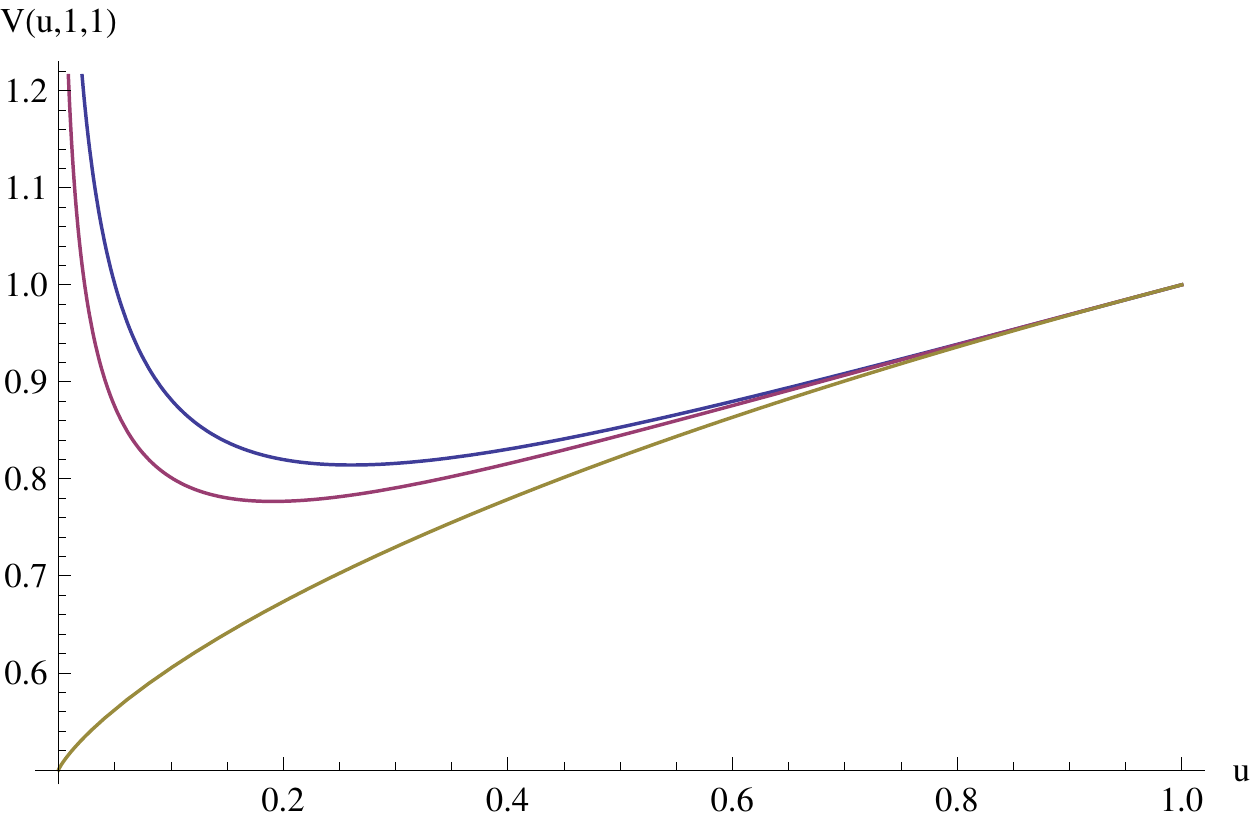}
\end{center}
\caption{$V^{(1)}(u,1,1)$, $V^{(2)}(u,1,1)$ and $V^{(3)}(u,1,1)$, normalized 
to unity at $(1,1,1)$. One loop is in green, two loops is in purple, and 
three loops is in blue.}
\label{fig:vu11}
\end{figure}
\begin{figure}
\begin{center}
\includegraphics[width=4.5in]{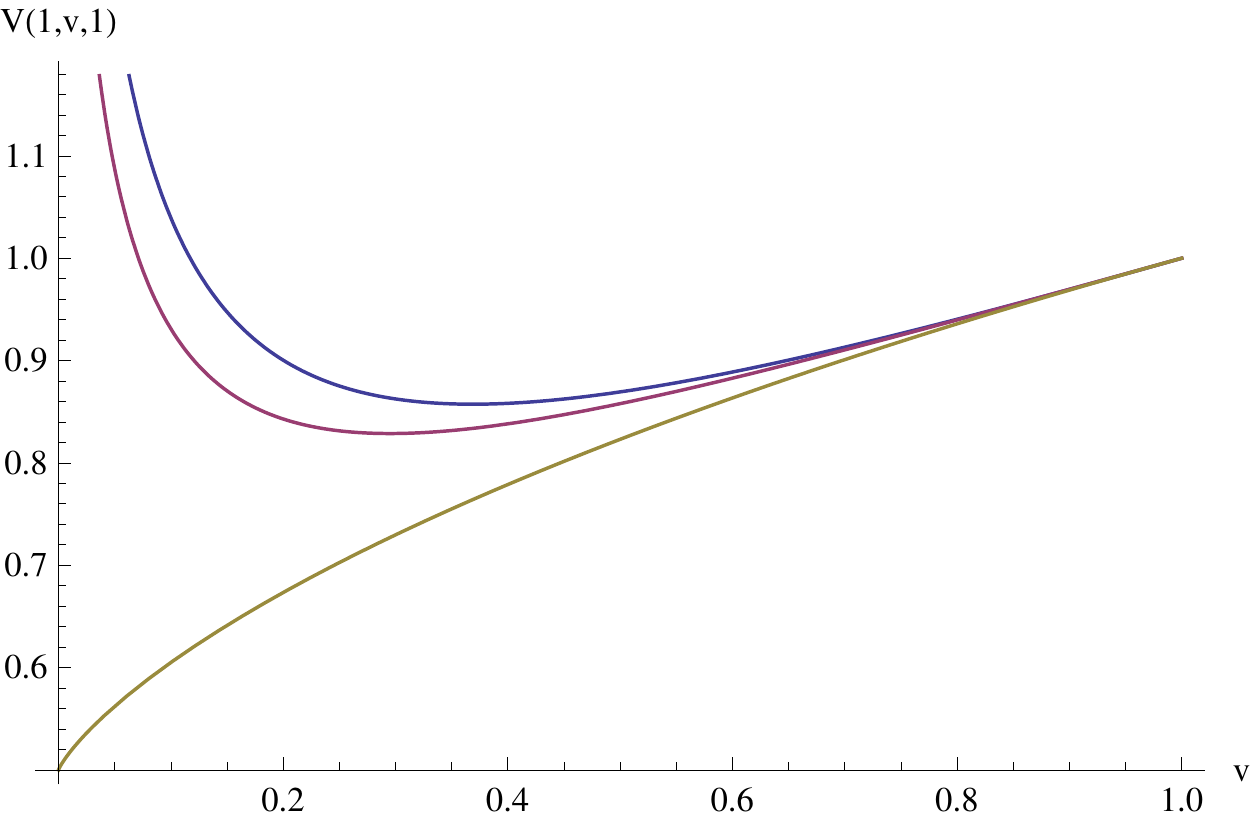}
\end{center}
\caption{$V^{(1)}(1,v,1)$, $V^{(2)}(1,v,1)$ and $V^{(3)}(1,v,1)$,
normalized to unity at $(1,1,1)$. One loop is in green, two loops is in 
purple, and three loops is in blue.}
\label{fig:v1u1}
\end{figure}
\begin{figure}
\begin{center}
\includegraphics[width=4.5in]{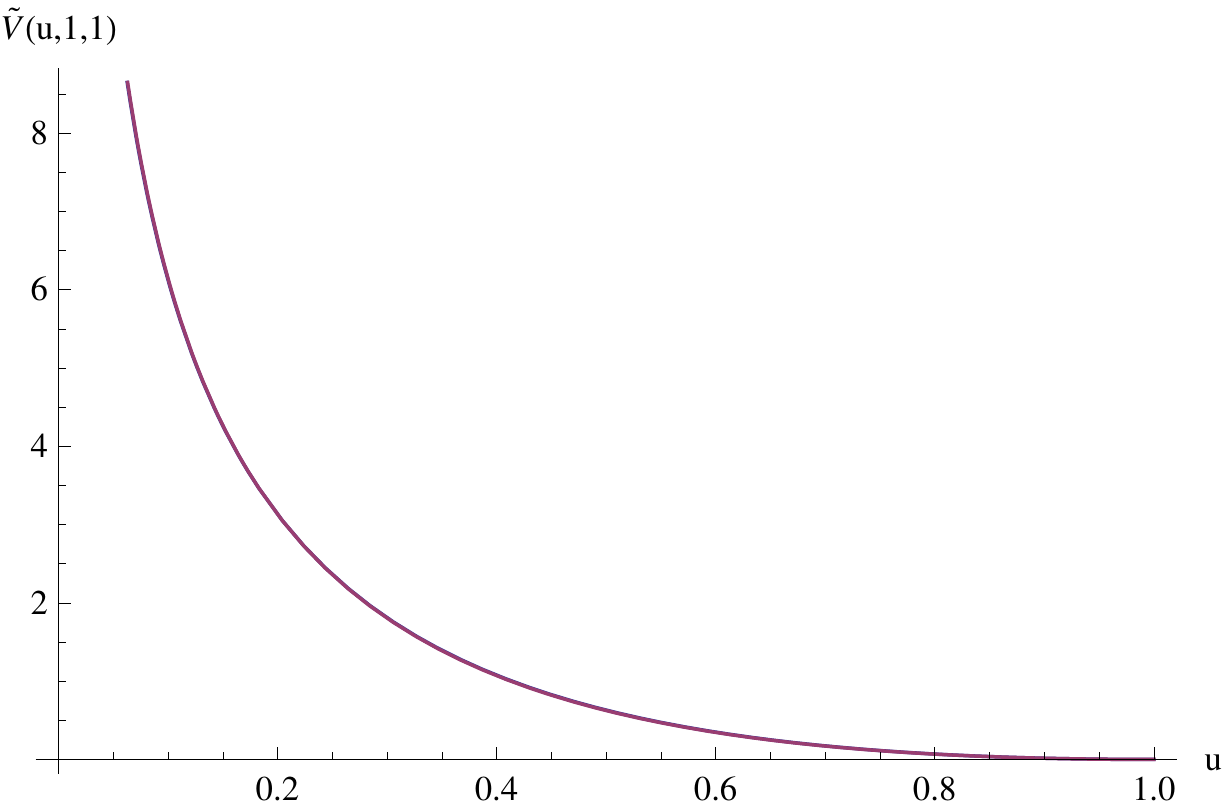}
\end{center}
\caption{$\tilde{V}^{(2)}(u,1,1)$ and $\tilde{V}^{(3)}(u,1,1)$,
normalized so their $u\rightarrow 0$ limit is $\ln^2 u$ with unit
coefficient.  Two loops is in purple and three loops is in blue.
At this scale, the lines are indistinguishable.}
\label{fig:vtu11}
\end{figure}

In figures \ref{fig:vu11}, \ref{fig:v1u1}, and \ref{fig:vtu11} 
we plot the functions $V$ and $\tilde{V}$ through three loops.
The even functions are normalized so that they are all equal to 
one at $(1,1,1)$.  The parity-odd functions vanish at $(1,1,1)$,
so we can't normalize them there.  However, $\tilde{V}^{(2)}(u,1,1)$
and $\tilde{V}^{(3)}(u,1,1)$ are both proportional to $\ln^2 u$ as $u$
goes to zero.  Hence we instead normalize by the coefficient of that
divergence: $-\frac{1}{8}\zeta_2$ for $\tilde{V}^{(2)}$ and
$\frac{47}{32}\zeta_4$ for $\tilde{V}^{(3)}$. Remarkably, with this choice
of normalization, the odd functions are almost indistinguishable on the 
line $(u,1,1)$.

The small $u$ behavior of the parity-even $V$ functions on the lines $(u,1,1)$
and $(1,u,1)$ is milder than that on $(u,u,1)$ and $(u,1,u)$,
having at most $\ln^2 u$ behavior:
\bea
V^{(1)}(u,1,1) &\sim& - \frac{1}{2} \, \zeta_2 \,, \label{V1_u11_u0} \\
V^{(2)}(u,1,1) &\sim& \frac{1}{8} \, \zeta_2 \, \ln^2 u
- \frac{1}{2} \, \zeta_3 \, \ln u
+ \frac{31}{8} \, \zeta_4 \,, \label{V2_u11_u0} \\
V^{(3)}(u,1,1) &\sim& - \frac{27}{16} \, \zeta_4 \, \ln^2 u
+ \frac{1}{4} \, ( 11 \, \zeta_5 + 5 \, \zeta_2 \, \zeta_3 ) \, \ln u
- \frac{97}{4} \, \zeta_6 \,, \label{V3_u11_u0} \\
V^{(1)}(1,u,1) &\sim& - \frac{1}{2} \, \zeta_2 \,, \label{V1_1u1_u0} \\
V^{(2)}(1,u,1) &\sim& \frac{1}{4} \, \zeta_2 \, \ln^2 u
- \frac{1}{2} \, \zeta_3 \, \ln u + \frac{67}{16} \, \zeta_4 \,,
\label{V2_1u1_u0} \\
V^{(3)}(1,u,1) &\sim& - \frac{101}{32} \, \zeta_4 \, \ln^2 u
+ \frac{1}{4} \, ( 11 \, \zeta_5 + 5 \, \zeta_2 \, \zeta_3 ) \, \ln u
- \frac{3447}{128} \, \zeta_6 + \frac{1}{4} \, (\zeta_3)^2 \,. \label{V3_1u1_u0}
\eea
As mentioned above, the small $u$ behavior of the parity-odd $V$ functions
on the line $(u,1,1)$ is of the same order,
\bea
\tilde{V}^{(2)}(u,1,1) &\sim& - \frac{1}{8} \, \zeta_2 \, \ln^2 u
- \frac{5}{16} \, \zeta_4 \,, \label{Vt2_u11_u0} \\
\tilde{V}^{(3)}(u,1,1) &\sim& \frac{47}{32} \, \zeta_4 \, \ln^2 u
+ \frac{343}{128} \, \zeta_6
- \frac{1}{4} \, (\zeta_3)^2 \,. \label{Vt3_u11_u0} 
\eea
The ratio of the $\ln^2 u$ coefficient for $\tilde{V}^{(3)}$ to that
for $\tilde{V}^{(2)}$ is about $-7.7$, while for the constant term it
is about $-7$.  This numerical similarity accounts for some of the
indistinguishability of the two curves in figure~\ref{fig:vtu11},
but only at small $u$.

\subsection{The line $(u,u,u)$}
\label{lineuuusection}

When all three cross ratios are equal, the parity-odd function
vanishes by $u\lr w$ symmetry, $\tilde{V}(u,u,u)=0$.
In contrast to the behavior on the previous lines,
on the line $(u,u,u)$ the ratio function $V$ does not collapse to
ordinary HPLs.  We can still study its asymptotic behavior analytically,
and we can evaluate it numerically in order to inspect how zero crossings
of the ratio function change with loop order.  (For the remainder
function, these zero crossings, at each loop order and at strong coupling,
are all very close to $u=\frac{1}{3}$~\cite{Dixon2014voa}.)
 
Like all hexagon functions, the ratio function on the line $(u,u,u)$ can be
expressed~\cite{Dixon2014voa} in terms of the \emph{cyclotomic} HPLs
defined in ref.~\cite{Ablinger2011te}.  At the level of the symbol,
this correspondence is easy to see because on the line $(u,u,u)$ we have 
$u=y/(1+y)^2$ and $1-u=(1+y+y^2)/(1+y)^2$, where $y\equiv y_u$.
Therefore the symbol entries are all drawn from the set
$\{y,\, 1+y,\, 1+y+y^2 \}$.  The latter two elements of this set
are the second and third cyclotomic polynomials in $y$,
with roots $e^{i\pi}$ and $e^{\pm 2\pi i/3}$.

\begin{figure}
\begin{center}
\includegraphics[width=4.5in]{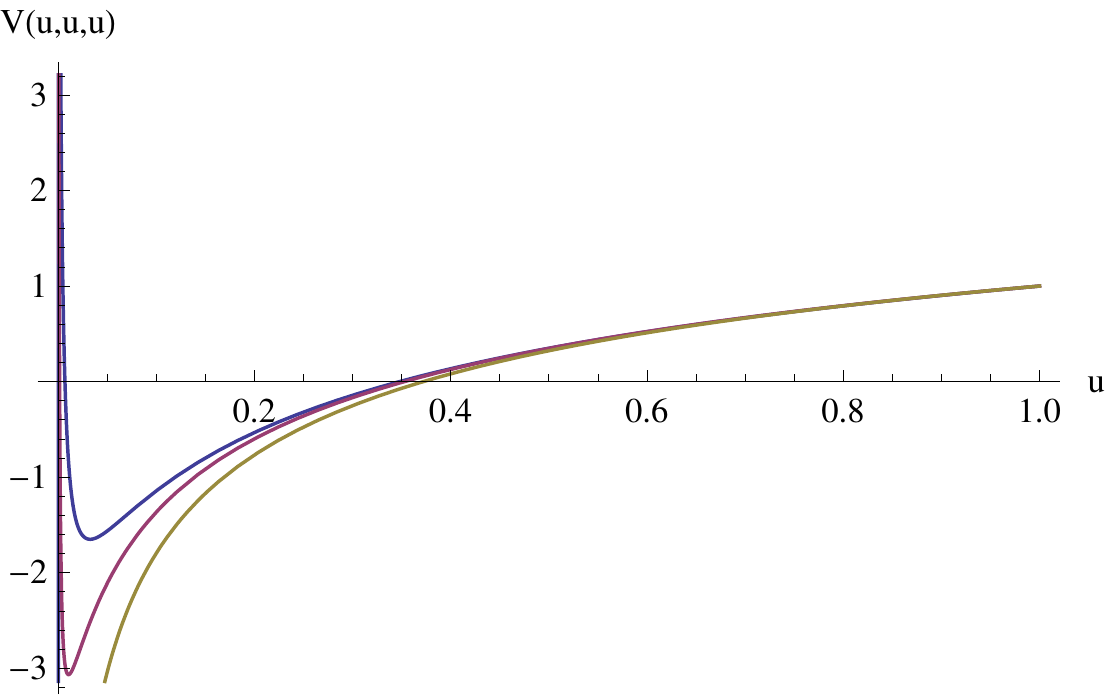}
\end{center}
\caption{$V^{(1)}(u,u,u)$, $V^{(2)}(u,u,u)$ and $V^{(3)}(u,u,u)$, 
normalized to unity at $(1,1,1)$. One loop is in green, two loops is
in purple, and three loops is in blue.}
\label{fig:vuuu}
\end{figure}

Here we do not make explicit use of the cyclotomic polylogarithm
correspondence.  In order to obtain a numerical representation, we simply
series expand to high orders (of order 100 terms) about $u=0,1,\infty$.
Such series representations have overlapping domains of convergence.
In figure \ref{fig:vuuu} we plot $V(u,u,u)$ for each loop order, normalized
to unity at $(1,1,1)$.
The point at which $V(u,u,u)$ crosses the zero line, in the neighborhood
of $u=\frac{1}{3}$, decreases gradually with increasing loop order.
We define the crossing values $u_0^{(L)}$ by the condition
$V^{(L)}(u_0^{(L)},u_0^{(L)},u_0^{(L)})=0$.  They are given by
\be
u_0^{(1)} = 0.372098\ldots, \qquad
u_0^{(2)} = 0.352838\ldots, \qquad
u_0^{(3)} = 0.347814\ldots. \qquad
\label{zerocrossing}
\ee
As in the case of the line $(u,u,1)$, there are oscillations
and additional zero crossings at higher loop order.  The
two-loop result has a zero crossing near $0.0015$.  The three-loop
function crosses near $0.007$ and again near $1.3\times 10^{-6}$.

These zero crossings are again dictated by the small $u$ asymptotic
behavior,
\bea
V^{(1)}(u,u,u) &\sim& \frac{1}{2} \, \ln^2 u 
+ \frac{1}{2} \, \zeta_2 \,, \label{V1_uuu_u0} \\
V^{(2)}(u,u,u) &\sim& \frac{1}{16} \, \ln^4 u
- \frac{3}{2} \, \zeta_2 \, \ln^2 u + \frac{1}{2} \, \zeta_3 \, \ln u
- \frac{53}{16} \, \zeta_4 \,, \label{V2_uuu_u0} \\
V^{(3)}(u,u,u) &\sim& \frac{1}{288} \, \ln^6 u
- \frac{41}{96} \, \zeta_2 \, \ln^4 u
+ \frac{1}{8} \, \zeta_3 \, \ln^3 u
+ \frac{419}{32} \, \zeta_4 \, \ln^2 u
- \Bigl( 2 \, \zeta_5 + \frac{3}{4} \, \zeta_2 \, \zeta_3 \Bigr) \, \ln u
\nonumber\\&&\null
+ \frac{2589}{128} \, \zeta_6 - \frac{1}{4} \, (\zeta_3)^2 \,. \label{V3_uuu_u0}
\eea
The leading-log behavior at each order has exactly the same coefficients as 
did the small-$u$ expansion on the line $(u,u,1)$.  The subleading-log
coefficients are different, however.

\subsection{The plane $u+v+w=1$}

The plane $u+v+w=1$ intersects the positive octant in an equilateral
triangle.  This triangle is bounded by the lines corresponding to the
collinear limits: $v=0$, $u+w=1$, and cyclic permutations of this
line.  The remainder function $R_6^{(3)}$ comes very close to
vanishing on this equilateral triangle~\cite{Dixon2013eka}, which may
not be so surprising, given that the remainder function is identically
zero on all three edges of the triangle. In contrast, the collinear
limits of the ratio function involve two different permutations of
$V$.  For this reason, the behavior of the ratio function on the plane
$u+v+w=1$ can be much less uniform than is the remainder function.
Both $V$ and $\tilde{V}$ show an interesting range of behavior,
and their zero-crossing surfaces slice through this plane.

\begin{figure}
\begin{center}
\includegraphics[width=5.5in]{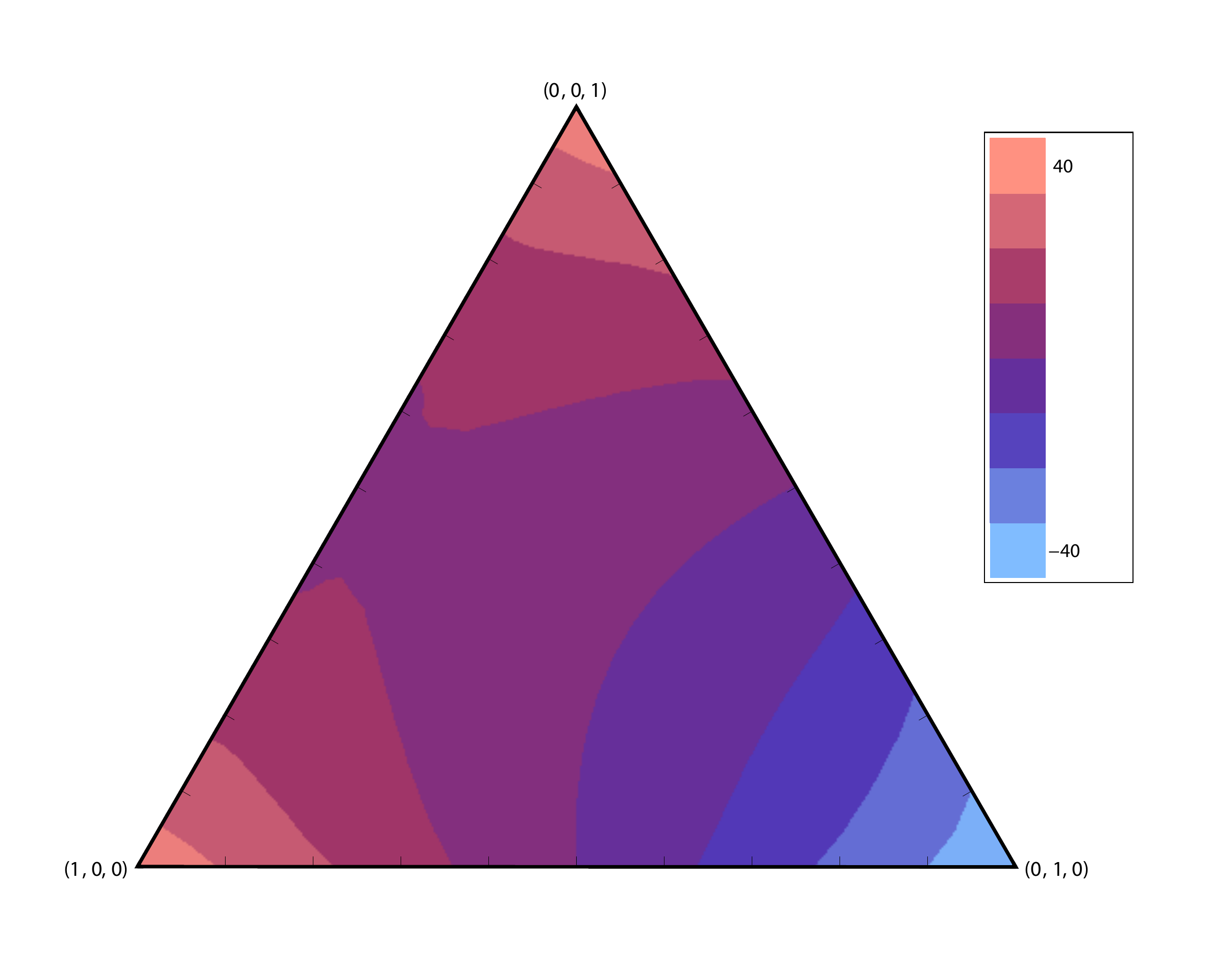}
\end{center}
\caption{$V^{(3)}(u,v,w)$ evaluated on the plane where $u+v+w=1$.
The corners are labeled with their $(u,v,w)$ values.}
\label{fig:vtriangle}
\end{figure}

Figure~\ref{fig:vtriangle} plots $V^{(3)}(u,v,w)$ on the equilateral
triangle.  In this region, the function reaches its highest values
near the triangle's vertices at $u=1$ and $w=1$, and its lowest values
near the vertex at $v=1$.  The function crosses zero on a curve in
between; the vanishing curve is not far from the circle of radius
$\frac{1}{2}$ centered at $(0,1,0)$.  From \eqn{OPE6134}, we can see that
$V^{(L)}(v,w,u)$ diverges like $\ln^L T$ in the collinear limit with
$w \approx T^2 \to 0$.  In fact, all permutations of $V^{(L)}$ diverge
like $\ln^L T$ in the near-collinear limit.  Therefore $V^{(3)}$
actually becomes infinite on each edge of the equilateral triangle in
figure~\ref{fig:vtriangle}.  However, it diverges extremely slowly, so
that the divergence is not apparent at all in the plot.

What is visible in the plot is the symmetry of $V^{(3)}$ under $u\lr w$,
which exchanges the lower-left and top corners of the triangle.
It is also clear from the plot that on the lower edge of the triangle
$V^{(3)}$ is odd under reflection about the edge's midpoint.  By symmetry,
the same reflection-odd property holds along the upper-right edge of the
triangle.  Using the $u\lr w$ symmetry of $V(u,v,w)$, this property
is just a consequence of the collinear vanishing
constraint~(\ref{collvanish}),
\be
V(u,v,w) + V(v,u,w)\ \to\ 0, \qquad \hbox{as $w\to0$,~~$v\to1-u$.}
\label{bottomedgevanish}
\ee
So the fact that the vanishing surface intersects two of the edges
of the equilateral triangle at their midpoints is no surprise.

\begin{figure}
\begin{center}
\includegraphics[width=5.5in]{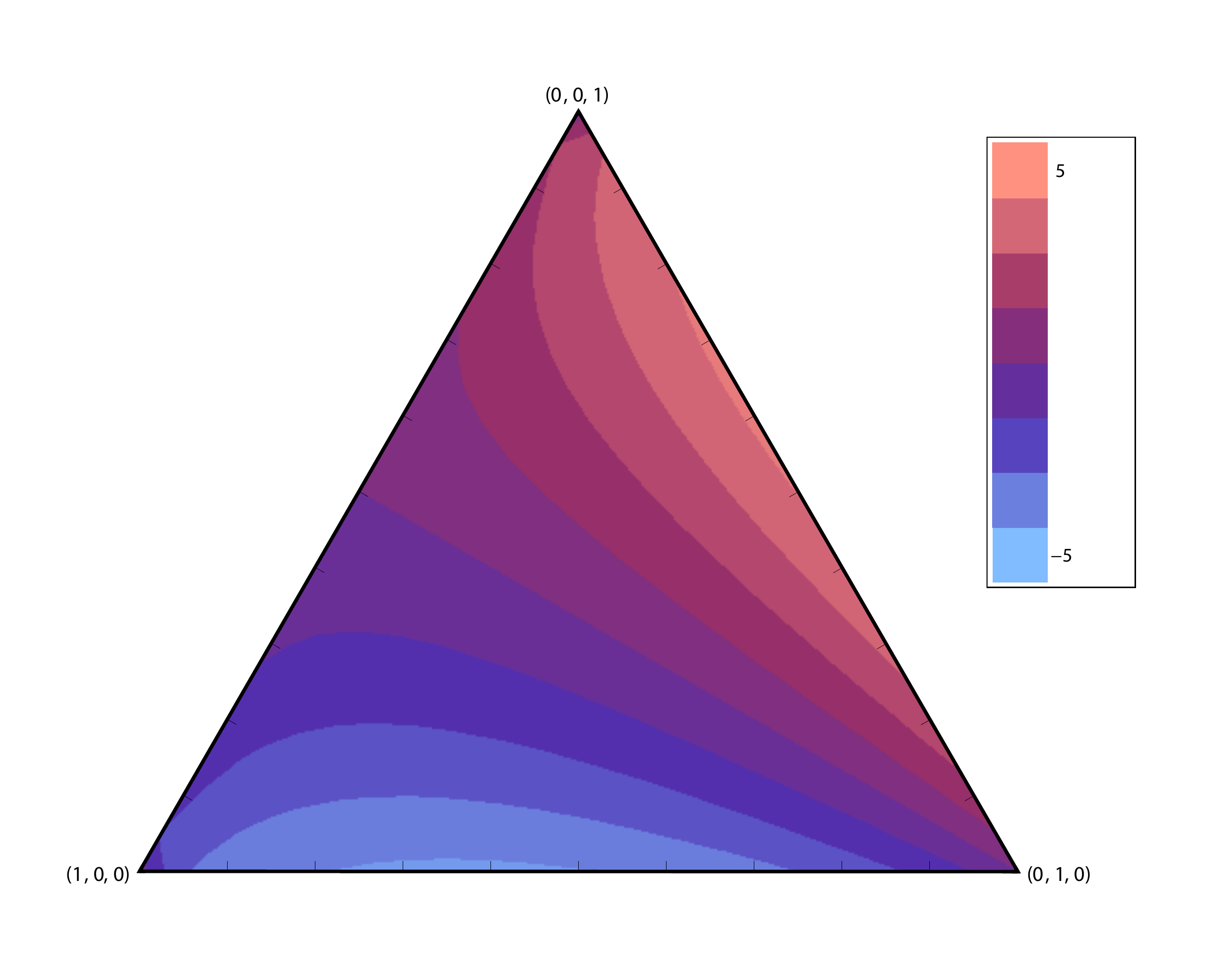}
\end{center}
\caption{$\tilde{V}^{(3)}(u,v,w)/i$ evaluated on the plane where $u+v+w=1$.
The corners are labeled with their $(u,v,w)$ values.}
\label{fig:vttriangle}
\end{figure}

The parity-odd function $\tilde{V}^{(3)}(u,v,w)$ is plotted on the same
equilateral triangle in figure~\ref{fig:vttriangle}.  Parity-odd pure
functions are pure imaginary when $\Delta < 0$, as in this region,
so we divide $\tilde{V}^{(3)}$ by $i$ before plotting it.
This function is antisymmetric under the exchange $u\lr w$ and therefore
it vanishes on the line $u=w$. It is positive for large $w$ and negative
for large $u$.  Like any parity-odd function, $\tilde{V}^{(3)}(u,v,w)$
vanishes in the collinear limits, on the edges of the triangle.
However, this vanishing happens so slowly that it is not evident in the plot.

The sign of $\tilde{V}^{(3)}/i$ depends on the value of $(y_u,y_v,y_w)$, not
just $(u,v,w)$.  For each point $(u,v,w)$, there are two points
$(y_u,y_v,y_w)$, related by flipping the sign of $\sqrt{\Delta}$ in
\eqn{yfromu}.  This sign flip inverts the three $y_i$.  On the plane
$u+v+w=1$, $\sqrt{\Delta}$ is imaginary, and the $y_i$ are pure phases,
satisfying $y_u y_v y_w = -1$.  The sign flip conjugates the three phases.
The branch of $\tilde{V}^{(3)}/i$ plotted in figure~\ref{fig:vttriangle}
is for positive $\sqrt{\Delta}/i$, corresponding to negative imaginary
parts for all three $y_i$.

\subsection{Planes in $v$}

In order to get a complete view of the ratio function's behavior, it
is useful to plot it as a function of $u$ and $w$ for successive values
of $v$.

\begin{figure}
\begin{center}
\includegraphics[width=5.5in]{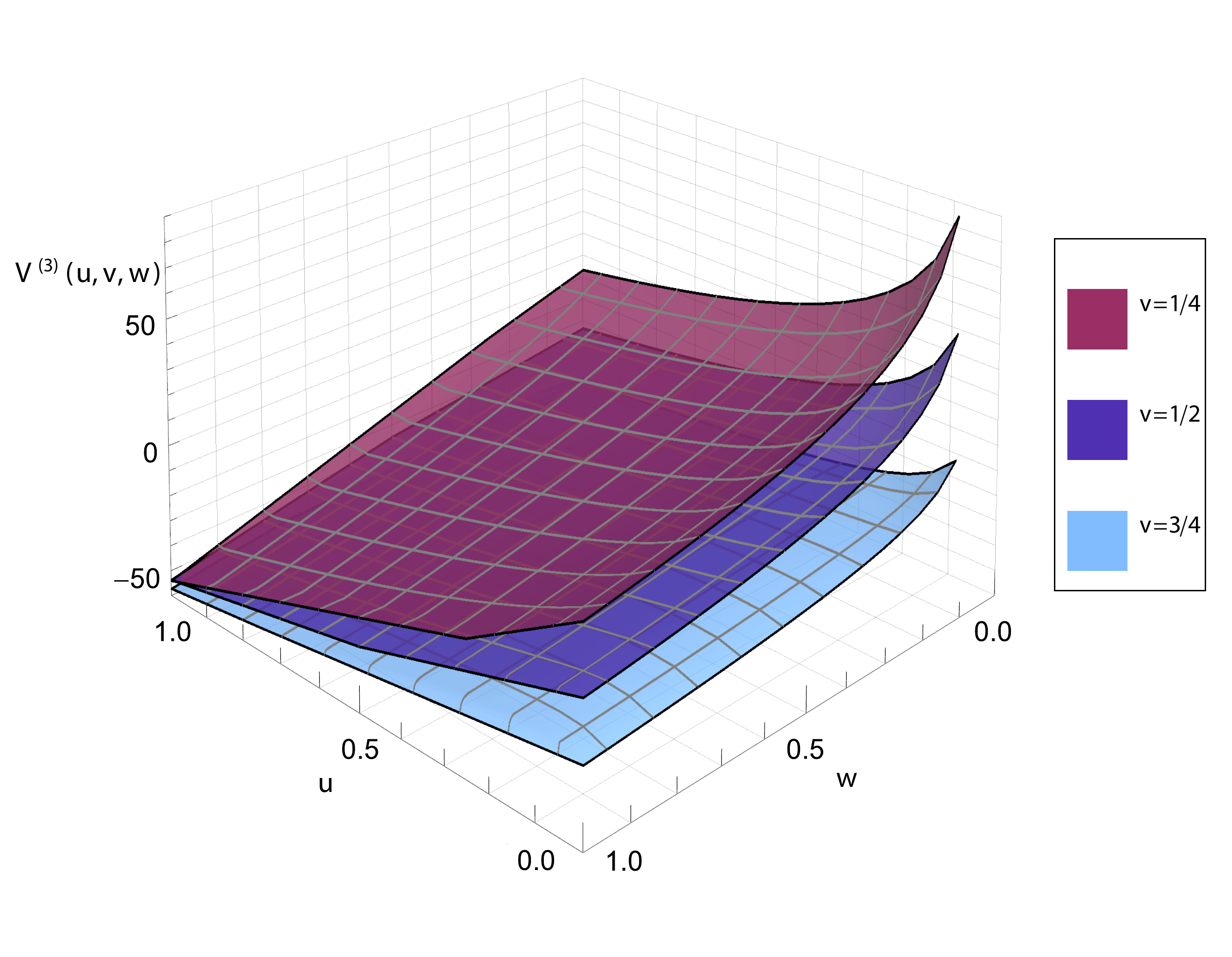}
\end{center}
\caption{$V^{(3)}(u,v,w)$ evaluated on successive planes in $v$.}
\label{fig:vstack}
\end{figure}

In figure~\ref{fig:vstack}, we plot $V^{(3)}(u,v,w)$ on the planes
$v=\frac{3}{4}$, $v=\frac{1}{2}$, and $v=\frac{1}{4}$.
A similar plot has been made for the remainder function $R_6^{(3)}$,
as figure 8 of ref.~\cite{Dixon2013eka}.  (The roles of $v$ and $w$
are reversed in that plot, but of course that is irrelevant for the
remainder function, since it is totally symmetric.)
Much as in the case of the remainder function $R_6^{(3)}$,
the function $V^{(3)}(u,v,w)$ looks monotonic in $v$, but actually the
$v=\frac{1}{2}$ and $v=\frac{1}{4}$ planes intersect near $u=w=1$.

\begin{figure}
\begin{center}
\includegraphics[width=5.5in]{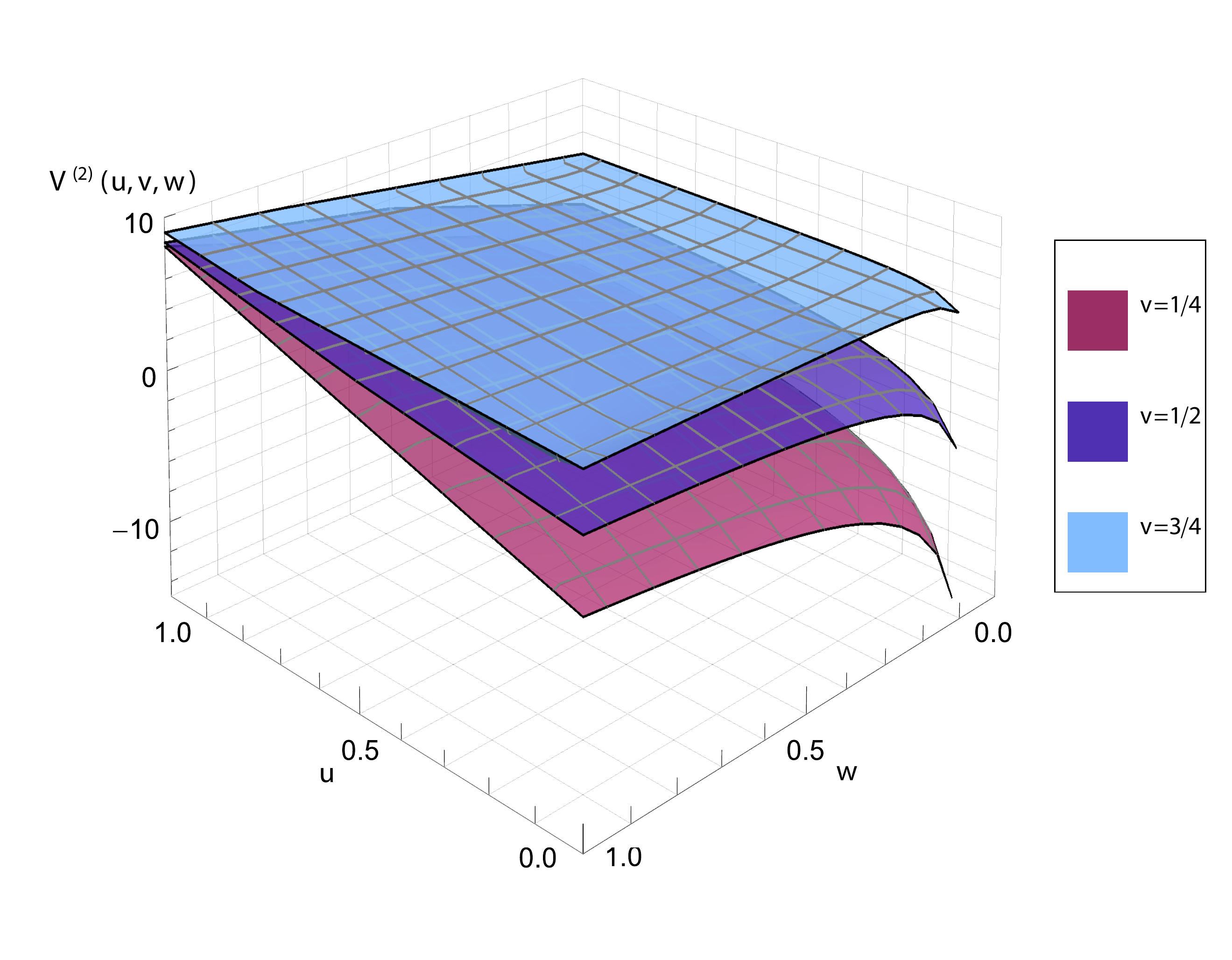}
\end{center}
\caption{$V^{(2)}(u,v,w)$ evaluated on successive planes in $v$.}
\label{fig:v2stack}
\end{figure}

The functions $V^{(3)}(u,v,w)$ and $V^{(2)}(u,v,w)$ cross zero on different
surfaces.  The difference in zero-crossing locations means that plotting
the ratio of $V^{(3)}(u,v,w)$ to $V^{(2)}(u,v,w)$ is relatively
uninformative. Instead, in figure \ref{fig:v2stack} we plot $V^{(2)}(u,v,w)$
on the same planes in $v$ for comparison.

\begin{figure}
\begin{center}
\includegraphics[width=5.5in]{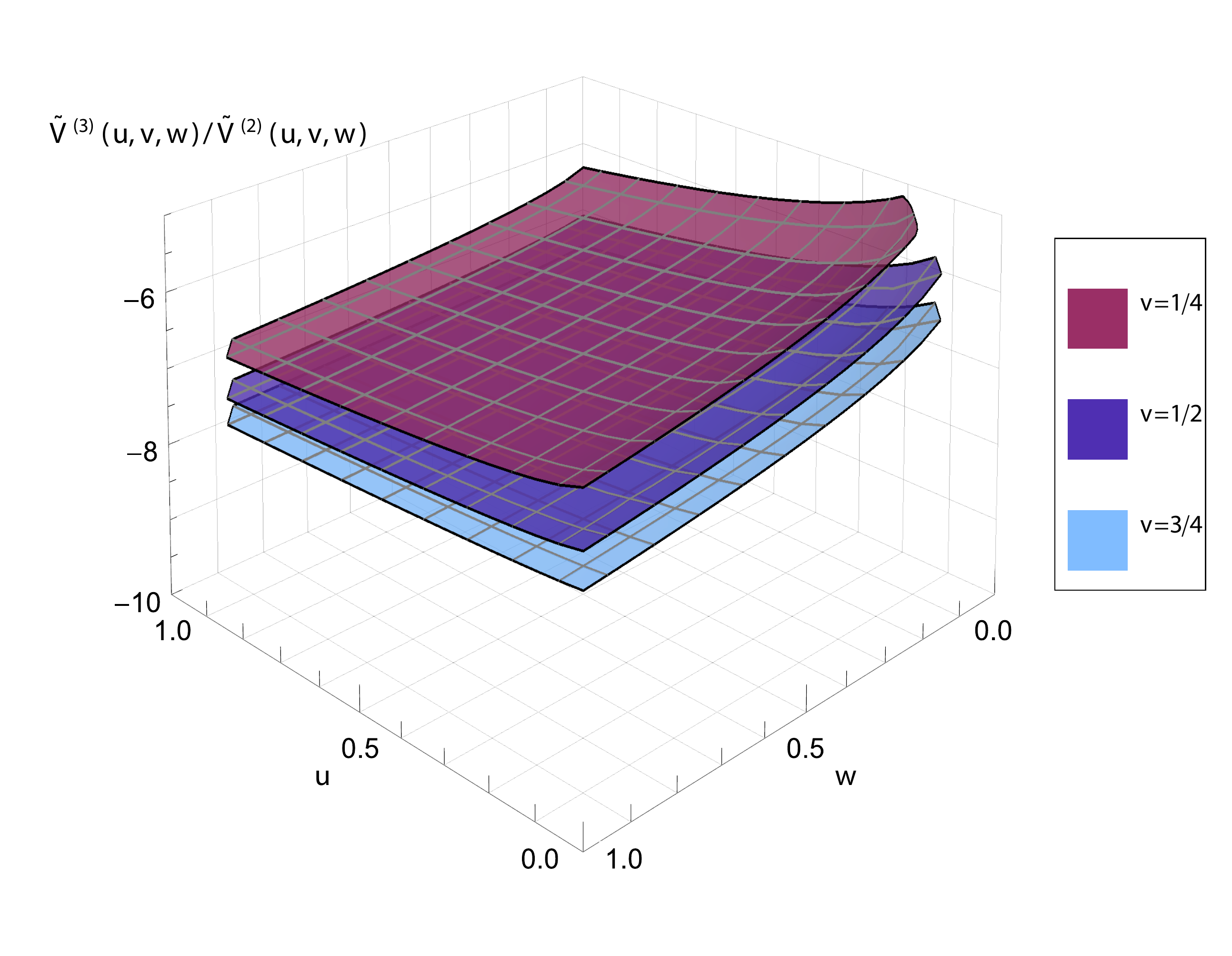}
\end{center}
\caption{The ratio $\tilde{V}^{(3)}(u,v,w)/\tilde{V}^{(2)}(u,v,w)$
evaluated on successive planes in $v$.}
\label{fig:vtstack}
\end{figure}

In contrast, $\tilde{V}^{(3)}(u,v,w)$ and $\tilde{V}^{(2)}(u,v,w)$ are
constrained to cross zero at the exact same place, on the plane
$u=w$. Parity-odd functions are either real or imaginary based on the sign of
$\Delta(u,v,w)$, and they vanish on the surface $\Delta=0$.  For these
reasons, it is simpler to plot the ratio between two odd functions
than to plot one odd function alone. Omitting points for which $u=w$,
and for which $\Delta$ vanishes, we plot the ratio
$\tilde{V}^{(3)}(u,v,w)/\tilde{V}^{(2)}(u,v,w)$ in figure
\ref{fig:vtstack}.  Given the vanishings of both numerator and
denominator within the region of the plot, it is remarkable that the ratio
$\tilde{V}^{(3)}/\tilde{V}^{(2)}$ stays within a fairly limited range and
has no dramatic behavior.  On the other hand, it is clear that it is not
totally constant in $u$, $v$, or $w$.

\section{Relation between $V$ and coproduct elements of $R_6$}
\label{curiousrelsect}

In \sect{coprels}, we observed that the function $U$ had surprising
simplicity, possessing only five independent $\{n-1,1\}$ coproduct components.
In this section, we will describe another interesting empirical observation.
This one relates the even-parity NMHV function at $L$ loops, $V^{(L)}$,
to the elements of the $\{2L,1,1\}$ coproduct component of the remainder
function at one higher loop, $R_6^{(L+1)}$.  The relation was originally noticed
while inspecting the behavior of both functions on the diagonal line
$(u,u,u)$, for the purpose of making the plot in \sect{lineuuusection}.
However, it can be extended to a relation that holds throughout
the $(u,v,w)$ space.

We denote the $\{2L,1,1\}$ component of the coproduct of $R_6^{(L+1)}$
by $\Delta_{2L,1,1}(R_6^{(L+1)})$. Its elements
can be represented as,
\be
\label{R64nm211anz}
\Delta_{2L,1,1}(R_6^{(L+1)}) = \sum_{s_i,s_j\in \Su} [R_6^{(L+1)}]^{s_i,s_j}
\otimes \ln s_i \otimes \ln s_j \,,
\ee
and they each have weight $2L$.
We found the following relation:
\be
V^{(L)}(u,v,w) = [R_6^{(L+1)}]^{Z,Z} + E^{(L+1)} + \frac{1}{8} \, \gamma_K^{(L+1)} \,,
\label{curiousrel}
\ee
where we define the ``$Z,Z$'' linear combination of coproduct elements
of a hexagon function $X$ to be,
\be
X^{Z,Z} \equiv - X^{v,v} - X^{1-v,v} + X^{y_u,y_u} + X^{y_w,y_w} - 3 \, X^{y_v,y_v} 
+ 2 \, \bigl( X^{y_w,y_v} + X^{y_u,y_v} \bigr) - X^{y_u,y_w} - X^{y_w,y_u} \,.
\label{ZZ}
\ee
Recall that the cusp anomalous dimension is given through four loops
by,
\be
\gamma_K(a) = \sum_{L=1}^\infty a^L\,\gamma_K^{(L)} = 4a - 4 \zeta_2\,a^2
+ 22\zeta_4\,a^3
- 4 \biggl( \frac{219}{8} \zeta_6 + (\zeta_3)^2 \biggr) a^4 
+ \mathcal{O}(a^5)\,.
\label{cuspdef}
\ee
The extra term $E$ is only needed so far at zero loops (which is
definitely a special case), and at four loops, where it is proportional to
the square of the $P$-odd $D=6$ hexagon integral:
\be
E^{(1)} = \frac{1}{2} \,, \qquad
E^{(2)} = E^{(3)} = 0 \,, \qquad
E^{(4)} = \frac{1}{16} \, [ \tilde\Phi_6 ]^2 \,.
\label{Eterm}
\ee
For $L=0$, the relation~(\ref{curiousrel}) is simply
$1=0+\tfrac{1}{2}+\tfrac{1}{2}$.  

For $L=1$, it is straightforward
to compute the $\{2,1,1\}$ coproduct component of $R_6^{(2)}$ from the
form given in ref.~\cite{Dixon2011nj} in terms of classical polylogarithms
and the function $\Omega^{(2)}$, whose $\{3,1\}$ coproduct component is
given in ref.~\cite{Dixon2013eka}.   We find,
\be
- [R_6^{(2)}]^{v,v} - [R_6^{(2)}]^{1-v,v} = [R_6^{(2)}]^{y_v,y_v} + \frac{\zeta_2}{2}
\ =\ [R_6^{(2)}]^{y_v,y_u} + \frac{\zeta_2}{2}
\ =\ \frac{1}{4} \Bigl[ H_2^u + H_2^v + H_2^w + \ln u \, \ln w \Bigr] \,.
\label{R62_coprod211}
\ee
Because $R_6^{(2)}(u,v,w)$ is totally symmetric in $u,v,w$, we can obtain
all the other coproduct elements entering $[R_6^{(2)}]^{Z,Z}$ by permuting
the ones given in \eqn{R62_coprod211}.  It is then easy to check that
the right-hand side of \eqn{curiousrel} for $L=1$ adds up to yield $V^{(1)}$
as given in \eqn{Voneloop}.

To check the relation~(\ref{curiousrel}) for $L=2$, we use the formulae
for the elements $R_6^{(3),u}$ and $R_6^{(3),y_u}$
of the $\{5,1\}$ coproduct component of $R_6^{(3)}$ 
given in ref.~\cite{Dixon2013eka}, as well as the final-entry relation
$R_6^{(3),1-u} = -R_6^{(3),u}$.  These formulae are given in terms of
the basis of weight-five hexagon functions, whose $\{4,1\}$ coproducts
are also tabulated in ref.~\cite{Dixon2013eka}.
This information makes it straightforward to extract the $\{4,1,1\}$
coproduct component of $R_6^{(3)}$ from the $\{5,1\}$ coproduct component:
\bea
~[R_6^{(3)}]^{v,v} &=& \frac{1}{32} \Bigl[ 
   3\, \Bigl( \Omega^{(2)}(u,v,w) + \Omega^{(2)}(v,w,u) \Bigr)
   + 2 \, \Omega^{(2)}(w,u,v)
   + \hbox{HPLs} + 16\,\zeta_4 \Bigr]\,,
\label{R63vv}\\
~[R_6^{(3)}]^{1-v,v} &=& \frac{1}{32} \Bigl[ 
     \Omega^{(2)}(u,v,w) + \Omega^{(2)}(v,w,u)
   + 2 \, \Omega^{(2)}(w,u,v)
   + \hbox{HPLs} + 24\,\zeta_4 \Bigr]\,,
\label{R63omvv}\\
~[R_6^{(3)}]^{y_v,y_v} &=& - \frac{1}{32} \Bigl[ 
     9 \, \Bigl( \Omega^{(2)}(u,v,w) + \Omega^{(2)}(v,w,u) \Bigr)
   + 6 \, \Omega^{(2)}(w,u,v)
   + \hbox{HPLs} - 24\,\zeta_4 \Bigr]\,,
\label{R63yvyv}\\
~[R_6^{(3)}]^{y_v,y_u} &=& - \frac{1}{32} \Bigl[ 
     8 \, \Omega^{(2)}(u,v,w)
   + 7 \, \Bigl( \Omega^{(2)}(v,w,u) + \Omega^{(2)}(w,u,v) \Bigr)
   + \hbox{HPLs} - 36\,\zeta_4 \Bigr]\,.~~~~~~~
\label{R63yvyu}
\eea
The portions of the expressions containing harmonic polylogarithms
are fairly lengthy, so we do not present them here.
Using these results and their various permutations, we can assemble the
right-hand side of \eqn{curiousrel} and verify that it agrees with $V^{(2)}$
as given in \eqn{V2}.

For $L=3$, we first check the first derivative of \eqn{curiousrel} with
respect to $u$, $v$ and $w$.  We do this using the fact that the derivatives
of each of the elements of the $\{6,1,1\}$ coproduct component of $R_6^{(4)}$
can be expressed in terms of the $\{5,1,1,1\}$ coproduct elements,
using the general \eqn{eq:diff_basis}.
The $\{5,1,1,1\}$ coproduct elements of $R_6^{(4)}$
can in turn be written in terms of the weight-five basis of
hexagon functions~\cite{Dixon2014voa}.  On the left-hand side of
\eqn{curiousrel}, we compute the derivative of $V^{(3)}(u,v,w)$
using the $\{5,1\}$ coproduct component for $U^{(3)}$ presented in
appendix~\ref{Ucoprodsubappendix}, together with the relation
between $V^{(3)}$ and $U^{(3)}$ given in \eqn{VfromU} or \eqn{Uh3}.
Expanding both sides of the derivative of \eqn{curiousrel} in terms
of the basis of weight-five functions, they agree perfectly.
Having checked the first derivative, we should check the relation
at one point in order to establish that the constant of integration
is also correct.  It is convenient to choose the point $(u,v,w)=(1,1,1)$.
At this point, we have
\bea
~[R_6^{(4)}]^{v,v}(1,1,1) &=& \frac{73}{8} \, \zeta_6 - \frac{1}{2} (\zeta_3)^2 \,,
\label{R64vv111}\\
~[R_6^{(4)}]^{1-v,v}(1,1,1) &=& 0 \,, \label{R64omvv111}\\
~[R_6^{(4)}]^{y_v,y_v}(1,1,1) &=& -\frac{607}{16} \, \zeta_6 \,,
\label{R64yvyv111}\\
~[R_6^{(4)}]^{y_v,y_u}(1,1,1) &=& -\frac{607}{16} \, \zeta_6 \,.
\label{R64yvyu111}
\eea
Using these relations, plus \eqn{V3_111} for $V^{(3)}(1,1,1)$ and
\eqn{cuspdef} for the four-loop cusp anomalous dimension, 
as well as the fact that $\tilde\Phi_6(1,1,1) = 0$, it is easy to
verify that \eqn{curiousrel} holds for $L=3$ at $(u,v,w)=(1,1,1)$.

There are a number of linear relations among the $\{2L,1,1\}$
coproduct elements, which follow from integrability,
{\it i.e.}~from the consistency of mixed partial derivatives
of the original weight-$2(L+1)$ function $R_6^{(L+1)}$.
These integrability relations make it possible
to rewrite \eqn{curiousrel} in various ways.  In \eqn{curiousrel},
we used these relations
to eliminate the ``off-diagonal'' even-even coproduct elements,
$[R_6^{(L+1)}]^{v,u}$, $[R_6^{(L+1)}]^{1-v,u}$, and permutations thereof.
It is possible that using the integrability relations in a different
way might lead to a version of \eqn{curiousrel} that is more
revealing of its origin.  No matter how it is rewritten, though,
the appearance of the cusp anomalous dimension in an equation
that holds throughout the full $(u,v,w)$ space of cross ratios
is very interesting.

The appearance of the extra $[ \tilde\Phi_6 ]^2$ term at four loops
is presumably related to the fact that we are using a logarithmic
definition of the remainder function, \eqn{Rdef}.
The ratio function is not defined by taking any logarithms.
Let's define a modified remainder function $\bar{R}_6$ by
\be 
\frac{A^{\rm MHV}}{A^{\rm BDS}} = \bar{R}_6 = \exp(R_6).
\label{Rbardef}
\ee
Then $\bar{R}_6$ and $V$ are on the same footing.
The zero-loop value of $\bar{R}_6$ differs from that of $R_6$:
$\bar{R}_6^{(0)} = 1$, while $R_6^{(0)} = 0$.  (This shift does not
affect \eqn{curiousrel}, of course.)
Otherwise, $R_6$ and $\bar{R}_6$ are identical until four loops,
at which point they are related by,
\be
\bar{R}_6^{(4)} = [\exp(R_6)]^{(4)}
= R_6^{(4)} + \frac{1}{2} \Bigl[ R_6^{(2)} \Bigr]^2 \,.
\label{fourloopshift}
\ee
We can rewrite \eqn{curiousrel} in terms of coproducts of $\bar{R}_6$
instead of $R_6$.  In order to do that, according to \eqn{fourloopshift}
we need to compute the $Z,Z$ coproduct of $[ R_6^{(2)} ]^2$. We find that
\bea
\Bigl\{ \Bigl[ R_6^{(2)} \Bigr]^2 \Bigr\}^{Z,Z}
&=& \frac{1}{8} \, [ \tilde\Phi_6 ]^2 + 2 \, [ R_6^{(2)} ]^{Z,Z} \, R_6^{(2)}
\label{R62sqrelA}\\
&=& \frac{1}{8} \, [ \tilde\Phi_6 ]^2
  + 2 \, \left( V^{(1)} - \frac{1}{8} \gamma_K^{(2)} \right) \, R_6^{(2)} \,.
\label{R62sqrelB}
\eea
This relation implies that we can rewrite \eqn{curiousrel} in terms
of $\bar{R}_6$ as,
\be
V^{(L)}(u,v,w) = [\bar{R}_6^{(L+1)}]^{Z,Z}
   + \bar{E}^{(L+1)} + \frac{1}{8} \, \gamma_K^{(L+1)} \,,
\label{curiousrelbar}
\ee
where
\be
\bar{E}^{(1)} = \frac{1}{2} \,, \qquad
\bar{E}^{(2)} = \bar{E}^{(3)} = 0 \,, \qquad
\bar{E}^{(4)} = - [ \bar{R}_6^{(2)} ]^{Z,Z} \, \bar{R}_6^{(2)} \,.
\label{Ebarterm}
\ee

Next, we further ``improve'' \eqn{curiousrelbar} by removing
the $\bar{E}$ term.  We do this by considering not $V$ but
$V \bar{R}_6 = V\,\exp(R_6)$, much as we did in \sect{multiparticle} when
studying the multi-particle factorization properties.
As discussed in \sect{multiparticle}, $V \bar{R}_6$ is a pure NMHV
quantity, as the finite part of the MHV amplitude has been cleared
out of the denominator.  First, let's multiply \eqn{curiousrelbar}
by $a^{L+1}$ and sum over $L$ to obtain,
\be
a \Bigl( V - \frac{1}{2} \Bigr) - \frac{1}{8} \, \gamma_K 
= \bar{R}_6^{Z,Z} + \Bigl( \bar{E} - \frac{a}{2} \Bigr) \,,
\label{curiousA}
\ee
where $\bar{E}-a/2$ vanishes until four loops.
Now multiply by $\bar{R}_6$:
\be
\biggl[ a \, \Bigl( V - \frac{1}{2} \Bigr) - \frac{1}{8} \, \gamma_K \biggr]
\, \bar{R}_6 = \bar{R}_6^{Z,Z} \, \bar{R}_6 
            + \Bigl( \bar{E} - \frac{a}{2} \Bigr) \bar{R}_6 \,.
\label{curiousB}
\ee
Through four loops, using \eqn{Ebarterm}, we have,
\bea
\bar{R}_6^{Z,Z} \, \bar{R}_6 
&=& \bar{R}_6^{Z,Z} + a^4 [\bar{R}_6^{(2)}]^{Z,Z} \, \bar{R}_6^{(2)}
+ {\cal O}(a^5), \label{RRZZexpand}\\
\Bigl( \bar{E} - \frac{a}{2} \Bigr) \bar{R}_6
&=& - a^4 [\bar{R}_6^{(2)}]^{Z,Z} \, \bar{R}_6 ^{(2)} + {\cal O}(a^5),
\label{Ebarexpand}
\eea
so the explicit $a^4$ terms cancel in the sum.

We are left with,
\be
\biggl[ a \, \Bigl( V - \frac{1}{2} \Bigr) - \frac{1}{8} \, \gamma_K \biggr]
\, \bar{R}_6 = \bar{R}_6^{Z,Z} \,,
\label{curiousfinal}
\ee
which is valid at least through order $a^4$.
Except for the factor of $1/2$, this equation is a 
relation for the difference between the
NMHV and MHV amplitudes, $V \, \bar{R}_6 - \bar{R_6}$.
The right-hand side looks naively like a second-order differential
operator, but of course the coproduct operation is not the same
as taking a derivative.  Nevertheless, it might be useful
to try to prove~\eqn{curiousfinal} using the
$\bar{Q}$ differential equation found in the super-Wilson loop
approach~\cite{BullimoreSkinner,CaronHuot2011kk}.

Given this interesting relation for the $P$-even function $V$,
we investigated whether it was possible to write the $P$-odd part
of the ratio function, $\tilde{V}$, as a linear combination
of the $P$-odd $\{6,1,1\}$ coproduct elements of $R_6$ at one higher loop.
At one loop, or weight two, both sides of such a relation vanish trivially,
because there are no $P$-odd weight-2 hexagon functions.
For $\tilde{V}^{(2)}$, we found multiple solutions; however, the space
of $P$-odd weight-4 hexagon functions is quite small, so most
of the solutions are presumably accidental.  Because $\tilde{V}$ is not
itself totally physical, it is better to consider the difference of two
cyclic permutations, $\tilde{V}(u,v,w)-\tilde{V}(w,u,v)$.
Note that this combination is symmetric under the exchange $u\lr v$.

At three loops, we tried to write $\tilde{V}^{(3)}$ as a generic linear
combination of the odd elements
of the $\{6,1,1\}$ coproduct component of $R_6^{(4)}$, imposing
$u\lr v$ symmetry and taking into account the integrability relations among
the elements.  We could not find a solution for $\tilde{V}^{(3)}$
without also introducing the odd coproducts of $R_6^{(3)}$, multiplied
by a single logarithm.  Allowing for this, we found:
\bea
\tilde{V}^{(3)}(u,v,w) - \tilde{V}^{(3)}(w,u,v) &=&
4 \Bigl[ 2 \, R_6^{(4) \, y_w,w} + R_6^{(4) \, 1-w,y_w}
         - R_6^{(4) \, y_u,w} - R_6^{(4) \, y_v,w} \Bigr]\nonumber\\
&&\null
- 2 \, H_1^w \Bigl[ R_6^{(3) \, y_u} + R_6^{(3) \, y_v} - R_6^{(3) \, y_w} \Bigr]
\nonumber\\ 
&&\null
+ \tilde\Phi_6 \, \biggl[ \frac{1}{4} \, H_1^w \, \Omega^{(1)}
           - \frac{1}{2} \Bigl( H_3^w - H_{2,1}^w + H_1^w \, H_2^w \Bigr) \biggr]
\,,
\label{Vt3sol}
\eea
where $\tilde\Phi_6 = - 4 \, R_6^{(2) \, y_w}$.
Again integrability relations allow one to rewrite this linear
combination in different ways.

This solution is nice in that it descends smoothly to one loop lower,
in which the final term is absent:
\bea
\tilde{V}^{(2)}(u,v,w) - \tilde{V}^{(2)}(w,u,v) &=&
4 \Bigl[ 2 \, R_6^{(3) \, y_w,w} + R_6^{(3) \, 1-w,y_w}
         - R_6^{(3) \, y_u,w} - R_6^{(3) \, y_v,w} \Bigr]\nonumber\\
&&\null
- 2 \, H_1^w \Bigl[ R_6^{(2) \, y_u} + R_6^{(2) \, y_v} - R_6^{(2) \, y_w} \Bigr] \,.
\label{Vt2sol}
\eea
However, the structure of this relation does not yet seem as simple
as the one for the parity-even part $V$.

At the moment, the ultimate significance of these relations is still
quite unclear. It would be interesting to investigate their meaning in
the near-collinear and multi-Regge limits, where the OPE approach of
Basso, Sever and Vieira, and the recent work of Basso, Caron-Huot and Sever,
respectively, provide information at much higher loop order, 

Recently, BSV have investigated a double-scaling limit in which $T\to0$
but $TF$ is held fixed. In this limit, only
gluonic flux-tube excitations contribute~\cite{BSVIV}.
This limit corresponds to taking $v\rightarrow0$ with
$u$ and $w$ held fixed.  In this limit, the letters of the symbols for
hexagon functions (after extracting powers of $\ln v$)
can be shown to collapse to a simple five parameter set,
$\{u,w,1-u,1-w,1-u-w\}$.  This means that hexagon functions approach a
subset of the  2dHPL function space introduced by Gehrmann and
Remiddi~\cite{GehrmannRemiddi} in order to solve for the master integrals
for the process $\gamma^* \rightarrow qg\bar{q}$ at two loops.
(The 2dHPLs also allow for the letter $(u+w)$, which does not appear here.)
BSV have a simple rule for an insertion factor $h_a(u)$ (where $u$ is
the rapidity) that relates NMHV to MHV Wilson loops. At leading
order, the insertion factor leads to a relation for the 1111 component
of the NMHV Wilson loop in terms of a second-order Laplacian operator acting
on the MHV Wilson loop.  This relation looks superficially
similar to \eqn{curiousfinal}, although the NMHV side of the relation
involves two permutations each of $V$ and $\tilde{V}$, and it is clear that
the relation will have to be modified at higher loop orders.

In general, there might be other, cleaner ways to rewrite the parity-even
and parity-odd coproduct relations found in this section, which might
better reveal their origin.  We shall leave such investigations
for future work.


\section{Conclusions and outlook}
\label{conclusions}

In this paper we successfully extended the bootstrap program,
initiated in ref.~\cite{Dixon2011pw}, to calculate the three-loop six-point
NMHV ratio function in planar $\mathcal{N}=4$ super Yang-Mills theory.
We began with an ansatz for the coproduct of the desired functions, built
out of the hexagon functions introduced in ref.~\cite{Dixon2013eka}. By
constraining this ansatz with the known behavior of the ratio function
in various kinematic limits, we were able to uniquely determine the
NMHV coefficient functions $V$ and $\tilde{V}$ through three loops.

At three loops, we began with a 412-parameter ansatz. After applying
several constraints, including the vanishing of the cyclicly
symmetric part of $\tilde{V}^{(3)}$, a final-entry condition drawn from the
$\bar{Q}$ differential equation from ref.~\cite{CaronHuot2011kk}, the
vanishing of spurious poles, and vanishing in the collinear limit, we
had 92 parameters remaining.

We were then able to fix those parameters with near-collinear data
obtained from the work of Basso, Sever and Vieira~\cite{Basso2013vsa}.
The first-order $T^1$ correction in the near-collinear
limit, from single flux excitations~\cite{Basso2013aha}, was sufficient to
fix all but two parameters in our ansatz. Those two parameters can be
fixed if we also incorporate BSV's recently published results for
the contributions of two flux excitations, at order
$T^2$~\cite{Basso2014koa}. The rest of the order $T^2$ results
then serves as an extensive check on BSV's results.

Alternatively, the remaining two parameters can be fixed by examining
the multi-Regge limits of the amplitude. By generalizing the
predictions of ref.~\cite{Lipatov2012gk} beyond the leading-logarithmic
limit, we were able to fix the form of $V^{(3)}$ and $\tilde{V}^{(3)}$
independently of BSV's $T^2$ data, letting the $T^2$ comparison
serve as an entirely independent check.
Using the NLLA and NNLLA functions derived from the
two- and three-loop NMHV ratio functions we are able to find all contributions
to the MRK limit at any loop order, up to NNLLA in the imaginary part
and N${}^3$LLA in the real part. These results will serve as important input
for the calculation of the ratio function at higher loops.

With access to an NMHV amplitude at this loop order, we are uniquely
positioned to investigate multi-particle factorization behavior at
three loops. In constructing the multi-particle factorization function
we find remarkable simplicity.  We conjecture that our results
should be straightforwardly generalizable beyond six points.

In investigating multi-particle factorization, we found remarkable
relations in the coproduct entries of the ratio function, relations
that go beyond those predicted by Caron-Huot and
He~\cite{CaronHuot2011kk,SimonSongPrivate}. While we do not yet
understand the source of these relations, if they continue to higher
loops they might serve as useful constraints on further
bootstraps.  Similarly, the simplicity of the function $U$ along
certain kinematic lines (and in particular the status of $\Delta U$ as
a palindrome) suggest deeper properties.

By plotting $V$ and $\tilde{V}$ on a variety of lines and planes, we
have observed how its quantitative behavior changes with loop
order. While the overall behavior is not nearly as consistent between
loop orders as it was for the remainder function in ref.~\cite{Dixon2013eka},
we do find that the ratio between three loops and
two loops at least stays in a confined range over much of the space,
being particularly tightly constrained for $\tilde{V}$. Time will tell
if this behavior becomes more regular at higher loop orders.

In general, the success of the hexagon function program for the
four-loop remainder function~\cite{Dixon2014voa} indicates that the
same program should be viable for the four-loop NMHV ratio
function. Deriving the NMHV ratio function at four loops would allow
us to confirm the trends observed at three loops, with an eye towards
understanding their origins.

More generally, we have conjectured that the relative constant ratios
of successive loop orders for the remainder function and the ratio function
(in suitable regimes) are a byproduct of the convergence of perturbation
theory in the planar ${\cal N}=4$ theory.
This possibility, discussed in ref.~\cite{Dixon2014voa},
could be investigated in more detail using BSV's
approach to the OPE. Since the quantities they calculate are fully
non-perturbative, it may be possible to look at their behavior at
higher orders and thereby gain an understanding of why quantities like
the remainder function and $\tilde{V}$ have such clean inter-loop
ratios even at comparatively low loop order. Such an understanding
could lead to a merging of the two approaches, with the goal of
understanding amplitudes in planar $\mathcal{N}=4$ super Yang-Mills
for any value of the coupling and any kinematics.


\vskip0.5cm
\noindent {\large\bf Acknowledgments}
\vskip0.3cm

We are grateful to Benjamin Basso, Amit Sever and Pedro Vieira
for very helpful discussions, in particular for checking our
results for the $T^2$ terms in the OPE against theirs.  We thank
Benjamin Basso for suggesting how the NMHV and MHV impact factors
might be related.   We also thank Nima Arkani-Hamed, Zvi Bern,
John Joseph Carrasco, James Drummond, Claude Duhr,
Georgios Papathanasiou and Jaroslav Trnka for illuminating conversations.
We thank the Simons Center for Geometry and Physics for its warm
hospitality when this project was initiated.
This research was supported by the US Department of Energy under
contract DE--AC02--76SF00515.

\vfill\eject


\appendix

\section{Coproduct elements of $U$ and $\tilde{V}$}
\label{Ucoprod}

Because of the coproduct relations for $U$ and $\tilde{V}$,
and their (anti)symmetry under $u \lr w$, only four independent 
$\{n-1,1\}$ coproduct elements need to be specified in each case.
We take these four components to be $u$, $v$, $y_u$ and $y_v$.
(In the case of the even function $U$, we should also specify
the constant of integration by giving the value of the function at a
particular point, say $(u,v,w)=(1,1,1)$, which we do elsewhere
in this article.)

\subsection{$U$}
\label{Ucoprodsubappendix}

For the function $U$, the other $\{n-1,1\}$ coproduct elements
are given in terms of $U^u$, $U^v$, $U^{y_u}$ and $U^{y_v}$ as follows:
\bea
U^w(u,v,w) &=& U^u(w,v,u) \,, \label{U_w}\\
U^{1-u}(u,v,w) &=&\null - U^u(u,v,w) - U^v(u,v,w) \,, \label{U_omu}\\
U^{1-v}(u,v,w) &=& 0 \,,  \label{U_omv}\\
U^{1-w}(u,v,w) &=& U^{1-u}(w,v,u) \,, \label{U_omw}\\
U^{y_w}(u,v,w) &=& U^{y_u}(u,v,w) \,.
\label{U_yw}
\eea
In the rest of this subsection, we give the four independent
coproduct elements for $U$ through three loops.

The one-loop independent coproduct elements are trivial,
given \eqn{U1} for $U^{(1)}$:
\bea
~[U^{(1)}]^{u} &=& -\frac{1}{2} \ln(uw/v) \,, \label{U1_u}\\
~[U^{(1)}]^{v} &=& \frac{1}{2} \ln(uw/v) \,, \label{U1_v}\\
~[U^{(1)}]^{y_u} &=& 0 \,, \label{U1_yu}\\
~[U^{(1)}]^{y_v} &=& 0 \,. \label{U1_yv}
\eea

The two-loop independent coproduct elements can be computed
from \eqn{U2} for $U^{(2)}$, but we list them here for convenience:
\bea
~[U^{(2)}]^{u} &=& \frac{1}{8} \, \Bigl[
 - 2 \, H_3^u + 4 \, H_{2,1}^u + 2 \, H_{2,1}^v 
                 - 2 \, H_3^w - 4 \, H_{2,1}^w
 + ( 3 \, \ln u + \ln(v/w) ) \, H_2^u
\nonumber\\&&\hskip0.5cm\null
 + \ln(uv/w) \, H_2^v
 - ( 3 \, \ln(u/v) + \ln w ) \, H_2^w
 - \ln^2 v \, \ln u 
\nonumber\\&&\hskip0.5cm\null
 - \ln^2 w \, ( 3 \, \ln u - \ln v )
 + 3 \, \ln u \, \ln v \, \ln w
 + 2 \, \zeta_2 \, ( \ln u - 5 \, \ln(v/w) ) \Bigr] \,, \label{U2_u}\\
~[U^{(2)}]^{v} &=& \frac{1}{4} \, \Bigl[ H_3^u + H_{2,1}^u + H_3^w + H_{2,1}^w
 - \ln(v/w) \, \Bigl( H_2^u + \frac{1}{2} \, \ln^2 u \Bigr)
\nonumber\\&&\hskip0.5cm\null
 - \ln(v/u) \, \Bigl( H_2^w + \frac{1}{2} \, \ln^2 w \Bigr)
 - 4 \, \zeta_2 \, \ln(uw/v) \Bigr] \,, \label{U2_v}\\
~[U^{(2)}]^{y_u} &=& \frac{1}{8} \, \tilde\Phi_6(u,v,w) \,, \label{U2_yu}\\
~[U^{(2)}]^{y_v} &=& 0 \,. \label{U2_yv}
\eea

The independent parity-even $\{5,1\}$ coproduct elements of $U^{(3)}$ are
\bea
&&[U^{(3)}]^u = \frac{1}{32} \biggl\{
  - M_1(w,u,v) + M_1(u,w,v)
  - \frac{128}{3} \, ( \Qep(v,w,u) - \Qep(v,u,w) )
\nonumber\\&&\null
  - \ln(u/w) \, ( \Omegauvw + \Omegavwu )
  - ( 3 \, \ln u - 4 \, \ln v + 5 \, \ln w ) \, \Omegawuv
\nonumber\\&&\null
 + 24 \, H_5^u - 4 \, H_{4,1}^u + 10 \, H_{3,2}^u + 96 \, H_{3,1,1}^u
 + 22 \, H_{2,2,1}^u
 - 72 \, H_{2,1,1,1}^u - 2 \, H_2^u \, ( 3 \, H_3^u + 5 \, H_{2,1}^u )
\nonumber\\&&\null
 - \frac{3}{2} \, \ln u \, \Bigl( 24 \, H_4^u - 20 \, H_{3,1}^u
                                + 28 \, H_{2,1,1}^u - (H_2^u)^2 \Bigr)
 + 4 \, \ln^2 u \, ( 4 \, H_3^u - 3 \, H_{2,1}^u ) - 2 \, \ln^3 u \, H_2^u
\nonumber\\&&\null
 - 96 \, H_{3,1,1}^v - 32 \, H_{2,2,1}^v - 16 \, H_{2,1,1,1}^v
 + 16 \, H_2^v \, H_{2,1}^v
 - 4 \, \ln v \, \Bigl( 4 \, H_{3,1}^v + 2 \, H_{2,1,1}^v - (H_2^v)^2 \Bigr)
\nonumber\\&&\null
 + \frac{2}{3} \, \ln^3 v \, H_2^v
 + 24 \, H_5^w - 28 \, H_{4,1}^w - 10 \, H_{3,2}^w + 240 \, H_{3,1,1}^w
 + 74 \, H_{2,2,1}^w + 8 \, H_{2,1,1,1}^w
\nonumber\\&&\null
 + 2 \, H_2^w \, ( 3 \, H_3^w - 19 \, H_{2,1}^w )
 - \frac{1}{2} \, \ln w \, \Bigl( 24 \, H_4^w - 68 \, H_{3,1}^w
                       - 20 \, H_{2,1,1}^w + 19 \, (H_2^w)^2 \Bigr)
 + 4 \, \ln^2 w \, H_{2,1}^w
\nonumber\\&&\null
 + \frac{2}{3} \, \ln^3 w \, H_2^w
 - 4 \, ( H_2^u - H_2^w + 2 \, ( \ln^2 u + \ln^2 w ) ) \, H_{2,1}^v
 - \frac{1}{2} \ln(u/w) \Bigl( 4 \, H_4^v + 40 \, H_{3,1}^v + 4 \, H_{2,1,1}^v
\nonumber\\&&\null
 - 11 \, (H_2^v)^2 \Bigr)
 - \frac{1}{2} \, ( \ln^3 u - \ln^3 w ) \, H_2^v
 + 2 \, \ln v \, \biggl( 6 \, ( H_4^u - H_4^w )
    - 18 \, H_{3,1}^u - 14 \, H_{3,1}^w - 2 \, H_{2,1,1}^u
\nonumber\\&&\null
    - 14 \, H_{2,1,1}^w + 4 \, \Bigl( (H_2^u)^2 + (H_2^w)^2 \Bigr)
     - ( H_2^u - H_2^w ) \, H_2^v
     - \ln u \, ( H_{2,1}^u - H_3^v + 3 \, H_{2,1}^v )
\nonumber\\&&\null
     - \ln^2 u \, ( H_2^u + 3 \, H_2^v )
     + \ln w \, ( 8 \, H_3^w - 3 \, H_{2,1}^w - H_3^v - H_{2,1}^v )
     - \ln^2 w \, ( H_2^w + H_2^v ) \biggr)
\nonumber\\&&\null
 - \frac{1}{4} \, \ln^2 v \, \Bigl(
    24 \, H_3^u + 2 \, \ln u \, ( 3 \, H_2^u + 4 \, H_2^v ) - \ln^3 u
   + 8 \, H_3^w - 64 \, H_{2,1}^w - 2 \, \ln w \, ( 3 \, H_2^w - 4 \, H_2^v )
\nonumber\\&&\null
   + \frac{19}{3} \ln^3 w \Bigr)
 - \frac{2}{3} \ln^3 v ( H_2^u - H_2^w - 2 \, \ln^2 u )
 - \frac{10}{3} ( H_2^w \, H_3^u - H_2^u \, H_3^w )
 + \frac{14}{3} ( H_2^w \, H_{2,1}^u - H_2^u \, H_{2,1}^w )
\nonumber\\&&\null
 + \frac{1}{6} \, \ln u \, \Bigl( 72 \, H_4^w + 156 \, H_{3,1}^w
     + 168 \, H_{2,1,1}^w - 45 \, (H_2^w)^2 + 20 \, H_2^u \, H_2^w
     - 12 \, \ln w \, ( 8 \, H_3^w - 3 \, H_{2,1}^w
\nonumber\\&&\null
 - H_{2,1}^u )
     + \ln^2 w \, ( 12 \, H_2^w + 7 \, H_2^u ) \Bigr)
 - \frac{1}{6} \, \ln w \, \Bigl( 72 \, H_4^u - 228 \, H_{3,1}^u
     - 24 \, H_{2,1,1}^u + 51 \, (H_2^u)^2
\nonumber\\&&\null
     + 20 \, H_2^u \, H_2^w + \ln^2 u \, ( 7 \, H_2^w - 12 \, H_2^u ) \Bigr)
 - \frac{1}{4} \, \ln^3 u \, ( 2 \, H_2^w - \ln^2 w )
 + \frac{1}{12} \, \ln^3 w \, ( 6 \, H_2^u - 67 \, \ln^2 u )
\nonumber\\&&\null
 - \frac{1}{3} \ln^2 w \, ( 23 H_3^u - 13 H_{2,1}^u )
 - \frac{1}{3} \ln^2 u \, ( H_3^w - 35 H_{2,1}^w )
 + \frac{1}{6} \ln u \, \ln w \, \Bigl( 144 H_{2,1}^v
    + 33 \ln(u/w) \, H_2^v \Bigr)
\nonumber\\&&\null
 + \ln v \, \Bigl( 2 \, \ln u \, ( 2 \, H_3^w - 15 \, H_{2,1}^w )
    + 2 \, \ln w \, ( 6 \, H_3^u - H_{2,1}^u )
    + 2 \, \ln u \, \ln w \, ( H_2^u + 6 \, H_2^v - H_2^w )
\nonumber\\&&\null
    + 2 \, ( \ln^2 u \, H_2^w - \ln^2 w \, H_2^u )
    - \ln^3 u \, \ln w + 10 \, \ln^2 u \, \ln^2 w 
    + \frac{11}{3} \, \ln u \, \ln^3 w \Bigr)
\nonumber\\&&\null
 - \frac{1}{4} \, \ln^2 v \, \Bigl( 6 \, ( \ln u \, H_2^w - \ln w \, H_2^u )
                + 25 \, \ln^2 u \, \ln w + 7 \, \ln u \, \ln^2 w \Bigr)
 + \frac{2}{3} \, \ln^3 v \, \ln u \, \ln w
\nonumber
\eea
\bea
&&\null
 - \zeta_2 \, \biggl[ H_3^u - 12 \, H_{2,1}^u - 3 \, \ln u \, H_2^u - 3 \, \ln^3 u
    + 32 \, H_{2,1}^v + 16 \, \ln v \, H_2^v + \frac{4}{3} \, \ln^3 v
    - H_3^w - 84 \, H_{2,1}^w
\nonumber\\&&\hskip1.0cm\null
    - 29 \, \ln w \, H_2^w + \frac{1}{3} \, \ln^3 w
    + 2 \, ( 14 \, \ln v - 15 \, \ln w ) \, H_2^u
    - 4 \, \ln^2 u \, ( \ln v - \ln w ) 
\nonumber\\&&\hskip1.0cm\null
    - \ln u \, ( 34 \, H_2^w + 44 \, \ln^2 w - 18 \, H_2^v + 20 \, \ln^2 v
              - 56 \, \ln v \, \ln w )
    + 12 \, \ln v \, ( 3 \, H_2^w + \ln^2 w )
\nonumber\\&&\hskip1.0cm\null
    - 2 \, \ln w \, ( 9 \, H_2^v + 2 \, \ln^2 v ) \biggr]
 - 2 \, \zeta_3 \, \Bigl[ 4 \, ( H_2^u - H_2^w ) + 3 \, ( \ln^2 u - \ln^2 w )
                   \Bigr]
\nonumber\\&&\null
 - 2 \, \zeta_4 \, \Bigl[ 35 \, \ln u - 160 \, \ln v + 157 \, \ln w \Bigr]
\biggr\} \,,
\label{U3_u}
\eea
and
\be
[U^{(3)}]^v = A^v(u,v,w) + A^v(w,v,u),
\label{U3_v}
\ee
where
\bea
A^v(u,v,w) &=& - \frac{1}{8} \biggl\{
   \Bigl( \frac{1}{2} \, \ln v - \ln u \Bigr) \Omega^{(2)}(w,u,v)
  + 6 \, H_5^u - 4 \, H_{4,1}^u + 66 \, H_{3,1,1}^u + 20 \, H_{2,2,1}^u
\nonumber\\ &&\hskip0.7cm\null
  - 4 \, H_{2,1,1,1}^u - 10 \, H_2^u \, H_{2,1}^u
  - \ln u \, \Bigl( 6 \, H_4^u - 12 \, H_{3,1}^u + 2 \, H_{2,1,1}^u
                   + 2 \, (H_2^u)^2 \Bigr)
\nonumber\\ &&\hskip0.7cm\null
  + \ln^2 u \, ( 2 \, H_3^u - H_{2,1}^u ) - \frac{1}{3} \, \ln^3 u \, H_2^u
\nonumber\\ &&\hskip0.7cm\null
  - \ln(v/w) \, \Bigl( 8 \, H_{3,1}^u + 4 \, H_{2,1,1}^u - 2 \, \ln u \, H_3^u
                     - 2 \, (H_2^u)^2 \Bigr)
\nonumber\\ &&\hskip0.7cm\null
  + \ln^2(v/w) \, \Bigl( - H_3^u + 4 \, H_{2,1}^u + \ln u \, H_2^u
                         - \frac{1}{3} \, \ln^3 u \Bigr)
\nonumber\\ &&\hskip0.7cm\null
  + \zeta_2 \, \biggl[ 20 \, H_{2,1}^u + 8 \, \ln u \, H_2^u
                     + \frac{2}{3} \, \ln^3 u
               - 8 \, \ln(v/w) \, \Bigl( H_2^u + \frac{1}{2} \, \ln^2 u \Bigr) 
               \biggr]
\nonumber\\ &&\hskip0.7cm\null
  - 32 \, \zeta_4 \, ( 2 \, \ln u - \ln v ) \biggr\} \,.
\label{P_uvw}
\eea

The parity-odd coproducts of $U^{(3)}$ are given by,
\bea
[U^{(3)}]^{y_u} &=&
\frac{1}{32} \biggl\{ 3 \, H_1(u,v,w) + H_1(v,w,u) + H_1(w,u,v)
- \frac{11}{4} \, J_1(u,v,w)
\nonumber\\&&\hskip0.7cm\null
- \frac{1}{4} \bigl( J_1(v,w,u) + J_1(w,u,v) \bigr)
+ \tilde\Phi_6(u,v,w) \Bigl[ \ln^2 u + \ln^2 w + \ln^2 v
\nonumber\\&&\hskip1.5cm\null
   + 2 \Bigl( \ln u \, \ln w - \ln(uw) \, \ln v \Bigr)
   - 22 \zeta_2 \Bigr] \biggr\} \,,
\label{U3yuAppendix}\\
~[U^{(3)}]^{y_v} &=& \frac{1}{8} \, H_1(u,v,w) \,.
\label{U3yvAppendix}
\eea
%

\subsection{$\tilde{V}$}

For the function $\tilde{V}$, the other $\{n-1,1\}$ coproduct elements
are given in terms of $\tilde{V}^u$, $\tilde{V}^v$, $\tilde{V}^{y_u}$
and $\tilde{V}^{y_v}$ as follows:
\bea
\tilde{V}^w(u,v,w) &=& - \tilde{V}^u(w,v,u) \,, \label{Vt_w}\\
\tilde{V}^{1-u}(u,v,w) &=&\null - \tilde{V}^u(u,v,w) \,, \label{Vt_omu}\\
\tilde{V}^{1-v}(u,v,w) &=& - \tilde{V}^v(u,v,w) \,,  \label{Vt_omv}\\
\tilde{V}^{1-w}(u,v,w) &=& - \tilde{V}^{1-u}(w,v,u) \,, \label{Vt_omw}\\
\tilde{V}^{y_w}(u,v,w) &=& - \tilde{V}^{y_u}(w,v,u) \,.
\label{Vt_yw}
\eea
In the rest of this subsection, we give the four independent
coproduct elements for $\tilde{V}$ through three loops.

The one-loop function $\tilde{V}^{(1)}$ vanishes.
The two-loop function $\tilde{V}^{(2)}$ is given in \eqn{Vt2}.
Its $\{3,1\}$ coproduct elements are,
\bea
~[\tilde{V}^{(2)}]^u &=& \frac{1}{8} \, \tilde\Phi_6(u,v,w) \,, \label{Vt2_u}\\
~[\tilde{V}^{(2)}]^v &=& 0 \,, \label{Vt2_v}\\
~[\tilde{V}^{(2)}]^{y_u} &=& \frac{1}{4} \, \biggl[
    H_3^u - H_{2,1}^v - H_3^w
   - \frac{1}{2} \, \ln(u/w) 
           \, \Bigl( H_2^u + H_2^v + H_2^w + \ln v \, \ln w \Bigr)
\nonumber\\&&\hskip0.5cm\null
   - \frac{1}{2} \, \ln v \, ( H_2^u + H_2^v - H_2^w )
   + \zeta_2 \, \ln(uv/w) \biggr] \,, \label{Vt2_yu}\\
~[\tilde{V}^{(2)}]^{y_v} &=& \frac{1}{4} \, \Bigl[
   H_3^u - H_{2,1}^u - \ln u \, H_2^u
 - H_3^w + H_{2,1}^w + \ln w \, H_2^w
\nonumber\\&&\hskip0.5cm\null
  - \ln(u/w) \, \Bigl( H_2^v + \frac{1}{2} \, \ln u \, \ln w
                      - 2 \, \zeta_2 \Bigr) \Bigr] \,. \label{Vt2_yv}
\eea

The independent parity-odd $\{5,1\}$ coproduct
elements of $\tilde{V}^{(3)}$ are given by,
\bea
[\tilde{V}^{(3)}]^u &=& \frac{1}{96} \biggl\{
- H_1(u,v,w) + H_1(v,w,u) + 3 \, H_1(w,u,v)
- \frac{23}{4} \, J_1(u,v,w)
\nonumber\\&&\hskip0.7cm\null
- \frac{13}{4} \, J_1(v,w,u) - \frac{3}{4} \, J_1(w,u,v)
- 6 \, \ln u \, \Bigl( F_1(u,v,w) - F_1(w,u,v) \Bigr)
\nonumber\\&&\hskip0.7cm\null
+ 3 \, \tilde\Phi_6(u,v,w) \, \Bigl[ 3 \, \ln^2 u + \ln^2 v + \ln^2 w
\nonumber\\&&\hskip01.5cm\null
              + 2 \, ( H_2^u + H_2^v + H_2^w - \ln u \, \ln w )
              - 26 \, \zeta_2 \Bigr]
 \biggr\} \,,
\label{Vt3_u}\\
~[\tilde{V}^{(3)}]^v &=& \frac{1}{96} \biggl\{
 2 \, \Bigl( H_1(v,w,u) - H_1(w,u,v) \Bigr)
+ \frac{5}{2} \, \Bigl( J_1(v,w,u) - J_1(w,u,v) \Bigr)
\nonumber\\&&\hskip0.7cm\null
- 6 \ln v \Bigl( F_1(u,v,w) - F_1(w,u,v)
               - \ln(u/w) \, \tilde\Phi_6(u,v,w) \Bigr) \biggr\} \,.
\label{Vt3_v}
\eea

\vfill\eject

The independent parity-even coproducts of $\tilde{V}^{(3)}$ are given by,
\bea
&&[\tilde{V}^{(3)}]^{y_u} = \frac{1}{96} \biggl\{
 M_1(w,v,u) - M_1(v,w,u) + 3 \, ( M_1(w,u,v) - M_1(u,w,v) )
\nonumber\\&&\null
 - \frac{64}{3} \, \Bigl( 2 \, ( \Qep(w,v,u) - \Qep(w,u,v) ) 
                + 7 \, ( \Qep(v,w,u) - \Qep(v,u,w) ) \Bigr)
\nonumber\\&&\null
 + ( 3 \ln u + \ln v - 4 \ln w ) \, \Omegauvw
 - ( 3 \ln u - \ln v - 2 \ln w ) \, ( \Omegawuv - \Omegavwu )
\nonumber\\&&\null
 - 72 \, H_5^u + 72 \, H_{4,1}^u + 15 \, H_{3,2}^u + 36 \, H_{3,1,1}^u
 + 27 \, H_{2,2,1}^u - 9 \, H_2^u \, H_{2,1}^u
 + 3 \, \ln u \, \Bigl( 12 \, H_4^u - 4 \, H_{3,1}^u
\nonumber\\&&\null
                      + 3 \, H_{2,1,1}^u - (H_2^u)^2 \Bigr)
 - 3 \, \ln^2 u \, H_3^u - \frac{3}{2} \, \ln^3 u \, H_2^u
 - 12 \, H_{4,1}^v - 10 \, H_{3,2}^v + 168 \, H_{3,1,1}^v + 58 \, H_{2,2,1}^v
\nonumber\\&&\null
 + 56 \, H_{2,1,1,1}^v + 6 \, H_2^v \, ( H_3^v - 7 \, H_{2,1}^v )
 + \frac{1}{2} \, \ln v \, \Bigl( 24 \, H_4^v + 36 \, H_{3,1}^v + 68 \, H_{2,1,1}^v
                 - 31 \, (H_2^v)^2 \Bigr)
\nonumber\\&&\null
 - 8 \, \ln^2 v \, ( H_3^v - H_{2,1}^v ) + \frac{2}{3} \, \ln^3 v \, H_2^v
 + 72 \, H_5^w - 60 \, H_{4,1}^w - 5 \, H_{3,2}^w + 84 \, H_{3,1,1}^w
 + 11 \, H_{2,2,1}^w 
\nonumber\\&&\null
 - 8 \, H_{2,1,1,1}^w - 3 \, H_2^w \, ( 2 \, H_3^w + 3 \, H_{2,1}^w )
 - \frac{1}{2} \, \ln w \, \Bigl( 96 \, H_4^w - 84 \, H_{3,1}^w
                 + 38 \, H_{2,1,1}^w  - (H_2^w)^2 \Bigr)
\nonumber\\&&\null
 + \ln^2 w \, ( 11 \, H_3^w - 8 \, H_{2,1}^w ) - \frac{7}{6} \, \ln^3 w \, H_2^w
 - \frac{1}{4} \, \ln^3 u \, ( 5 \, \ln^2 v - 2 \, H_2^v )
\nonumber\\&&\null
 + \frac{1}{12} \, \ln^2 u \, \Bigl( 3 \, \ln^3 v
        + 2 \, \ln v \, ( 35 \, H_2^v + 24 \, H_2^u )
                     + 44 \, H_3^v + 164 \, H_{2,1}^v \Bigr)
\nonumber\\&&\null
 + \ln u \, \biggl( \ln^2 v \, \Bigl( 4 \, H_2^v + \frac{1}{6} \, H_2^u \Bigr)
       + \ln v \, \Bigl( - 18 \, (H_3^u + H_3^v) + 16 \, H_{2,1}^u
                         + 12 \, H_{2,1}^v \Bigr)
       + 18 \, H_4^v 
\nonumber\\&&\null
       + 32 \, H_{3,1}^v + 10 \, H_{2,1,1}^v
       - \frac{37}{2} \, (H_2^v)^2 - \frac{20}{3} \, H_2^v \, H_2^u \biggr)
 + \frac{1}{6} \, \ln^3 v \, H_2^u 
 + \frac{1}{3} \, \ln^2 v \, ( 13 \, H_3^u + 7 \, H_{2,1}^u )
\nonumber\\&&\null
 + \ln v \, \Bigl( 14 \, H_4^u + 16 \, H_{3,1}^u + 22 \, H_{2,1,1}^u
           - \frac{31}{2} \, (H_2^u)^2 - \frac{28}{3} \, H_2^v \, H_2^u \Bigr)
 + \frac{2}{3} \, H_2^u \, ( 5 \, H_3^v - 7 \, H_{2,1}^v )
\nonumber\\&&\null
 + \frac{26}{3} \, H_2^v \, ( H_3^u + H_{2,1}^u )
 - \ln^3 u \, ( \ln^2 w - H_2^w )
 + \frac{1}{12} \, \ln^2 u \, \Bigl( 15 \, \ln^3 w
                     + 2 \, \ln w \, ( 19 \, H_2^w - 24 \, H_2^u )
\nonumber\\&&\null
                     - 26 H_3^w + 70 H_{2,1}^w \Bigr)
 + \frac{1}{3} \ln u \, \biggl( 4 \ln^2 w \, ( H_2^u + 6 \, H_2^w )
     + 6 \ln w \, \Bigl( 9 H_3^u - 8 H_{2,1}^u
                          - 12 \, ( H_3^w- H_{2,1}^w ) \Bigr)
\nonumber\\&&\null
     + 72 H_4^w - 6 H_{3,1}^w + 96 H_{2,1,1}^w
     - \frac{69}{2} (H_2^w)^2 - 7 \, H_2^u \, H_2^w \biggr)
 - \frac{13}{6} \ln^3 w \, H_2^u
 + \frac{1}{6} \ln^2 w \, ( H_3^u + 13 \, H_{2,1}^u )
\nonumber\\&&\null
 - \frac{1}{3} \, \ln w \, \Bigl( 42 \, H_4^u + 66 \, ( H_{3,1}^u + H_{2,1,1}^u )
           - 51 \, (H_2^u)^2 - H_2^u \, H_2^w \Bigr)
 - \frac{1}{3} \, H_2^u \, ( H_3^w + 49 \, H_{2,1}^w )
\nonumber\\&&\null
 + \frac{1}{3} H_2^w \, ( H_3^u + 37 H_{2,1}^u )
 + \frac{1}{12} \ln^3 v \, ( 7 \ln^2 w + 10 H_2^w )
 + \frac{1}{6} \ln^2 v \, \Bigl( 13 \ln^3 w
              + 6 \ln w \, ( 7 \, H_2^w - 4 \, H_2^v )
\nonumber\\&&\null
              - 48 H_3^w + 60 H_{2,1}^w \Bigr)
 - \frac{1}{2} \ln v \, \biggl( \ln^2 w \, ( 4 H_2^w - 25 H_2^v )
       + 12 \ln w \, \Bigl( 2 H_{2,1}^v - 3 H_3^v
                             - 4 \, ( H_3^w - H_{2,1}^w ) \Bigr)
\nonumber\\&&\null
       + 48 \, H_4^w - 20 \, H_{3,1}^w + 64 \, H_{2,1,1}^w - 19 \, (H_2^w)^2
       + 12 \, H_2^v \, H_2^w \biggr)
 - \frac{1}{3} \ln^3 w \, H_2^v + 2 \, \ln^2 w \, ( H_3^v + 7 \, H_{2,1}^v )
\nonumber
\eea
%
%
\bea
&&\null
 + \ln w \, \Bigl( - 18 \, H_4^v - 34 \, H_{3,1}^v - 10 \, H_{2,1,1}^v
             + 19 \, (H_2^v)^2 + 12 \, H_2^v \, H_2^w \Bigr)
 - 24 \, H_2^w \, H_{2,1}^v - 12 \, H_2^v \, H_3^w
\nonumber\\&&\null
 + 3 \, \ln^3 u \, \ln v \, \ln w
 - \frac{1}{4} \, \ln^2 u \, \Bigl( 4 \, \ln^2 v \, \ln w
             + \ln v \, ( 17 \, \ln^2 w + 18 \, H_2^w )
             + 24 \, \ln w \, H_2^v \Bigr)
\nonumber\\&&\null
 + \ln u \, \biggl( - \frac{1}{3} \ln^3 v \, \ln w
             - \frac{1}{4} \ln^2 v \, ( 3 \, \ln^2 w - 10 \, H_2^w )
     + \ln v \, \Bigl( \frac{4}{3} \ln^3 w
               - 2 \, \ln w \, ( 8 \, H_2^w + 3 \, H_2^u
\nonumber\\&&\null
 + 13 \, H_2^v )
               + 4 \, H_3^w - 24 \, H_{2,1}^w \Bigr)
     + \frac{13}{2} \, \ln^2 w \, H_2^v
     - 2 \, \ln w \, ( 2 \, H_3^v + 21 \, H_{2,1}^v )
     - 12 \, H_2^v \, H_2^w \biggr)
\nonumber\\&&\null
 - \frac{3}{2} \, \ln w \, \Bigl( 4 \, \ln^2 v \, H_2^u
     - \ln v \, ( 9 \, H_2^u \, \ln w + 4 \, H_{2,1}^u )
     - 8 \, H_2^u \, H_2^v \Bigr)
\nonumber\\&&\null
 + \zeta_2 \, \biggl[
   - \frac{57}{2} \, H_3^u + H_3^v + \frac{55}{2} \, H_3^w
   - 18 \, H_{2,1}^u + 92 \, H_{2,1}^v + 46 \, H_{2,1}^w
\nonumber\\&&\hskip0.9cm\null
   + \frac{1}{2} \, H_2^u \, ( 45 \, \ln u + 92 \, \ln v - 104 \, \ln w )
   + H_2^v \, ( 58 \, \ln u + 53 \, \ln v - 60 \, \ln w )
\nonumber\\&&\hskip0.9cm\null
   + \frac{1}{2} \, H_2^w \, ( 76 \, \ln u - 12 \, \ln v - 7 \, \ln w )
   - \frac{3}{2} \, \ln^3 u - \frac{1}{3} \, \ln^3 v + \frac{35}{6} \, \ln^3 w
\nonumber\\&&\hskip0.9cm\null
   - 8 \, \Bigl( \ln^2 u \, \ln(v/w) + \ln^2 v \, \ln(u/w) \Bigr)
   - 4 \, \ln^2 w \, ( 4 \, \ln u + 11 \, \ln v )
   + 72 \, \ln u \, \ln v \, \ln w \biggr]
\nonumber\\&&\null
 - \zeta_3 \, \Bigl[ 66 \, H_2^u - 8 \, H_2^v - 58 \, H_2^w
   + 27 \, \ln^2 u - 6 \, \ln^2 v - 21 \, \ln^2 w \Bigr]
\nonumber\\&&\null
 - 2 \, \zeta_4 \, \Bigl[ 117 \, \ln u + 123 \, \ln v - 114 \, \ln w
  \Bigr] \biggr\} \,,
\label{Vt3_yu}
\eea
and
\be
[\tilde{V}^{(3)}]^{y_v} = B^{y_v}(u,v,w) - B^{y_v}(w,v,u),
\label{Vt3_yv}
\ee
where
\bea
&&B^{y_v}(u,v,w) \nonumber\\&&\null
= \frac{1}{96} \biggl\{
 - 3 \, \bigl( M_1(u,v,w) + M_1(v,w,u) \bigr) + 2 \, M_1(w,u,v)
 + \frac{320}{3} \, \bigl( Q_{\rm ep}(u,v,w) + Q_{\rm ep}(w,u,v) \bigr)
\nonumber\\&&\hskip0.9cm\null
 - \bigl( \ln u - 6 \, \ln v + 5 \, \ln w \bigr) \, \Omega^{(2)}(u,v,w)
 + \ln u \, \Omega^{(2)}(w,u,v)
 - 72 \, H_5^u + 48 \, H_{4,1}^u - 5 \, H_{3,2}^u
\nonumber\\&&\hskip0.9cm\null
 + 84 \, H_{3,1,1}^u + 47 \, H_{2,2,1}^u + 64 \, H_{2,1,1,1}^u
 + 3 \, H_2^u \, \bigl( 4 \, H_3^u - 11 \, H_{2,1}^u \bigr)
 + \ln u \, \Bigl( 60 \, H_4^u - 24 \, H_{3,1}^u
\nonumber\\&&\hskip0.9cm\null
 + 53 \, H_{2,1,1}^u - 16 \, (H_2^u)^2 \Bigr)
 - \ln^2 u \, ( 19 \, H_3^u - 16 \, H_{2,1}^u )
 + \frac{11}{6} \, \ln^3 u \, H_2^u
 - \ln^3 u \, \ln^2 v
\nonumber\\&&\hskip0.9cm\null
 + \frac{1}{6} \, \ln^2 u \, ( 25 \, H_3^v + H_{2,1}^v - 7 \, \ln v \, H_2^v )
 + \frac{1}{3} \, H_2^u \, ( 25 \, H_3^v - 11 \, H_{2,1}^v ) 
\nonumber\\&&\hskip0.9cm\null
 + 2 \, \ln u \, \Bigl( 14 \, H_4^v + 19 \, H_{3,1}^v + 22 \, H_{2,1,1}^v
               - \frac{65}{4} \, (H_2^v)^2
               - \ln v \, ( 18 \, H_3^v - 16 \, H_{2,1}^v )
               + 4 \, \ln^2 v \, H_2^v \Bigr)
\nonumber\\&&\hskip0.9cm\null
 - \frac{1}{4} \, \ln^3 v \, ( \ln^2 u + 2 \, H_2^u )
 + \frac{1}{6} \, \ln^2 v \, ( 35 \, H_3^u + 47 \, H_{2,1}^u
                             + 16 \, \ln u \, H_2^u )
\nonumber
\eea
\bea 
&&\hskip0.9cm\null
 + 2 \, \ln v \, \Bigl( - 3 \, H_4^u + 17 \, H_{3,1}^u - 11 \, H_{2,1,1}^u
          - \frac{7}{2} \, (H_2^u)^2 + 3 \, \ln u \, ( H_3^u - 2 \, H_{2,1}^u )
          - 2 \, \ln^2 u \, H_2^u \Bigr)
\nonumber\\&&\hskip0.9cm\null
 + \frac{1}{3} \, H_2^v \, \Bigl( 7 \, \ln^3 u + 11 \, H_3^u + 35 \, H_{2,1}^u
                 - ( 29 \, \ln u + 13 \, \ln v ) \, H_2^u \Bigr)
\nonumber\\&&\hskip0.9cm\null
 - \frac{1}{12} \, \ln^3 u \, ( 19 \, \ln^2 w - 14 \, H_2^w )
 - \frac{11}{2} \, \ln^2 u \, \ln w \, H_2^w
\nonumber\\&&\hskip0.9cm\null
 + 2 \, \ln u \, \Bigl( - \frac{19}{4} \, (H_2^w)^2
       + 3 \, \ln w \, ( H_3^w - 2 \, H_{2,1}^w ) + \ln^2 w \, H_2^w \Bigr)
\nonumber\\&&\hskip0.9cm\null
 + 2 \, \ln w \, ( 3 \, H_4^u - 22 \, H_{3,1}^u + 11 \, H_{2,1,1}^u )
 + 2 \, \ln^2 w \, ( 5 \, H_3^u + 2 \, H_{2,1}^u )
\nonumber\\&&\hskip0.9cm\null
 + 6 \, H_2^w \, ( 2 \, H_3^u - 4 \, H_{2,1}^u - 3 \, \ln u \, H_2^u  )
 + \frac{1}{4} \, \ln^2 v \, \ln u \, ( 6 \, H_2^w - 13 \, \ln^2 w )
\nonumber\\&&\hskip0.9cm\null
 + \frac{3}{2} \, H_2^v \, \ln w \, ( 8 \, H_2^u - 13 \, \ln^2 u )
\nonumber\\&&\hskip0.9cm\null
 + \frac{1}{3} \, \ln v \, \Bigl( - 5 \, \ln^3 u \, \ln w
         - 12 \, \ln^2 u \, H_2^w
         + 6 \, \ln u \, ( 4 \, H_3^w + 9 \, H_{2,1}^w + 5 \, \ln w \, H_2^w )
         \Bigr)
\nonumber\\&&\hskip0.9cm\null
 + \zeta_2 \, \biggl[ - \frac{53}{2} \, H_3^u + 46 \, H_{2,1}^u
    + H_2^u \, \Bigl( \frac{113}{2} \, \ln u + 20 \, \ln v - 54 \, \ln w \Bigr)
    + 98 \, \ln u \, H_2^v - \frac{37}{6} \, \ln^3 u
\nonumber\\&&\hskip1.8cm\null
    + \ln^2 u \, ( 8 \, \ln v + 52 \, \ln w )
    - 16 \, \ln u \, \ln^2 v \biggr]
\nonumber\\&&\hskip0.9cm\null
 - \zeta_3 \, ( 50 \, H_2^u + 15 \, \ln^2 u )
 - 474 \, \zeta_4 \, \ln u    \biggr\} \,.  \label{Byv}
\eea
%


\end{document}